\documentclass[aps,superscriptaddress,nofootinbib,showpacs,eqsecnum,preprint,tightenlines]{revtex4}
\usepackage{hyperref}
\usepackage{epsfig,rotating}
\usepackage{amsmath,amssymb}
\usepackage{dsfont}
\numberwithin{equation}{section}

\newcommand{\vp}{\vec{p}}

\newcommand{\vk}{\vec{k}}

\newcommand{\ta}{\tilde{a}}
\newcommand{\tFp}{\widetilde{F}}
\newcommand{\tfu}{\widetilde{f}}
\newcommand{\td}{ {\delta}}
\newcommand{\tdfi}{\widetilde{\phi}}

\newcommand{\kpm}{\frac{\vec{k}\cdot \vec{p}}{m}}

\newcommand{\be}{\begin{equation}}
\newcommand{\ee}{\end{equation}}
\newcommand{\bea}{\begin{eqnarray}}
\newcommand{\eea}{\end{eqnarray}}

\begin{document}

\title{Small scale aspects of warm dark matter :\\ power spectra and  acoustic oscillations.}

\author{Daniel Boyanovsky}
\email{boyan@pitt.edu} \affiliation{Department of Physics and
Astronomy, University of Pittsburgh, Pittsburgh, PA 15260}

\author{Jun Wu}\email{juw31@pitt.edu}  \affiliation{Department of Physics and
Astronomy, University of Pittsburgh, Pittsburgh, PA 15260}

\date{\today}

\begin{abstract}
We provide a semi-analytic study of the small scale aspects of the power spectra
of warm dark matter (WDM) candidates that decoupled while relativistic with arbitrary
distribution functions. These are characterized by two widely different scales
$k_{eq} \sim 0.01\,(\mathrm{Mpc})^{-1}$ and $k_{fs}= \sqrt{3}\,k_{eq}/2\,\langle V^2_{eq} \rangle^\frac{1}{2} $
 with $\langle V^2_{eq} \rangle^\frac{1}{2} \ll 1 $ the velocity dispersion at matter radiation equality.
  Density perturbations evolve through three stages: radiation domination when the particle is
  relativistic and non-relativistic and   matter domination. An early ISW effect during the first
   stage leads to an enhancement of density perturbations   and a  plateau in the transfer function
    for $k \lesssim k_{fs}$. An effective fluid description emerges at small scales
      which includes the effects of free streaming in initial conditions and inhomogeneities.
      The transfer function features \emph{WDM-acoustic oscillations} at scales $k \gtrsim 2 \,k_{fs}$.
      We study the power spectra for two models of sterile neutrinos with $m \sim \,\mathrm{keV}$
       produced non-resonantly, at the QCD and EW scales respectively. The latter case yields
       acoustic oscillations on mass scales $\sim 10^{8}\,M_{\odot}$.
       Our results reveal a \emph{quasi-degeneracy} between the mass,
       distribution function and decoupling temperature  suggesting caveats on the constraints on the
       mass of a sterile neutrino from current WDM N-body simulations and   Lyman-$\alpha$ forest
       data. A simple analytic interpolation of the power spectra between large and small scales and its numerical implementation is given.
\end{abstract}

\pacs{98.80.Cq; 98.80.-k; 98.80.Bp}

\maketitle

\section{Introduction}\label{sec:intro}

In the \emph{concordance} $\Lambda\mathrm{CDM}$ standard cosmological
model  dark matter (DM) is composed of primordial particles which are
cold and collisionless\cite{primack}.  In this  cold dark matter (CDM) scenario  particles
feature negligible small velocity dispersion  leading to a power spectrum that favors small
scales.
  Structure formation proceeds in a hierarchical ``bottom up''
approach: small scales become non-linear and collapse first and
their merger and accretion leads to structure on larger scales, dense clumps that survive the
merger process form satellite galaxies.

Large scale simulations seemingly yield  an over-prediction of
satellite galaxies\cite{moore2} by almost an order of magnitude
larger than
 the number of satellites that have been observed in Milky-Way
sized galaxies\cite{kauff,moore,moore2,klyp,will}. Simulations
within the $\Lambda$CDM paradigm also yield a density profile in
virialized (DM) halos that
increases monotonically towards the
center\cite{frenk,dubi,moore2,bullock,cusps} and features a cusp, such as the
Navarro-Frenk-White (NFW) profile\cite{frenk} or more general
  central density profiles $\rho(r) \sim r^{-\beta}$ with
$1\leq \beta \lesssim 1.5$\cite{moore,frenk,cusps}. These density
profiles accurately describe clusters of galaxies but there is an
accumulating body of observational
evidence\cite{dalcanton1,van,swat,gilmore,salucci,battaglia,deblok,cen,wojtak}
that suggest that the     central regions of (DM)-dominated
dwarf spheroidal satellite (dSphs) galaxies
 feature smooth cores instead of cusps as predicted by (CDM).  This difference is known as the core-vs-cusp problem\cite{deblok}.
 Salucci et.al.\cite{salucci2} reported that the mass distribution of spiral disk galaxies can be best fit by a cored Burkert-type
 profile.

 In ref.\cite{cen} a ``galaxy size'' problem has been reported,
 where large scale simulations at $z=3$ yield galaxies that are too
 small, this problem has been argued to be related to that of the
 missing dwarf galaxies.

 Thus there seems to be emerging evidence that the $\Lambda CDM$ paradigm for structure formation \emph{may} have problems at small scales\cite{primack2}.

Warm dark matter (WDM) particles were
invoked\cite{mooreWDM,turokWDM,avila} as possible solutions to the
discrepancies both in the over abundance of satellite galaxies and
as a mechanism to smooth out  the cusped   density profiles
predicted by (CDM) simulations into the  cored profiles that fit the
observations in   (dShps). (WDM) particles feature a range of
velocity dispersion in between the (CDM) and hot dark matter
leading to free streaming scales that smooth  out small scale
features and could be consistent with core radii of the (dSphs). If
the free streaming scale of these particles is  smaller than the
scale of galaxy clusters, their large scale structure properties are
indistinguishable from (CDM) but may affect the  \emph{small} scale
power spectrum\cite{bond} so as to provide an explanation of the
smoother inner profiles of (dSphs), fewer satellites and the size of
galaxies at $z=3$\cite{cen}.

Furthermore recent numerical results hint to more evidence of  possible small scale discrepancies with
the $\Lambda CDM$ scenario: another over-abundance problem, the ``emptiness of voids'' \cite{tikhoklypin} and the spectrum of ``mini-voids''\cite{tikho} both may be explained by
a WDM candidate. Constraints from the luminosity function of Milky Way satellites\cite{maccio} suggest a lower limit of $\sim 1\,\mathrm{keV}$ for a WDM particle, a result consistent with
 Lyman-$\alpha$\cite{lyman,lyman2,vieldwdm}, galaxy power spectrum\cite{abakousha}  and lensing observations\cite{maccio2}. More recently, results from the Millenium-II simulation\cite{sawala} suggest that the $\Lambda CDM$ scenario \emph{overpredicts} the abundance of massive $\gtrsim 10^{10}\,M_{\odot}$ haloes, which is
corrected with a WDM candidate of $m\sim 1\,\mathrm{keV}$.  This body of emerging evidence in
favor of WDM as possible solutions to these potential small scale problems of the $\Lambda CDM$ scenario warrants deeper understanding of their small scale clustering properties.

A model independent analysis suggests that dark matter
particles with a mass in the $\mathrm{keV}$ range is a suitable (WDM)
candidate\cite{hectornorma,hecsal}, and sterile   neutrinos with masses in the $\sim \mathrm{keV}$ range are compelling
(WDM)
candidates\cite{dw,colombi,este,abadwdm,shapo,kusenko,kuse2,kusepetra,petra,kuserev,micharev}.
These neutrinos can decay into an active-like neutrino and an X-ray
photon\cite{xray}, and recent  astrophysical evidence in favor of   a
$5~\mathrm{keV}$ line has been presented in ref.\cite{lowekuse} (see also\cite{boyarski5}). The
analysis in ref.\cite{palazzo} suggests upper mass limits for a
sterile neutrino   in the range $\sim
6-10~\mathrm{keV}$. Possible  \emph{direct}
detection signals of such candidates have been recently assessed in
ref.\cite{kusedirect}.

 A   property of a dark
matter candidate relevant for structure formation    is its distribution function after
decoupling\cite{hogan,coldmatter,darkmatter,boysnudm}. It depends
on the production mechanism and the (quantum) kinetics of its
evolution from production to decoupling. There are different production mechanisms of
  sterile neutrinos\cite{dw,colombi,este,abadwdm,shapo,kuse2,boysnudm}, leading in general to non-thermal distribution
  functions.
 There is some  tension between
the X-ray\cite{xray} and Lyman-$\alpha$ forest\cite{lyman,lyman2,vieldwdm}
data if sterile neutrinos are produced via the Dodelson-Widrow
(DW)\cite{dw} non-resonant mixing  mechanism, leading to the
suggestion\cite{palazzo} that these may not be the dominant (DM)
component. Constraints from the Lyman-$\alpha$ forest spectra are
particularly important because of its sensitivity to the suppression
of the power spectrum by free-streaming in the linear
regime\cite{lyman,lyman2,vieldwdm}. The most recent constraints from the
Lyman-$\alpha$ forest\cite{lyman2,vieldwdm} improve upon previous ones, but
rely on the(DW)\cite{dw} model for the distribution function of
sterile neutrinos, leaving open the possibility of evading these
tight constraints with non-equilibrium distribution functions from
other production mechanisms, such as those studied in
refs.\cite{boysnudm,jun}.

The gravitational clustering properties of collisionless (DM) in the
linear regime are described by the   power spectrum of gravitational
perturbations. Free streaming of collisionless (DM)  leads to a
suppression of the transfer function on length scales smaller than
the free streaming scale via Landau
damping\cite{bond,bondszalay,kt}. This scale is   determined by the
decoupling temperature, the particle's mass and the distribution
function at decoupling\cite{freestream}.

\vspace{2mm}

\textbf{Goals:} The most accurate manner to obtain the transfer function for DM
perturbations is to use the publicly available computer codes for cosmic microwave background
(CMB) anisotropies\cite{cmbfast,camb,cmbeasy}, with modifications that would allow to include the
different distribution functions of the WDM particles.
 These codes include baryons, radiation, neutrinos and DM and yield very
 accurate numerical results. The drawbacks in using these codes for WDM particles are that they do
  not readily yield to an understanding of what aspects of a distribution function influence the small
  scale behavior, and must be modified for the individual   WDM
  candidates because their distribution functions  are ``hard-wired'' in the codes.

The goals of this article are twofold: i) to provide a semi-analytic understanding of the main physical
processes that determine the transfer function   of WDM candidates at
\emph{small scales that entered the horizon well before matter-radiation equality} for
 \emph{arbitrary} distribution functions, ii) to provide a relatively simple formulation of
  the power spectrum   that allows a   straightforward numerical
  implementation, valid  for arbitrary distribution functions.   In order to achieve these goals we must necessarily invoke several approximations: a) we neglect the contribution from baryons, b) we also neglect anisotropic stresses resulting from the free streaming of ultrarelativistic standard model \emph{active} neutrinos. These approximations entail that the results of the transfer functions will be trustworthy up to $10-15\%$ accuracy. However, the main purpose of this work is \emph{not} to obtain the WDM transfer function to a few percent accuracy, but to provide a semi-analytic ``tool'',  to study the main features of the transfer function \emph{at small scales}  for a particular WDM candidate given its distribution function determined by the microscopic process of production and decoupling.  If the transfer function features important small scale properties that could potentially lead to substantial changes in structure formation, this would warrant more accurate study with the CMB codes and eventual inclusion into N-body simulations.

In this article we study the transfer function for WDM density and gravitational perturbations by solving the collisionless Boltzmann equation in a \emph{radiation and matter} dominated cosmology including the perturbations from the radiation fluid for arbitrary distribution function of the WDM particle, thus the results (within the acknowledged possible errors) are valid for $z > 2$.

\vspace{2mm}

\textbf{Strategy:}

WDM particles with a mass in the $\sim \mathrm{keV}$ range typically decouple
from the primordial plasma when they are relativistic. For example sterile neutrinos produced
 non-resonantly via the Dodelson-Widrow (DW)\cite{dw} mechanism or by the
  decay of scalar or vector bosons (BD)\cite{kuse2,kusepetra,boysnudm,jun}
   decouple at the QCD or Electroweak  (EW) scale respectively.
   Therefore these species decouple when they are still relativistic
   in the radiation dominated era and become non-relativistic when $T \approx m \approx \mathrm{keV}$ when the
size of the comoving horizon $\eta \lesssim \mathrm{Mpc}$.

Therefore we anticipate that there are \emph{three stages} of
evolution for density perturbations: I) when the particle is still
relativistic, this is a radiation dominated (RD) stage, II) when the
particle is non-relativistic but still during the (RD) era, III)
when matter perturbations dominate the gravitational potential (the
particle is non-relativistic in this era).  When the WDM particles
are relativistic, their contribution to the total radiation
component is negligible because their effective number of degrees of
freedom is $\ll 1$ (see below). Therefore during   stages I) and II)
the gravitational potential is completely determined by the
radiation fluid. Our strategy is to solve the Boltzmann equation for
WDM density perturbations in the three stages. In stages I) and II)
the gravitational potential is completely determined by the
radiation fluid and the Boltzmann equation is solved by considering
the gravitational potential as a \emph{background} determined by the
Einstein-Boltzmann equation for the radiation fluid. In stage III)
when matter perturbations dominate, the 00-Einstein equation for
small scale perturbations is equivalent to Poisson's equation. The
initial conditions for the Boltzmann equation are given deep in the
(RD) era when the relevant modes are well outside the horizon. In
this work we consider adiabatic initial conditions determined by the
primordial perturbations seeded during inflation.  The main strategy
is to use the solution of the integration of the Boltzmann equation
in a previous stage as the \emph{initial condition} for the next
stage.   During stage I) suppression by free streaming is
independent of the distribution function and the free streaming
scale grows with the comoving horizon. However we find that modes
that enter the horizon when the particle is relativistic with
wavelengths up to the sound horizon are amplified via an early
integrated Sachs-Wolfe effect (ISW) as a consequence of the time
dependence of the gravitational potential produced by acoustic
oscillations of the radiation fluid. The evolution of WDM density
perturbations at the end of this stage determine the initial
conditions for stage II) when the particle becomes non-relativistic
but still the gravitational potential is determined by the
perturbations in the radiation fluid. During this stage the free
streaming scale depends only logarithmically on the comoving
horizon. Whereas CDM perturbations grow logarithmically during this
stage (Meszaros effect), WDM perturbations are suppressed by a free
streaming function that depends on the distribution function of the
decoupled WDM particle. In stage III) when WDM perturbations
dominate the gravitational potential, density perturbations obey a
self-consistent Boltzmann-Poisson integral equation which we analyze
in a systematic expansion valid for small scales.

\textbf{The main results are}:

\begin{itemize}

\item{There are two relevant scales: $k_{eq} \sim 0.01\,(\mathrm{Mpc})^{-1}$ which is the wavevector of modes that enter the Hubble radius at matter-radiation equality, and the free streaming scale $$ k_{fs} = \frac{\sqrt{3}\,k_{eq}}{2\,\langle V^2_{eq}\rangle^\frac{1}{2}} $$ where $\langle V^2_{eq}\rangle^\frac{1}{2}$ is the mean square root velocity dispersion of the WDM particle at \emph{matter-radiation equality}. For a WDM candidate with $m \sim \mathrm{keV}$  produced non-resonantly and    decoupling either at the electroweak or QCD scale $k_{fs} \gtrsim 10^3 \,k_{eq}$. The free streaming length scale $1/k_{fs}$ is proportional to the distance traveled by a  non-relativistic particle with average velocity $\langle V^2_{eq}\rangle^\frac{1}{2}$ from matter-radiation equality until today, \emph{and} it also determines the size of the (comoving) horizon (conformal time) when the WDM particle transitions from relativistic to non-relativistic: $$\eta_{NR} = \frac{\sqrt{3}}{\sqrt{2} \, k_{fs}}\,.$$ This means that perturbations with $k > k_{fs}$ enter the horizon when the WDM particle is still relativistic and undergo suppression by relativistic free streaming between the time of horizon entry until $\eta_{NR}$.  }

    \item{During the   (RD) era  acoustic oscillations in the radiation fluid determine the gravitational potential $\phi$. The time dependence of $\phi$ induces an early ISW that results in an \emph{enhancement} of the amplitude of WDM density perturbations for wavelengths larger than the sound horizon of the radiation fluid at $\eta_{NR}$, namely $\eta_{NR}/\sqrt{3}$,  but  those with $k\eta_{NR}/\sqrt{3} \gg 1$ are suppressed by relativistic free streaming. }

   \item{In stage III), we turn the Boltzmann-Poisson equation into a self-consistent
   differential integral equation that admits a systematic Fredholm series solution.
    Its  leading term is the Born approximation and lends itself to a simple and straightforward
     numerical analysis for arbitrary distribution functions. This approximation is equivalent
     to a fluid description
     \emph{ but with an inhomogeneity and initial conditions completely determined by the past
      history during stages I) and II)}.
       The resulting fluid equation is a
       WDM generalization of    Meszaros equation\cite{mesaros,groth,peeblesbook}.
        The solutions describe \emph{WDM acoustic oscillations},
        the suppression by free streaming is manifest in the inhomogeneity
         and initial conditions.  }

         \item{  In the Born approximation we obtain
         a semi-analytic expression for the transfer function and compare
          it to the CDM case. We also provide an  expression for
          the power spectra that interpolates between large and small scales
           and give a concise summary for its
            numerical evaluation for arbitrary distribution functions, mass and decoupling temperature.  }

       \item{We study  the transfer functions and power spectra for two different scenarios of sterile neutrinos
         produced non-resonantly: via the (DW) mechanism\cite{dw} and via boson
         decay\cite{boysnudm,jun}. The transfer functions are very different \emph{even for the same mass}.
         The ISW amplification of density perturbation enhances the transfer function for wavelengths
         larger than the free streaming length and is more pronounced for the colder species (BD).
         WDM acoustic oscillations are manifest in both transfer functions for $k \gtrsim 2 k_{fs}$.
         We analyze the main physical aspects of these oscillations and suggest
         that their amplification by non-linear gravitational collapse \emph{may} lead to clumpiness
         on mass scales $\sim 10^{8}\,M_{\odot}$ for the colder species . We compare the results from the semi-analytic
         approach for (DW) sterile neutrinos with the transfer function obtained by numerical integration of
         Boltzmann codes in refs.\cite{vieldwdm,hansendwdm,abadwdm}. The results of the Born approximation agree to $< 5\%$ with the numerical fit to the transfer function provided in ref.\cite{vieldwdm} in the region where the fit is valid.}

     \item{ An important   corollary of the study is a \emph{quasi-degeneracy}:  not only the value of the mass
     but also the detailed form of the distribution function
along with the decoupling temperature (in the form of the number of relativistic degrees of freedom
at decoupling) determine the transfer function. Two particles of the
\emph{ same mass but very different distribution functions and decoupling temperatures},
  may feature very
different power spectra. Conversely, two WDM particles of
\emph{different masses} and different distribution functions may
feature similar power spectra on a wide range of scales.

This is studied in detail here through a comparison between sterile
neutrinos produced by the two mechanisms (DW,BD). This result
suggests a caveat in the constraints on the  mass of the (WDM)
particle from current (WDM) simulations and the Lyman-$\alpha$
forest data.}

\end{itemize}

This work differs from that in ref.\cite{komaneutrino} that analyzes (standard model active)
neutrinos as (WDM) but in an Einstein-Desitter cosmology, in two main aspects: i) our study includes
 the (RD) era and the transition to matter domination (MD) including the time dependence of the gravitational
  potential, which is a source of an early ISW effect during stage I) and the history during stages I) and II),
   and ii) we study   non-thermal distribution functions. The inclusion of stages I) and II) during the
(RD) era also distinguishes this work from that in ref.\cite{darkmatter,boysnudm}.

\section{Preliminaries}\label{sec:prelim}

We consider a radiation and matter dominated cosmology:
\be H^2=  \frac{\dot{a}^2}{a^4} = H^2_0 \left[\frac{\Omega_r}{a^4}+ \frac{\Omega_m}{a^3}\right] = \frac{H^2_0 \Omega_m}{a^4}\left[a+a_{eq}\right] \label{H2}\ee
 where the dot stands for derivative with respect to conformal time ($\eta$), the scale factor is normalized to $a_0=1$ today,  and
\be a_{eq} = \frac{\Omega_r}{\Omega_m} \simeq \frac{1}{3229}
\,.\label{aeq}\ee Introducing \be \ta = \frac{a}{a_{eq}}\,,
\label{tildea}\ee it follows that \be \frac{d\,\ta}{d\eta} =
\left[\frac{H^2_0 \Omega_m}{a_{eq}}
\right]^{\frac{1}{2}}\,\left[1+\ta\right]^{\frac{1}{2}}
\label{dotta}\ee leading to \be \eta = \frac{2}{\left[\frac{H^2_0
\Omega_m}{a_{eq}} \right]^{\frac{1}{2}}}\Big[\sqrt{1+\ta}-1 \Big]
\equiv 288.46 \,\Big[\sqrt{1+\ta}-1 \Big]
 \,(\mathrm{Mpc})\,, \label{eta}\ee where we have used $\Omega_m h^2 = 0.134
 $\cite{WMAP5}.
At matter-radiation equality  we define \be k_{eq} \equiv H_{eq}\,a_{eq} = \sqrt{2}\left[\frac{H^2_0 \Omega_m}{a_{eq}} \right]^{\frac{1}{2}}  =
 \frac{9.8 \times 10^{-3}}{\mathrm{Mpc}} \label{keq}\ee corresponding to the comoving wavevector that enters the Hubble radius at matter-radiation equality. Furthermore from (\ref{eta})  we find the comoving size of the horizon at matter-radiation equality,
  \be \eta_{eq} = \frac{2 \sqrt{2}(\sqrt{2}-1)}{H_{eq}a_{eq}} \simeq \frac{1.172}{H_{eq}a_{eq}} \simeq 120\,\mathrm{Mpc}\label{etaeq}\ee from which we obtain
  \be \ta= \frac{\eta}{\eta^*} \Bigg[1 + \frac{\eta}{4\eta^*} \Bigg] ~~;~~ \eta^*
 =  \frac{\eta_{eq}}{2(\sqrt{2}-1)} = \frac{\sqrt{2}}{k_{eq}} \,. \label{taofeta}\ee

During radiation domination \be \ta \approx \Big(\frac{\eta}{\eta^*} \Big) \ll 1 \,, \label{RD}\ee and in this regime
\be \eta \simeq \ta ~144.23~(\mathrm{Mpc})   \,.\label{etaRD}\ee

 During matter-radiation domination, a comoving wavevector $k$ enters the (comoving)  Hubble radius when $k = H a$ corresponding
 to a value of the scale factor
\be \ta_k = \frac{1+ \sqrt{1+8 \big( \frac{k}{k_{eq}}\big)^2}}{4 \big( \frac{k}{k_{eq}}\big)^2} \,.\label{tak}\ee

We are interested in small scale properties for perturbations with comoving wavelenghts
$100 \,\mathrm{pc} \leq \lambda \equiv 2\pi/k  < 10\, \mathrm{Mpc}$ corresponding to $k\gg k_{eq}$. For these modes, which have entered the horizon during the radiation dominated era,  it follows that \be \ta_k \sim \sqrt{2}~ \frac{k_{eq}}{k} \ll 1 \,.\label{sma}\ee

A  weakly interacting massive particle (WIMP) of mass $m \sim 100 ~\mathrm{GeV}$ that  undergoes chemical freeze-out at $T_{ch} \sim m/20 $ and thermal decoupling
at $T_{d} \sim 10 ~ \mathrm{MeV}$ when $\ta_d \sim 10^{-7}$, and $\eta_d \sim 10~\mathrm{pc}$, i.e, deep in the (RD) era, is non-relativistic at decoupling. Scales $\lesssim \eta_d$ where inside the horizon when
the DM particle  was still coupled to the cosmological plasma and acoustic oscillations of the photon
fluid are imprinted on the transfer function at these very small scales\cite{loeb}.
 However, larger scales were outside the horizon and their perturbations are frozen, they enter
 the horizon after decoupling and their evolution is described by the collisionless Boltzmann equation.

On the other hand, sterile neutrinos with mass $m \sim \mathrm{keV}$ decoupled thermally
at much higher temperature ($\sim 150~\mathrm{MeV}$ for   (DW)\cite{dw}, $\sim 100 ~\mathrm{GeV}$
for production via scalar or vector boson decay\cite{boysnudm,jun}),
 and become non-relativistic at $T \sim m \sim ~\mathrm{keV}$, namely for $\ta \sim 10^{-3}$.
  In terms of conformal time, $m\sim \mathrm{keV}$ sterile neutrinos become non-relativistic at
   \be \eta_{NR} \sim 0.2 ~\mathrm{Mpc}\, \Bigg(\frac{\mathrm{keV} }{m}\Bigg) \,,\label{etaNR}\ee
    so that for $\eta \ll \eta_{NR}$ this (DM) candidate is
relativistic and non-relativistic for $\eta > \eta_{NR}$.
Therefore, for DM candidates that decoupled for temperatures $T_d \gtrsim 10 \,\mathrm{MeV}$
all modes of cosmological
relevance for (comoving) scales $\lambda \gtrsim 50 \,\mathrm{pc}$ may be studied in
the linear regime via the collisionless Boltzmann-Vlasov equation.
  A firmer estimate will be provided in section (\ref{subsec:freestream}).

For WDM particles with $m \sim \mathrm{keV}$  we see from eqns. (\ref{sma}, \ref{keq}) that comoving scales $\lambda  \gtrsim 0.2 \,\mathrm{Mpc}$ entered the horizon when the DM particle is non-relativistic, whereas smaller scales entered during the radiation dominated stage when the WDM particle is \emph{relativistic}. Therefore comoving scales smaller than that of cluster of galaxies became sub-horizon during (RD) when the WDM particle is still
 \emph{relativistic}.
This is important because free streaming changes   from the relativistic to the non-relativistic case: during the relativistic stage  the   free streaming length is of the order of the    horizon,  but much smaller during the non-relativistic stage (see below).

Hence  as anticipated above, there are three  distinct stages of
evolution of density perturbations for WDM particles with $m \sim
\mathrm{keV}$ and scales smaller than $0.2-1\,\mathrm{Mpc}$:
\begin{itemize}
\item{\textbf{I})
 (RD), relativistic   $\eta < \eta_{NR}$,}
  \item{\textbf{ II}) (RD), non-relativistic $\eta_{eq}> \eta > \eta_{NR}$, }
  \item{\textbf{III}) matter domination (MD), non-relativistic for $\eta \geq  \eta_{eq}$. }
 \end{itemize}

\section{Evolution of perturbations: the Boltzmann equation}\label{sec:BE}

We follow the notation of Ma and Bertschinger\cite{ma} (see
also\cite{dodelson,giova,lythbook,weinbergbook,ruthbook}), and
consider only scalar perturbations in the conformal Newtonian gauge
(longitudinal gauge) with a perturbed metric   \bea g_{00} & = &
-a^2(\eta)\Big[1+2\psi(\vec{x},\eta) \Big] \label{g00}\\ g_{ij} & =
& a^2(\eta)\Big[1-2\phi(\vec{x},\eta)\Big] ~\delta_{ij} \,.
\label{gij}\eea The perturbed distribution function is given by \be
f(p,\vec{x},\eta) = f_0(p)+F_1(p,\vec{x},\eta) \label{f}\ee where $
f_0(p) $ is the unperturbed distribution function,    which after
decoupling obeys the collisionless Boltzmann equation in absence of
perturbations and $\vec{p},\vec{x}$ are   comoving momentum and
coordinates respectively. As discussed in
ref.\cite{coldmatter,darkmatter,boysnudm} the unperturbed
distribution function is of the form \be f_0(p) \equiv
f_0(y;x_1,x_2,\cdots) \label{fform}\ee where \be y =
\frac{p}{T_{0,d}} \label{y} \ee where $p$ is the comoving momentum
and $T_{0,d}$ is the decoupling temperature \emph{today},  \be
T_{0,d} = \Big(
\frac{2}{g_d}\Big)^{\frac{1}{3}}~T_{CMB}\,,\label{T0d}\ee with $g_d$
being the effective number of relativistic degrees of freedom at
decoupling, $T_{CMB}= 2.35 \times 10^{-4}~\mathrm{eV}$ is the
temperature of the (CMB) today, and  $x_i$ are dimensionless
couplings or ratios of mass scales.

Although our study will be carried out for arbitrary $f_0$, we will
analyze in detail two candidates for WDM: sterile neutrinos produced
by the Dodelson-Widrow (DW) (non-resonant) mechanism for which \be
f_0(p) = \frac{\beta}{e^y+1} \label{DWf}\ee where $\beta \simeq
  10^{-2}$\cite{dw}, and sterile neutrinos produced near the electroweak scale by the
decay of a scalar with a mass of the order of the EW scale or vector
bosons (BD), which are abundant at temperatures near the EW scale\cite{boysnudm,jun},

\be f_0(p) = \lambda ~\frac{g_{5/2}(y)}{\sqrt{y}} ~~;~~ g_{5/2}(y) = \sum_{n=1}^\infty
\frac{e^{-n\,y}}{n^{\frac{5}{2}}}  \label{SDf}\ee and $\lambda \sim 10^{-2}$\cite{boysnudm,jun}. We will
compare the results for the WDM distributions with that for  weakly interacting massive particles (WIMPs) which freeze-out with
a Maxwell-Boltzmann (MB) distribution,

\be f_0(p) = \mathcal{N}~e^{-\frac{y^2}{2x} } ~~;~~ x = \frac{m}{T_d} \label{MBf}\ee where
$m\sim 100~\mathrm{GeV}$, $T_d\sim 10~\mathrm{MeV}$ is the thermal decoupling temperature, and $\mathcal{N}$ is determined at
chemical freeze-out\cite{kolb}.

An important observation for   WDM candidates is that during the radiation dominated
era when these are relativistic, their contribution to the energy density is \be \rho = \frac{1}{a^4} \int
\frac{d^3p}{(2\pi)^3}~p\,f_0(p) \propto T^4(t) \times \Bigg\{\begin{array}{c} \beta
 ~~(\mathrm{DW}) \\
                                                                           \lambda
                                                                           ~~
                                                                           (\mathrm{BD})
                                                                         \end{array}
\label{radcont}\ee for sterile neutrinos produced by the
Dodelson-Widrow (DW) or scalar decay (BD) mechanisms. Namely these
WDM candidates contribute to the radiation component with an
effective number of degrees of freedom proportional to $
\beta,\lambda, \sim  10^{-2} $   and \emph{can be safely neglected in
their contribution to the radiation component}. The same argument justifies neglecting the anisotropic stress (quadrupole moment)
arising from the free streaming of these particles when they are relativistic.

Introducing spatial Fourier transforms in terms of comoving momenta $\vec{k}$ (we keep the same notation for the spatial Fourier transform of perturbations), the  linearized Boltzmann equation for perturbations is given by\cite{ma,dodelson,giova,lythbook,weinbergbook,ruthbook}

\be \dot{F_1}(\vk,\vp\,;\eta)+ i   \frac{k\,\mu\,p}{\epsilon(p,\eta)}~F_1 (\vk,\vp\,;\eta)+
 \Big(\frac{d~f_0(p)}{dp} \Big)\Big[p~\dot{\phi}(\vk,\eta) - ik\,\mu \, \epsilon(p,\eta)~\psi(\vk,\eta) \Big] =0 \label{BE}\ee where $\mu = \widehat{\mathbf{k}}\cdot \widehat{\mathbf{p}}$, dots stand for derivative with respect
 to conformal time $\eta$ and \be  \epsilon(p,\eta) = \sqrt{p^2+m^2\,a^2(\eta)} \label{eps}\ee is
  the conformal energy of the particle of mass $m$.  During (RD) and  (MD), the $00$ component
   of Einstein's equation in conformal Newtonian gauge is\cite{dodelson} \be    \phi(\vk,\eta)
 +3 \frac{\mathcal{H}}{k} \Bigg(\frac{1}{k}\,\dot{\phi}(\vk,\eta)+\frac{\mathcal{H}}{k} \psi(\vk,\eta) \Bigg)   =  -\frac{3}{4} \frac{k^2_{eq}}{k^2\, \ta^2} \Bigg[ \ta\,\Big(\frac{\delta \rho}{\rho}\Big)_m +   \Big(\frac{\delta \rho}{\rho}\Big)_r \Bigg]   \,, \label{Einstein00}\ee
  where \be \mathcal{H} = \frac{\dot{\ta}}{\ta} = a H  = k_{eq}\,\frac{[1+\ta]^\frac{1}{2}}{\sqrt{2}\,\ta} \label{hubrad} \ee is the inverse comoving Hubble radius, and
 \be  \delta \rho_{j}(\vk,\eta)  = \frac{1}{a^4} ~\int \frac{d^3p}{(2\pi)^3} ~ \epsilon(p,\eta)~F_{1,j}(\vk,\vp,\eta) ~~;~~ j=r,m\,.\label{delrho}\ee

 In what follows we neglect stress anisotropies leading to \be \phi(\vk,\eta) = \psi(\vk,\eta)\,,
 \label{nostress}\ee thereby neglecting the quadrupole moment from relativistic standard model (active)
 neutrinos. We also neglect the baryonic component in the
 matter contribution, a compromise that   allows us to pursue a semi-analytic understanding of the (DM)
 transfer function at small scales.   The remaining Einstein's equations are not necessary for the
 discussion that follows.
 In absence of stress anisotropy,  Einstein's equation (\ref{Einstein00}) can be written in another useful form,
 \be \frac{2}{3} \frac{k^2 \ta^2}{k^2_{eq}}\,\phi+(1+\ta) \Big(\ta \,
\phi\Big)^{'}   = -\frac{1}{2}\Bigg[ \ta\,\Big(\frac{\delta \rho}{\rho}\Big)_m +   \Big(\frac{\delta \rho}{\rho}\Big)_r \Bigg]
\label{Einstein002}\ee where \be {}^{'} \equiv \frac{d}{d\ta}\,.
\label{deratil}\ee

 The formal solution of the Boltzmann  equation (\ref{BE}) is

 \be F_1(\vk,\vp\,;\eta) = F_1(\vk,\vp\, ;\eta_i)~e^{-ik\,\mu\,l(p,\eta,\eta_i)} -
 p\,\Big(\frac{d\,f_0(p)}{dp} \Big)~\int_{\eta_i}^{\eta} d\tau ~ e^{-ik\,\mu\,l(p,\eta,\tau)}
 \Bigg[\frac{d\phi(\vk,\tau)}{d\tau}-i~\frac{k\,\mu}{V(p,\tau)}~ \phi(\vk,\tau) \Bigg]\label{BEsol}\ee where
 \be l(p,\eta,\eta') = \int_{\eta'}^{\eta}    V(p,\tau)~d\tau  ~~;~~V(p,\tau) = \frac{p}{\epsilon(p,\tau)}
  \label{lFS}\ee is the comoving \emph{free streaming} distance that  a particle travels between $\eta'$ and $\eta$ with \emph{physical} velocity
 $V(p,\tau) = p/\epsilon(p,\tau)$.

The solution (\ref{BEsol}) with (\ref{lFS})  is the starting point
of our analysis. The density and gravitational perturbations
produced by a WDM particle with $m\sim \mathrm{keV}$ that decouples
from the plasma when it is still relativistic are obtained by
evolving the solution (\ref{BEsol}) through the  three  stages : I)
when the DM particle is still relativistic during (RD), II)   when
the particle becomes non-relativistic for $\ta
 \gtrsim 10^{-3}$ but still during (RD), III) during
(MD) $\ta \geq 1$ (the DM particle is non-relativistic).

 During the
first two stages the perturbation  in the gravitational potential
$\phi$ in (\ref{BEsol}) is completely determined by the radiation
component to which the WDM candidate contributes negligibly as
discussed above. The difference between  stages I) and II)  is
manifest in the free streaming distance $l(p,\eta,\eta')$. During
stage III) the gravitational potential is determined by the DM
density perturbations self-consistently through Poisson's equation
(this is the advantage of the conformal Newtonian gauge). Our
strategy is to determine initial conditions deep in the radiation
era when the cosmologically relevant modes are still superhorizon,
and to evolve the solution (\ref{BEsol}) through each of these
stages, using the distribution function at the end of each stage as
the initial condition for the next stage, thereby propagating the
initial condition determined deep in the radiation era to
matter-radiation equality.

 \vspace{2mm}

 \subsection{Free streaming distance:}\label{subsec:freestream}

  The free streaming distance $l(p,\eta,\eta')$ can be obtained analytically with (\ref{taofeta}), the general result can be expressed in terms of elliptic functions, however it is
 unyielding and not very illuminating. It simplifies considerably in two relevant cases: for  radiation domination when $\eta \ll \eta_{eq}$ which includes the era when the DM candidate
 becomes non-relativistic,   and in the non-relativistic
 regime for $\eta \gg \eta_{NR}$ which includes the matter dominated era.

 \vspace{2mm}

 \textbf{Radiation domination (RD)}:

 \vspace{2mm}

Since $f_0(p)$ is a function of $y=p/T_{0,d}$ it is convenient to write $p=y T_{0,d}$ in
 $V(p)$. In the radiation dominated era $\eta \ll \eta_{eq}$ during which $a(\eta) \sim \eta/\eta^*$  we find
 \be k\, l(p,\eta,\eta') = \frac{\alpha\,  {y}}{2}     \, \ln\Bigg[ \frac{z+\sqrt{\frac{y^2 \alpha^2}{4}+z^2}}{z'+\sqrt{\frac{y^2 \alpha^2}{4}+z^{'\,2}}}\Bigg] ~~;~~z = k\eta \label{klfs}\ee where we have introduced \be \alpha = 2 \sqrt{2}~\frac{k\,T_{0,d}}{m\,k_{eq}a_{eq}} \simeq 2.15 \times 10^{-3}\, \Big(\frac{k}{k_{eq}}\Big)\,\Big(\frac{2}{g_d}\Big)^{\frac{1}{3}}\,\Big( \frac{\mathrm{keV}}{m} \Big)\simeq 0.22 \,k \,\Big(\frac{2}{g_d}\Big)^{\frac{1}{3}}\,\Big( \frac{\mathrm{keV}}{m} \Big)\times \big(\mathrm{Mpc}\big)\,.\label{alfa}\ee

 Since for the WDM distributions under consideration $y^2 f(y)$ is strongly peaked at
 $y \sim \sqrt{\overline{y^2}}$ where
 \be  \overline{y^2}  = \frac{\int_0^\infty y^4 f_0(y) dy}{\int_0^\infty y^2 f_0(y) dy} =  \left\{
 \begin{array}{l}
   \frac{105}{12}\,\frac{\zeta(7)}{\zeta(5)} \simeq 8.505 ~;~~\mathrm{for}~(\mathrm{BD}) \\ \\
  15\,\frac{\zeta(5)}{\zeta(3)} \simeq 12.939 ~;~\mathrm{for}~(\mathrm{DW}~or ~\mathrm{thermal}~\mathrm{fermion}) \\ \\
   3\,x =  3 \, \frac{m}{T_d}~;~~~~~~~~\mathrm{for}~(\mathrm{MB})
 \end{array} \right.
 \label{ybar}\ee
  it follows that for $z,z' \ll \sqrt{\overline{y^2}} \,\alpha$  the ultrarelativistic approximation
   $v(p,\eta) \sim 1$ is valid\footnote{The condition $z\ll \sqrt{\overline{y^2}} \,\alpha$
  is equivalent to $\langle p^2 \rangle \gg m^2 a^2(\eta)$, where the average is with $f_0(p)$. },
 and in this regime \be  l(p,\eta,\eta') =  (\eta-\eta')\,, \label{relimit} \ee which is
 the comoving free streaming distance traveled by an ultrarelativistic particle between $\eta$ and $\eta'$. In the opposite limit when the particle
 is non-relativistic but still in the radiation dominated era $z,z' \gg \sqrt{\overline{y^2}} \,\alpha$ it follows that
  \be k\, l(p,\eta,\eta') = \alpha\, \frac{y}{2}\, \ln\Bigg[ \frac{z }{z' }\Bigg] \,.
  \label{nrRD}\ee

  \vspace{2mm}

  \textbf{Non-relativistic WDM}

  \vspace{2mm}

   From the expression of the conformal energy (\ref{eps}) and the physical velocity $V(p,\tau) $ in
   (\ref{lFS}) we see that the particle is relativistic if $p \gg
   m\,a(\eta)$ and non-relativistic for $p \ll m\,a(\eta)$. Since
   the comoving momentum is integrated over and weighted by the
   distribution function, we define     \be \ta_{NR} = \frac{\langle p^2 \rangle^\frac{1}{2}}{m\, a_{eq}}\;,
    \label{aNR}\ee where the average is taken with the distribution $f_0(p)$ as the value of $\ta$ that determines the
      transition between the relativistic
     and non-relativistic regime, the particle is
  relativistic for $\ta\ll \ta_{NR}$ and non-relativistic for $\ta >
  \ta_{NR}$. When the particle is non-relativistic \be V(p,\eta) =
\frac{p}{m\,a(\eta)}\,.  \label{NRv}\ee therefore \be \ta_{NR} =
\langle  {V}^2(t_{eq}) \rangle^\frac{1}{2}\,. \label{taNR}\ee
Writing $p = y\,T_{0,d}$ we find  \be \langle {V}^2(t_{eq})
\rangle^\frac{1}{2} \simeq 7.59 \,\times 10^{-4}
        \,\sqrt{\overline{y^2}} \,\Big(\frac{\mathrm{keV}}{m} \Big)\,\Big(\frac{2}{g_d} \Big)^\frac{1}{3}
         \,. \label{Veq}\ee

  A weakly interacting massive particle (WIMP) (CDM ) of mass $\sim 100 \,\mathrm{GeV}$ and
  $T_d\sim 10\, \mathrm{MeV}$ features
   $\langle  {V}^2(t_{eq}) \rangle^\frac{1}{2} \simeq 4\,\times 10^{-8}$,
   whereas for  a WDM candidate with $m\sim \mathrm{keV}$
   we find $\langle  {V}^2(t_{eq}) \rangle^\frac{1}{2} \lesssim 10^{-3}$,
    namely all these DM candidates are  non-relativistic at $t_{eq}$
    with $\langle  {V}^2(t_{eq})\rangle \ll 1$. Since the
  WDM particle is non-relativistic at the epoch of matter-radiation equality $\ta_{NR}\ll 1$,
  we find from eqns. (\ref{taofeta},\ref{etaRD}) that
  \be \eta_{NR} = \frac{\sqrt{2}}{k_{eq}}\,\langle  {V}^2(t_{eq})
   \rangle^\frac{1}{2}  \,,  \label{etaNR} \ee for  $\eta \gg \eta_{NR}$ the particle is non-relativistic and relativistic for $\eta \ll \eta_{NR}$.

   Since in the non-relativistic stage the physical velocity is given by (\ref{NRv}), the integral
  in (\ref{lFS}) is easily performed by changing integration variable from $\eta \rightarrow \ta$, we find \be k\,l(p,\eta,\eta') =
   y\,\alpha\, [u -u']\,, \label{klmd}\ee where we introduced
 \be u(\eta) = \frac{1}{2}\, \ln\Bigg[ \frac{\sqrt{1+\ta(\eta)}-1}{\sqrt{1+\ta(\eta)}+1}\Bigg] =
 \frac{1}{2}\, \ln\Bigg[ \frac{\eta}{4\eta^*+
 \eta}\Bigg] ~~;~~ u_{NR} \leq u(\eta) \leq 0 \,,\label{udef}\ee where $\ta_{NR} = \ta(\eta_{NR})$,
 normalized $u(\eta)$ so that $u(\infty) =0$ and introduced
 \be u_{NR} =   \ln\Big[\frac{\sqrt{\ta_{NR}}}{2} \Big]\,.\label{uNR}\ee

  During the radiation era when the WDM particle is non-relativistic, $\ta \ll 1$ we
  find that \be k\,l(p,\eta,\eta') =  \frac{\alpha \, y}{2}\,\ln\Big[ \frac{\eta}{\eta'}\Big] \label{NRradd}\ee
   which reproduces the result (\ref{nrRD}). During the matter dominated era for $\ta \gg 1$ it follows that
 \be u(\eta) \sim   - \frac{1}{\sqrt{\ta(\eta)}} \sim   - \frac{2\eta^*}{ \eta }  \,. \label{mdu}\ee

 \vspace{2mm}

 \textbf{Free-streaming wavevector from fluid analogy}\label{subsec:freestream}

 In analogy with the Jean's wavevector in the fluid description of perturbations during matter domination,  we introduce the \emph{comoving}
  free-streaming wavevector \be k^2_{fs}(t) =  \frac{4\pi G \rho_m(t)}{\langle \vec{V}^2(t) \rangle} \, a^2(t) \label{kfs} \ee
  where \be \rho_m(t) = \frac{\rho_m(0)}{a^3(t)} ~~;~~ \langle \vec{V}^2(t) \rangle =
  \frac{\langle \vec{V}^2(0) \rangle }{a^2(t) } \label{V2}\ee  and
  the value of the velocity dispersion \emph{today} is
   \be \langle \vec{V}^2(0) \rangle =   \overline{y^2} \,\Bigg(\frac{T_{d,0}}{m} \Bigg)^2  \,. \label{sigdisp} \ee

      We note that \be k_{fs}(a_{eq}) \equiv
      \frac{2\pi}{\lambda_{fs}}
      = \frac{\sqrt{3}}{2}\, \frac{k_{eq} }{ \langle \vec{V}^2(t_{eq}) \rangle^\frac{1}{2} }\,, \label{keqkfs}\ee

      Therefore
 for these particles \be k_{fs}(a_{eq}) \gg k_{eq} \,.\label{comparakfskeq}\ee

 We \emph{define} the free streaming wavevector as \be k_{fs} \equiv k_{fs}(a_{eq}) =
 \frac{11.17}{\sqrt{\overline{y^2}}} \, \Big( \frac{m}{\mathrm{keV}}\Big)\,
 \Big( \frac{g_d}{\mathrm{2}}\Big)^\frac{1}{3}\,(\mathrm{Mpc})^{-1}\,.
 \label{kfsdef}\ee This scale will be seen to play a  fundamental role in the (DM)
 transfer function.

 For a $m \sim \mathrm{keV}$ sterile neutrino produced non-resonantly by boson decay (BD) that decoupled near the electroweak scale\cite{boysnudm} ($g_d \sim 100$), it follows that \be k^{BD}_{fs} \sim 14.12 \, (\mathrm{Mpc})^{-1}\,,\label{kfsBD}\ee whereas for a similar mass sterile neutrino produced non-resonantly via the (DW) mechanism near the QCD scale ($g_d \sim 30$) we find \be k^{DW}_{fs} \sim 7.7 \, (\mathrm{Mpc})^{-1}\,.\label{kfsDW}\ee and  for a WIMP of $m \sim 10 \, \mathrm{GeV}$ that decoupled thermally at $T_d \sim 10 \,\mathrm{MeV}$ one finds $k_{fs} \sim 10^6 \, (\mathrm{Mpc})^{-1}$. We will see later that $k_{fs}$  determines the scale of suppression of the transfer function.

  It is convenient   to introduce  \be  \kappa \equiv    \sqrt{\overline{y^2}} ~ \alpha  \equiv
  \,\frac{\sqrt{6}\,k}{k_{fs} }= \sqrt{6} \, \frac{\lambda_{fs}}{\lambda}= 2\,\sqrt{2} \, \frac{k}{k_{eq}}\,\langle \vec{V}^2(t_{eq}) \rangle^\frac{1}{2}\,.\label{kapadef} \ee
  where $\overline{y^2} $ is given by (\ref{ybar}) for the DM species considered
  here, and $\lambda_{fs} = 2\pi/k_{fs}$.  The dimensionless ratio $\kappa$ will  be important in the discussion of non-relativistic DM.

 From (\ref{klmd},\ref{udef}) we find \be k\,l(p,\eta_0,\eta_{eq}) \simeq y\alpha \ln\big[\sqrt{2}+1\big] \label{freestreamtoday}\ee where $\eta_0$ is the conformal time \emph{today}, namely $l(p,\eta_0,\eta_{eq})$ is the free streaming distance traveled by the non-relativistic WDM particle from matter-radiation equality until today. Combining this result with eqn. (\ref{keqkfs}) we find\footnote{The slight discrepancy with the result in ref.\cite{darkmatter} can be traced back to
 the difference between matter only and matter-radiation evolution.  } \be l(p,\eta_0,\eta_{eq}) \simeq 0.344 \,\frac{y}{\sqrt{\overline{y^2}}}~\lambda_{fs}\,. \label{relalfs}\ee From which it follows that
 during matter domination $\lambda_{fs}$, which is the equivalent of the Jeans length for \emph{collisionless} matter perturbations, is simply related to the free streaming distance traveled by  the non-relativistic particle moving with average comoving momentum $\sqrt{\langle p^2 \rangle}$ from the time of matter-radiation equality until today, namely  $\lambda_{fs} \approx 2.9~ l\big(\sqrt{\langle p^2 \rangle},\eta_0,\eta_{eq}\big)$.

  From (\ref{Veq}) and (\ref{uNR}) we find
  \be u_{NR} \simeq -4.27 + \frac{1}{2}\ln\Bigg[\frac{  \sqrt{\overline{y^2}} }{3}\,
  \Big( \frac{\mathrm{keV}}{m}\Big)\,\Big(\frac{50 }{g_d}\Big)^\frac{1}{3} \Bigg]\label{uNRvalue}\,,\ee
  where the argument of the logarithm is $\mathcal{O}(1)$ for $m \sim \mathrm{keV}$ sterile
  neutrinos produced via the (DW) or (BD) mechanisms.

 From eqns. (\ref{taofeta},\ref{etaNR}) we find the relation
  \be \eta_{NR} = \frac{\sqrt{2}}{k_{eq}}\,\langle \vec{V}^2(t_{eq})
   \rangle^\frac{1}{2} = \frac{\sqrt{3}}{\sqrt{2}\,k_{fs}} \,,  \label{etaNR2} \ee
    hence from the definition  of $ \kappa$, eqn. (\ref{kapadef}) and (\ref{etaNR}) it follows that \be k\,\eta_{NR} = \frac{\kappa}{2}  \,. \label{ketaid}\ee
    Therefore comoving modes that entered the horizon when the particle
    is still relativistic correspond to $ \kappa \gtrsim 2  \Rightarrow   k \gtrsim k_{fs} $
    whereas those that entered when the particle is non-relativistic correspond
     to $ \kappa \lesssim 2   \Rightarrow  k \lesssim k_{fs}$. The main corollary  is that
      the free streaming wavelength is \emph{of the order of the size of the horizon at
      the time when the (DM) particle transitions from being relativistic to non-relativistic}.

  This is important: when the particle is relativistic the free streaming distance grows
  with the comoving horizon $\eta$ and free streaming is most efficient to erase density
  perturbations, whereas when the
  particle is non-relativistic, the free streaming distance grows only with the logarithm of
  the comoving horizon and free streaming is less efficient to erase perturbations because the
   particle free streams with a small velocity. Therefore the dimensionless ratio
   $\kappa$
   indicates the regimes in which free streaming is more ($\kappa \gg 2$) or less ($\kappa \ll 2$)
    efficient to suppress density perturbations.

 \subsection{Initial conditions}\label{subsec:inicond}

 Initial conditions are determined deep in the radiation dominated era and when the
 wavelengths are well outside the horizon. We will only consider adiabatic  initial conditions for which \emph{all} the radiation components feature the same $ \delta \rho_r /\rho_r$ and (non-relativistic) matter perturbations obey \be \Bigg( \frac{\delta \rho}{\rho}\Bigg)_m = \frac{3}{4}\,\Bigg( \frac{\delta \rho}{\rho}\Bigg)_r \,. \label{adia}\ee For the radiation
 component temperature perturbations correspond to a perturbation in the distribution function
  \be F_{1,r}(\vk,\vp;\eta_i) = -\Theta(\vk,\eta_i) \,p \Big(\frac{df_{0,r}(p)}{dp}\Big)~~;~~
  \Theta(\vk,\eta_i)= \frac{\Delta T(\vk,\eta_i)}{T_0} \label{radinicon}\ee so that
   \be \Bigg( \frac{\delta \rho}{\rho}\Bigg)_{i,r} = 4\Theta(\vk,\eta_i) \,.\label{delrho}\ee
    For superhorizon perturbations when perturbations in the radiation component are nearly
    constant the temperature anisotropy is determined by the Newtonian
    potential\cite{dodelson,giova,ma}
    \be \Theta(\vk,\eta_i) = - \frac{1}{2}\,\phi_i(k) ~~; k\,\eta_i \ll 1 \,.\label{inisup}\ee
     Initial conditions for adiabatic perturbations of the \emph{matter}
 component \emph{also} correspond to \be F_{1,m}(\vk,\vp;\eta_i) = -\Theta(\vk,\eta_i) \,p
 \Big(\frac{df_{0,m}(p)}{dp}\Big) \label{inimat}\ee
 which leads to
 \be \frac{\delta \rho_m (\vk,\eta_i)}{\rho_m} =
  - \Theta(\vk,\eta_i)~\frac{\int p^3 \,\Big(\frac{d f_{0,m}(p)}{dp}\Big)\, dp}{\int p^2 \,f_{0,m}(p)\, dp}
  = 3\,\Theta(\vk,\eta_i) =
   \frac{3}{4}\, \Bigg( \frac{\delta \rho}{\rho}\Bigg)_{i,r} \label{delrhomat}\ee

 The subtlety for WDM candidates is that in setting up initial conditions for superhorizon
 fluctuations, small comoving scales are superhorizon when the WDM candidate is \emph{relativistic} and intermediate and large comoving scales are superhorizon when the
 particle has become non-relativistic. However, adiabatic initial conditions for \emph{all} modes are determined by (\ref{inimat}). Indeed, when the WDM candidate is relativistic such
 initial condition yields an energy density perturbation which is adiabatic for a radiation
 component and when the particle is non-relativistic it gives the corresponding relation (\ref{adia}).
  Therefore adiabatic initial conditions for all modes (superhorizon at the
 initial time $\eta_i$) for the WDM perturbations are
 \be F_1(\vk,\vp;\eta_i) = \frac{1}{2}\,\phi_i(k) \,p \,\Big(\frac{df_0(p)}{dp}\Big)~~;~~k\,\eta_i \ll 1\,, \label{iniwdm} \ee where
 $f_0(p)$ is the unperturbed distribution function for the DM candidate, and $\phi_i(k)$ is the primordial gravitational
 potential determined during inflation.

 In what follows it is convenient to \emph{define}
\be \tFp(\vk,\vp;\eta) = \frac{F_1(\vk,\vp;\eta)}{n_0}~~;~~\tfu (p) = \frac{f_0(p)}{n_0} \label{tildes} \ee where \be n_0 = \int \frac{d^3p}{(2\pi)^3}
 \,f_0(p)\,, \label{n0}\ee is the density of (DM) \emph{today}. Furthermore, we introduce  \be \td(\vk,\eta) = \int \frac{d^3p}{(2\pi)^3} \,\tFp(\vk,\eta)\,. \label{tildelta} \ee which becomes $\delta \rho_{m}/\rho_m$ after the DM particle becomes non-relativistic, its initial
 condition is  \be \td_i(k)\equiv \td(\vk,\eta_i) = -\frac{3}{2}\phi_i(k)~~;~~\mathrm{for}~k
 \eta_i \ll 1 \,. \label{tdini}\ee

 \subsection{Long wavelength perturbations:}\label{subsec:longwave}

 We begin by studying the evolution of $\phi(k,\eta)$ for
 long-wavelength modes that remain superhorizon throughout,  to establish the
 normalization of the transfer function.

  For $k \rightarrow 0$ the solution of the Boltzmann equation (\ref{BEsol}) becomes the \emph{same} for DM or radiation (relativistic) components namely
 \be \tFp(\eta) = \tFp(\eta_i) - \Bigg( p \frac{d\tfu}{d p}\Bigg) \left[ \phi(\eta)-\phi(\eta_i)\right] \,,\label{k0BE}\ee  where we have suppressed the argument $\vk$ since we consider only $k=0$ here.

 For the radiation component we write,
 following eqn. (\ref{F1rel}) \be \tFp_r(\eta) = -\Theta(\eta)\,\Bigg( p \frac{d\tfu}{d p}\Bigg) \label{radk0}\ee leading to the solution \be \Theta(\eta) = \phi(\eta) -\frac{3}{2}\phi_i \label{radk0sol} \ee where we used the initial condition
 (\ref{inisup}). For DM perturbations, from eqn. (\ref{tildelta}) we obtain \be \td(\eta)= 3\phi(\eta) - \frac{9}{2}\phi_i \label{deltaK0}\ee where we used the initial condition
 (\ref{tdini}).

For a DM particle that decouples while relativistic and during the
stage when it is still relativistic $\delta \rho/\rho \neq \td$.
However, for a WDM particle with  $m \sim \mathrm{keV}$ it follows
that $\delta \rho/\rho = \td$ for $\ta \gtrsim \ta_{NR} \sim 10^{-3}
$. Hence, for $\ta  \gtrsim \ta_{NR}$   the Einstein equation
(\ref{Einstein002})    becomes \be \frac{2}{3} \frac{k^2
\ta^2}{k^2_{eq}}\,\phi+(1+\ta)\left[\ta \, \phi^{'}+\phi\right] =
-\frac{1}{2}\left[\ta \,\td + 4\Theta\right] \label{Einstein003}\ee
where we have used (\ref{hubrad}).  Using the
solutions of the Boltzmann equations (\ref{radk0sol},\ref{deltaK0})  for $k=0$,
and defining $\tdfi = \phi/\phi_i$, we find \be \tdfi'+\tdfi
\left[\frac{5\,\ta+6}{2\,\ta\,(1+\ta)} \right] =
\frac{3}{4\ta}\left[\frac{3\,\ta + 4}{1+\ta} \right]
\label{tdfieq}\ee the solution of this equation is \be \tdfi(\ta) =
\frac{\sqrt{1+\ta}}{\ta^3} ~\int^{\ta}_{0}
\frac{3}{4y}\left[\frac{3\,y + 4}{1+y}
\right]\,\frac{y^3~dy}{\sqrt{1+y}}  +
\mathcal{C}\left[\frac{\sqrt{1+\ta}}{\ta^3} \right]\,,
\label{soltdfi}\ee and $\mathcal{C}$   is determined by giving
  $\tdfi(\ta_{NR})$. Since
$\ta_{NR} \leq 10^{-3}$ for the DM candidates studied here, we will
take $\ta_{NR} \rightarrow 0$ whence $\tdfi(\ta_{NR} \rightarrow 0) =1$, namely we are
assuming that the DM particle becomes non-relativistic when the
Newtonian potential   still has the primordial superhorizon value.
With this initial condition we find  \be \tdfi(\ta) = \frac{1}{10
\,\ta^3} \Bigg[16 \sqrt{1+\ta}+9\ta^3 +2\ta^2 - 8\ta
-16\Bigg]\,,\label{tdfita} \ee a result that agrees with those found
in refs.\cite{kodama,dodelson}. For $\ta \ll 1$ it follows that
$\tdfi(\ta) = 1- \ta/10 +\mathcal{O}(\ta^2)$ therefore the
approximation $\phi(\ta_{NR}) \simeq \phi(0) = \phi_i$ is very
reliable. $\tdfi(\ta)$ decreases monotonically from $\tdfi(0)=1$ to
$\tdfi(\infty) = 9/10$, and at matter-radiation equality $\tdfi(1)=
0.963$.

For $k\neq 0$ the transfer function for the Newtonian potential is
\emph{defined} as \be \tilde{\phi}(k;\ta \gg 1) \equiv \frac{9}{10}~
T (k)~~;~~ T(0)=1 \,.\label{Tofkdef}\ee

Whereas long wavelength perturbations in the gravitational potential
remain nearly constant, short wavelength perturbations fall off as a consequence of   suppression by free streaming.

 For
$k\ta \gg k_{eq}$ the first term in the left hand side of Einstein's
equation (\ref{Einstein002}) dominates, leading to Poisson's
equation \be   \phi(k,\ta)   = -\frac{3}{4}\frac{k^2_{eq}}{k^2
\ta^2}\Bigg[ \ta\,\Big(\frac{\delta \rho}{\rho}\Big)_m +
\Big(\frac{\delta \rho}{\rho}\Big)_r \Bigg]\,.  \label{poisson} \ee

 \section{Evolution of density perturbation during radiation domination.}

  Although   the Newtonian potential is determined by Einstein's equation (\ref{Einstein00})
   where the right hand side also has a contribution from the DM
 perturbations during the stage when they are relativistic, such contribution is
 negligible because of the perturbatively small effective number of degrees of freedom ($\beta,\lambda\sim 10^{-2}$)
 as discussed above.

 Hence, during the (RD) era $\ta \ll1$ the DM perturbations can be neglected,
 and the evolution of the perturbations is completely determined by the evolution of the
 radiation fluid. In this case there is an exact solution for the Newtonian
 potential\cite{dodelson,giova,lythbook,weinbergbook,ruthbook,kodama}
 \be \phi(z) = -3\,\phi_i(k)\Bigg[\frac{ \big(\frac{z}{\sqrt{3}}\big) \,\cos(\frac{z}{\sqrt{3}})-\sin(\frac{z}{\sqrt{3}})}{(\frac{z}{\sqrt{3}})^3} \Bigg] ~~;~~
 z= {k\,\eta}  \label{fiRD}\ee where $\phi_i$ the primordial value
  of the Newtonian potential determined during inflation. The solution (\ref{fiRD}) reflects the
 acoustic oscillations of the radiation fluid with speed of sound $c_s =1/\sqrt{3}$.

 \subsection{Relativistic DM: stage I} \label{subsec:RDM}
 During the(RD) stage in which the DM
 particle is still relativistic, namely for $k\eta \ll  \sqrt{\overline{y}^2}\,\alpha$ the
free streaming distance $l(p,\eta,\eta') = \eta-\eta'$ and $v(p,\eta) = 1 $, the integrand in (\ref{BEsol}) does not depend on $p$. In this case it proves convenient to write
\be \tFp(\vk,\vp\;\,\eta) = -\Theta(k,\mu ;\,\eta)\, p\,\Big(\frac{d\tfu(p)}{dp}\Big)\,, \label{F1rel}\ee
and we find \be \Theta(k,\mu ;\,\eta) = -\phi(z)+ e^{-i\mu \,z}\Bigg[\frac{1}{2}\,\phi_i(k)  +
  2 \int_0^z dz'\Big( \frac{d\phi(z')}{dz'} \Big)\,e^{ i\mu\, z'}\Bigg]~~;~~z=k\,\eta \,.\label{TetaRD}\ee

Expanding $\Theta(k,\mu;\,\eta) $ in Legendre polynomials, \be
\Theta(k,\mu;\,\eta) = \sum_{l=0}^\infty (-i)^l
\,(2l+1)\,\Theta_l(k;\,\eta) \mathcal{P}_l(\mu) \label{delleg}\ee we
obtain \be \Theta_l(k;\,\eta) = -\phi(z)\,\delta_{l,0}
+\frac{1}{2}\,\phi_i(k)\, j_l(z)+2\int_0^z dz'
\Big(\frac{d\phi(z')}{dz'}\Big) j_l(z-z') \,,\label{deltals}\ee
where we have taken $k\,\eta_i =0$. The last term describes an ISW
contribution akin to that in the temperature perturbations of
photons\cite{dodelson}. We note that if the mode remains outside the horizon all
throughout the evolution during the (RD) stage in which the DM
particle is relativistic, namely $k\,\eta=z \ll 1$, it follows that
\be \Theta_l(z) = - \frac{1}{2}\,\phi_i(k)\,\delta_{l,0}
+\mathcal{O}(z)\,. \label{suphorRDrel}\ee

The WDM density perturbation  \be \delta(k;\,\eta) = \frac{1}{2}
\int_{-1}^1 d\mu \,\int_0^\infty \tFp(\vk,\vp;\,\eta) \frac{p^2
dp}{4\pi^2} \,,\label{tildel} \ee  therefore during the RD era when
the DM perturbation is relativistic \be \delta(k;\eta) =
3\,\Theta_0(z)~~;~~\delta(k;\,\eta_i) = -\frac{3}{2}\, \phi_i(k) \,.
\label{RDdelta}\ee

\begin{figure}[h!]
\begin{center}
\includegraphics[height=7cm,width=7cm,keepaspectratio=true]{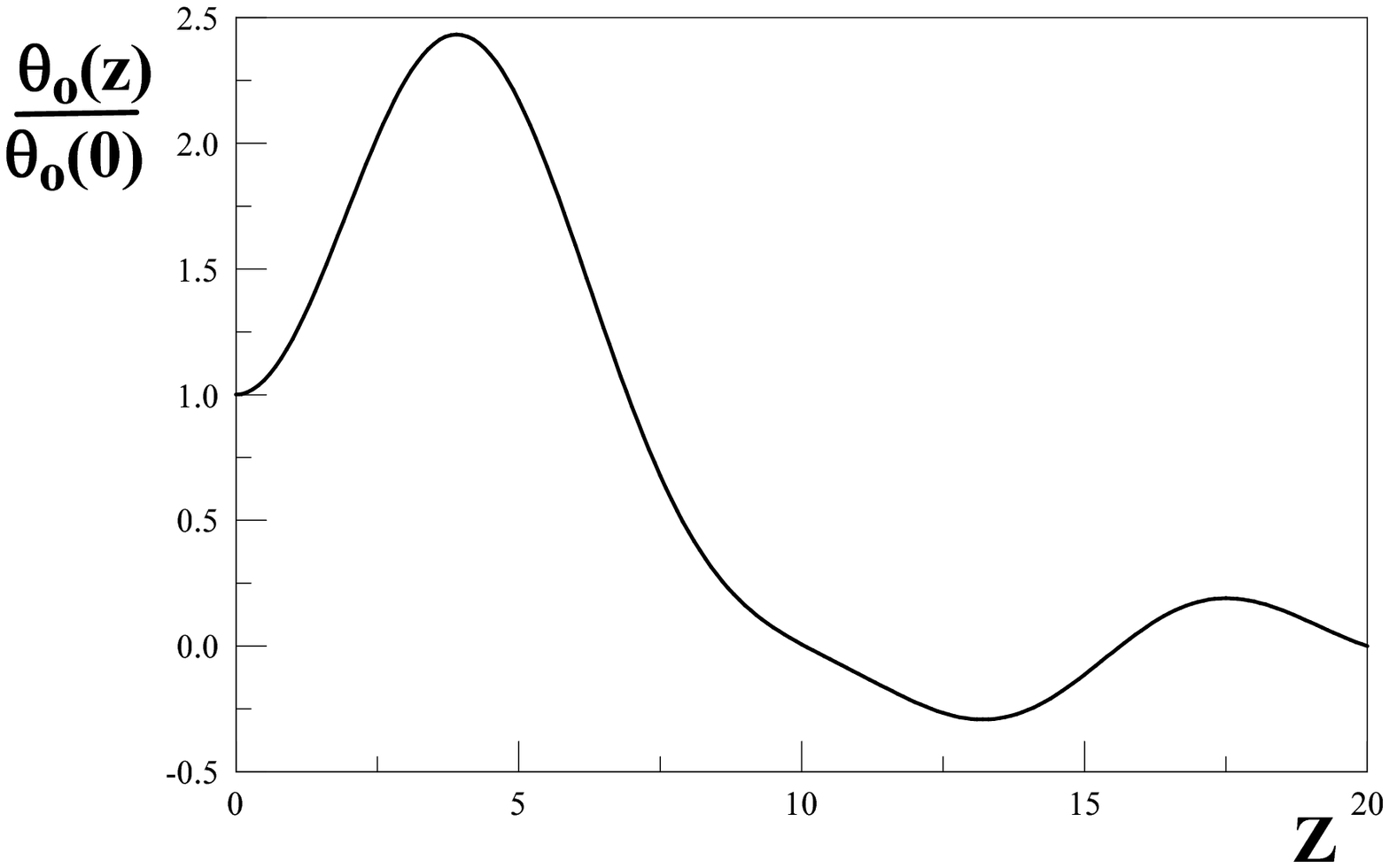}
\includegraphics[height=7cm,width=7cm,keepaspectratio=true]{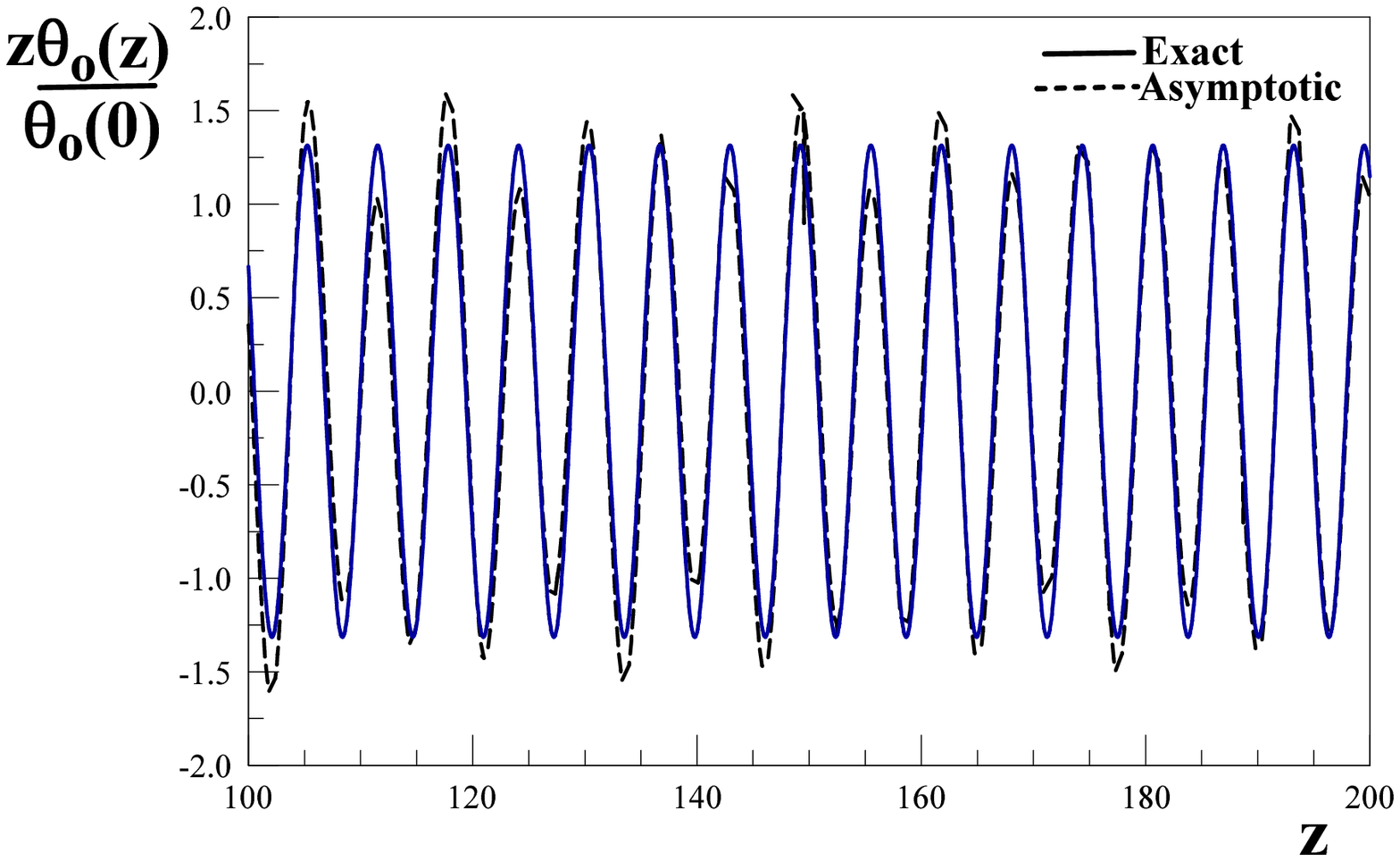}
\caption{Left panel $\frac{\Theta_0(z)}{\Theta_0(0)}$, right panel:
$z\Theta_0(z)/\Theta_0(0)$ compared to the asymptotic form
(\ref{deltalasy}) for the monopole. } \label{fig:deltao}
\end{center}
\end{figure}

The monopole $\Theta_0(z)$ begins to grow when it enters the horizon
as a consequence of the  ISW contribution, it reaches a maximun and
damps out as a consequence of (relativistic) free streaming. This is
understood from the following argument: at early time the derivative
of the Newtonian potential is negative and its modulus increases,
reaching a maximum approximately at the sound horizon $k\eta \simeq
\sqrt{3}\,\pi$, whereas the free streaming function $j_0(z-s)$ is
approximately constant for $z\sim s$, therefore the integrand
receives the largest contribution near the upper limit, and the
total integral peaks near the sound horizon. However, at later times
the integrand is strongly suppressed by free-streaming since
$d\phi/ds$ peaks near the sound horizon, but for $z \gg \pi
\sqrt{3}$ the free-streaming function suppresses the integrand.
  Fig. (\ref{fig:deltao}) displays $\Theta_0(z)/\Theta_0(0)$.

Although an analytic expression for the integrals in (\ref{deltals}) is not readily
available, we can obtain a reliable asymptotic expansion for $z\gg 1$. For this
purpose it is convenient to integrate by parts the derivative of the Newtonian potential,
for $z\gg 1$ the contributions near the upper limit of the integral $s\sim z$ vanish rapidly and the integral is dominated by the small $s$ region since the Newtonian potential $\propto 1/s^2$ for large s. Using the asymptotic expansion \be j_l(z) = \frac{\sin(z-\frac{l\pi}{2})}{z} +\mathcal{O}\big(\frac{1}{z^3}\big) \label{asyjl}\ee and
setting $z\rightarrow \infty$ in the upper limit of the integrals we find for $z \gg 1$ \be \Theta_l(z) \stackrel{z\gg 1}{=} 3 \phi_i(k)\, \frac{\sin\big[z-\frac{l\pi}{2}\big]}{z} \Bigg[ \frac{5}{2}- \sqrt{3}\,\ln \Bigg(\frac{\sqrt{3}+1}{\sqrt{3}-1} \Bigg) \Bigg]+ \mathcal{O}\Bigg(\frac{1}{z^2}\Bigg)\,,
\label{deltalasy}\ee this damped oscillatory behavior emerges for $z  \gtrsim 15$.

We note that the oscillations in (\ref{deltalasy}) do not feature the frequency corresponding to sound waves, the only remnant of the  acoustic oscillations of the radiation fluid in the asymptotic form is in the terms
featuring the $\sqrt{3}$ in the prefactor of the asymptotic form (\ref{deltalasy}).

  An important conclusion of this section is that during the stage
  in which the DM particle is \emph{relativistic} density
  perturbations \emph{do not depend } on the unperturbed
  distribution function and particle statistics.

  When the particle becomes non-relativistic, for modes $k\,\eta_{NR}\gg1$ the asymptotic form is still valid,
  and the monopole features oscillatory behavior \be \Theta_0(k) \propto \frac{\sin {\frac{\kappa}{2}}}{\kappa} \,. \label{oscirel}\ee This oscillatory behavior is a consequence of the acoustic oscillations of the radiation fluid, numerically we find that oscillations  arise for $\kappa/2 \gtrsim 15$ (see fig. (\ref{fig:deltao})).

\subsection{Non-relativistic DM: stages II  and III }
When the DM particle becomes non-relativistic (NR) $\epsilon(p,\eta) \sim m\,a(\eta)~;~ v(p,\eta)=p/m\,a(\eta)$.
 It proves convenient to change   from $\eta$ to a new variable $s$ defined
 by
\be ds = \frac{d\eta}{a(\eta)} \Rightarrow s(\eta)=
\frac{2\,u(\eta)}{\Big[H^2_0\,\Omega_m a_{eq}\Big]^\frac{1}{2}}=
\frac{2\sqrt{2}\,u}{k_{eq}a_{eq}} \label{sdef}\ee where $u(\eta)$ is
given by eqn. (\ref{udef}). The solution of the Boltzmann equation
for the normalized perturbation (\ref{tildes}) is \bea
\tFp(\vk,\vp;s) & = &    -\phi(\vk,s) \, \Big( p\,\frac{d\tfu}{dp}
\Big) + \int_{s_{NR}}^s ds'\,  \Bigg\{   i m a^2(s')\, \phi(\vk,s')
\Big(\vk\cdot \vec{\nabla}_p \tfu \Big)\Big[1+
\frac{p^2}{m^2\,a^2(s')}\Big] \, \Bigg\}\,  e^{-i \kpm (s-s')}
\nonumber \\  & +  & e^{-i \kpm (s-s_{NR})}\,\Big[
\tFp(\vk,\vp;\eta_{NR})+\phi(\vk,\eta_{NR})\,\Big(
p\,\frac{d\tfu}{dp} \Big)\Big]\,.  \label{BENR}\eea The initial
``time'' $s_{NR}=s(\eta_{NR})$ corresponds to the (conformal) time
at which the DM particle becomes non-relativistic. For WIMP's that
decoupled thermally for $T_d \ll m$ at conformal time $\eta_d \sim
10~\mathrm{pc}$ during the (RD) era,   $s_{NR}$ can be taken to be
$s_{NR} = s(\eta_d)$. Modes with comoving scales much larger than
$\eta_d$ where outside the horizon at $s_{NR}$, for these modes the
initial condition is given by eqn. (\ref{iniwdm}), namely \be
\tFp(\vk,\vp;\eta_{NR}) = \frac{1}{2}\,\phi_i(k) \,p
\,\Big(\frac{d\tfu(p)}{dp}\Big) \,. \label{iniNRWimp} \ee

On the other hand, WDM particles with  $m \sim \mathrm{keV}$ WDM
   decouple when they are still relativistic, namely $T_d \gg m$. For
these candidates comoving scales that enter the horizon during the
(RD) stage when the WDM particle is still relativistic evolve until
the particle becomes non-relativistic at $\eta=\eta_{NR}$ as
described in the previous section.  Therefore $s_{NR} =
s(\eta_{NR})$ and \be \tFp(\vk,\vp; \,\eta_{NR}) = -\Theta(k,\mu
;\,\eta_{NR})\, p\,\Big(\frac{d\tfu(p)}{dp}\Big)\,, \label{F1iniNR}\ee
where $ \Theta(k,\mu ;\,\eta_{NR}) $ is given by equations
(\ref{delleg},\ref{deltals}) with $\eta=\eta_{NR}$. Integrating eqn.
(\ref{BENR}) by parts in $s'$ and $\vp$, and neglecting the term
$(p/m\,a(s))^2 \ll 1$ in the non-relativistic limit, the evolution
of the density perturbation is given by \bea \tilde{\delta}(\vk,s) &
= & 3\phi(k,s)-k^2 \int_{s_{NR}}^s ds'   a^2(s')
\phi(k,s')\,(s-s')\,K (k,s-s')  \nonumber \\ & & + \int
\frac{d^3p}{(2\pi)^3}\,p\,\Big(\frac{d\tfu(p)}{dp}\Big)\, e^{-i
\frac{\vk\cdot\vp}{m}(s-s_{NR})}\,
 \, \mathcal{S}\big[\vk,\vp\,;\eta_{NR}\big]\,. \label{gilbertgen} \eea  where     \be K  (k,s-s') =  \int \frac{d^3p}{(2\pi)^3}\,
e^{-i \frac{\vk\cdot\vp}{m}(s-s')} \tfu(p)  \label{KerK}\ee
determines the suppression by non-relativistic free streaming and \be \mathcal{S}\big[\vk,\vp\,;\eta_{NR}\big] = \frac{3}{2}\phi_i(k) e^{-i\mu z_{NR}} +
  2\,i\mu \int_0^{z_{NR}}dz' \phi(z')\,e^{-i\mu(z_{NR}-z')}~~;~~z=k\eta \label{source} \ee is the result of evolution during stage I and determines the initial condition for the evolution during the non-relativistic stages II and III.

  Since $f_0$ only depends on $p$, using
 eqns. (\ref{sdef},\ref{alfa}) it follows that \be K  (k,s-s') \equiv K [\alpha(u-u')] = \frac{1}{N}  \int y^2 f_0(y) j_0[y\alpha(u-u')]\,dy~~;~~ N=   {\int y^2 f_0(y)\,dy}  \label{Kj0}\ee and $j_0$ is the spherical Bessel function.

 The first line  in (\ref{gilbertgen}) integrates the gravitational potential
during the stages in which the particle is non-relativistic. As
described above, there are two distinct epochs: when the
gravitational potential is dominated by perturbations in the
radiation fluid and when it is dominated by dark matter
perturbations. The crossover between the two stages occurs at a
scale $s^* \equiv s(a^*)$ that is determined self-consistently, for $s> s^*$ the matter perturbation dominates the gravitational potential.

It is
  convenient to separate the contributions to the gravitational potential from the DM and radiation components,
   writing in obvious notation $\phi(k,\eta)= \phi_r(k,\eta)+ \phi_m(k,\eta)$ where $\phi_r(k\eta)$
    is given by (\ref{fiRD}). The contribution
from DM is obtained from   Einstein's equation (\ref{Einstein002})
which for $a>a^*$ reduces to the  Poisson's equation for all scales
smaller
 than a few Mpc, namely  \be
\phi_m(k,\eta) =
-\frac{3}{4}\frac{k^2_{eq}}{k^2\,\ta}\,\td(\vk,s)\,.
\label{poissonmat} \ee For $s > s^*$ the integral in
(\ref{gilbertgen}) can be split up into the integral from $s_{NR}$
up to $s^*$ which is dominated by $\phi_r$ and corresponds to stage II, and the integral from
$s^*$ up to $s$ in which the gravitational potential is dominated by
the DM component (\ref{poissonmat}).

 Therefore for $s> s^*$, the density perturbation $\delta$
 obeys Gilbert's equation\cite{gilbert,bert,boysnudm,darkmatter} \be
 {\delta}(\vk,s)   = -
\frac{9}{4}\frac{k^2_{eq}}{k^2\,\ta}\,\td(\vk,s) +
\frac{3}{2}\,H^2_0\,\Omega_m \int_{s^*}^s ds' (s-s')K
(k,s-s')\,a(s')\,\td(\vk,s') + I[k,s] \,,\label{gileqn}\ee where the
inhomogeneity \bea  {I}[k,s] & = & 3\phi_r(k,s)-k^2
\int_{s_{NR}}^{s^*} ds'   a^2(s') \phi_r(k,s')\,(s-s')\,K (k,s-s')
\nonumber \\ & & + \int
\frac{d^3p}{(2\pi)^3}\,p\,\Big(\frac{d\tfu(p)}{dp}\Big)\, e^{-i
\frac{\vk\cdot\vp}{m}(s-s_{NR})}\,
 \, \mathcal{S}\big[\vk,\vp\,;\eta_{NR}\big]\,,\label{inh} \eea and $\phi_r$ is the radiation contribution to the gravitational potential given by (\ref{fiRD}). Thus the inhomogeneity incorporates the \emph{past history} during stages I and II.

\subsection{Kernels for CDM and WDM:}

The kernel $K(k,s-s')$ determines the suppression of WDM
perturbations by non-relativistic free streaming and depends on the distribution
function $\tfu(p)$. For WIMPs (CDM) $\tfu$ is the Maxwell-Boltzmann distribution function
 given by eqn. (\ref{MBf})  whereas for (DW) or (BD) WDM particles $\tfu(y)$
is given by eqn. (\ref{DWf}) or (\ref{SDf}) respectively.

\subsubsection{CDM: Maxwell-Boltzmann distribution function}

For CDM we find   \be K(k,s-s') = e^{-\frac{\kappa^2}{6}(u-u')^2}
\,, \label{Piscdm}\ee where $u(\eta)$ is defined by eqn.
(\ref{udef}), and from the definitions (\ref{kapadef},\ref{alfa}),
along with eqn. (\ref{ybar}), we find \be \kappa =
\frac{\sqrt{6}\,k}{k_{fs}}=0.38\, k  \, \left(
\frac{100~\mathrm{GeV}}{m} \right)^\frac{1}{2} \,\left(
\frac{10~\mathrm{MeV}}{T_d} \right)^\frac{1}{2}\,\left(
\frac{2}{g_d} \right)^\frac{1}{3}~\times (pc) \,.\label{gamwimp}\ee

\subsubsection{WDM: DW distribution function }

With the distribution function (\ref{DWf}) one finds\cite{bert,darkmatter,boysnudm} \be  K (k,s-s') =
K[Q] = \frac{4}{3\zeta(3)}\,\sum_{n=1}^\infty ~\frac{(-1)^{n+1} ~
n}{(n^2+Q^2)^2}   \label{pisdw}\ee  where \be Q= \alpha \,(u-u')
~~;~~ \alpha =   \frac{0.68~k}{k_{fs}} = 0.278\,\kappa
\label{argu}\ee

\subsubsection{WDM: BD distribution function }
With the distribution function (\ref{SDf}) one finds\cite{boysnudm} \be   K(k,s-s') =
K [Q] = \frac{\sqrt{2}}{\sqrt{3}\,\zeta(5)}~\sum_{n=1}^\infty
\frac{1}{\Big( \rho~n\Big)^\frac{5}{2}}\,\Bigg[1+\frac{n}{\rho}
\Bigg]^\frac{1}{2}\,\Bigg[\frac{2n+\rho}{n+\rho} \Bigg]
~~;~~\rho=\sqrt{n^2+Q^2}\,, \label{pisd}\ee  where in this case \be
Q=\alpha (u-u')~;~\alpha = \frac{0.84\,k}{k_{fs}}= 0.343\,\kappa
\label{Qalfabd}\ee

We note that in all the cases considered here, the kernels $K$ are functions of
the combination $\kappa^2(u-u')^2$.

The free streaming kernels are suppressed, either exponentially (MB)
or as high inverse powers (DW,BD) of the ratio  $k^2/k^2_{fs}$.

\section{Cold Dark Matter}\label{sec:cdm}

For a WIMP of $m \sim 100 ~\mathrm{GeV}$ decoupling at $T_d \sim
10~\mathrm{MeV}$ (for which $g_d \sim 10$)  comoving scales $\lambda
\gg \eta_d \sim 10\,\mathrm{pc}$ entered the horizon well after
decoupling and when the particle is non-relativistic, in which case
we can set $\eta_{NR} \sim 0$ and \be \Theta(k,\mu ;\,\eta_{NR}) =
\frac{1}{2}\, \phi_i(k)\label{largescaleD}\,. \ee For these
CDM particles,  $\lambda_{fs}\lesssim  1 \,\mathrm{pc}$ and for
comoving wavelengths $\lambda \gg 10 \,\mathrm{pc}$ it follows that
$\kappa \ll 1$ therefore $K \simeq 1 $, this amounts to setting
$\langle V^2_{eq}\rangle^\frac{1}{2}=0$, consistently with CDM. The
perturbation equation (\ref{gilbertgen}) simplifies to \be
\delta(\vk,s) = 3\phi(k,s)-k^2 \int_{s_{NR}}^s ds' a^2(s')
\phi(k,s')\,(s-s')-\frac{9}{2}\,\phi_i(k) \label{gilbertcdm} \ee
This equation can be recognized by taking $d^2/ds^2$ of both sides,
\be \frac{d^2}{ds^2}\left[\delta(\vk,s) - 3\phi(k,s)\right] = -k^2
a^2 \phi(k,s) \label{diff}\ee using $d/ds = a d/d\eta$ and
$\dot{a}/a =1/\eta$ during (RD)   we find \be \ddot{\delta} +
\frac{\dot{\delta}}{\eta} = 3 \ddot{\phi} + \frac{3}{\eta}\dot{\phi}
- k^2 \phi\,, \label{cdmrad}\ee which is the equation obeyed by CDM
perturbations during the (RD) era\cite{dodelson}.

\begin{figure}[h!]
\begin{center}
\includegraphics[height=3.5in,width=3.5in,keepaspectratio=true]{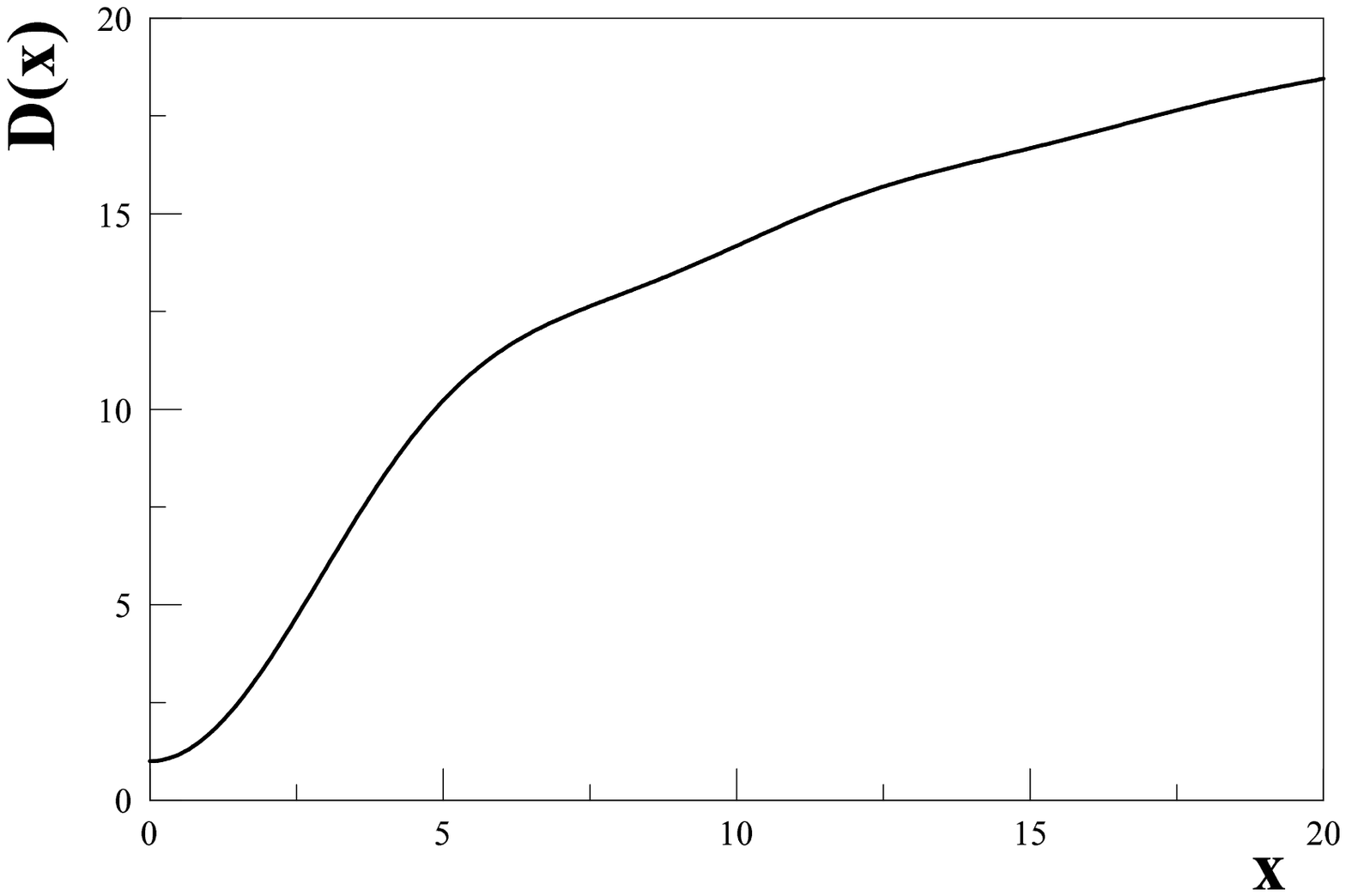}
\caption{$D(x)=\frac{\td(x)}{\td(0)}$ vs. $x=k\eta/\sqrt{3}$. }
\label{fig:dofx}
\end{center}
\end{figure}

During this   era when $\phi$ is determined by the radiation
fluid $a^2(\eta) = [H^2_0 \Omega_m a_{eq}]\eta^2~;~s(\eta) =
\ln(\eta)/[H^2_0 \Omega_m a_{eq}]^{\frac{1}{2}}+\mathrm{constant}$
and $\phi(k,\eta)$ is given by eqn. (\ref{fiRD}), and eqn.
(\ref{gilbertcdm}) becomes \be {\delta}(k,\eta) =
9\,\phi_i(k)\Bigg\{- \left[ \frac{x \cos(x)-\sin(x)}{x^3}\right]+
\int^x_{x_{NR}} dx' \ln\left( \frac{x}{x'}\right)\,
\frac{d}{dx'}\left( \frac{\sin(x')}{x'} \right) -\frac{1}{2} \Bigg\}
\label{RDCDMgil}\ee where $x= k\eta/\sqrt{3}$. For WIMPs and
perturbations with comoving scales $\lambda \gg 10\,\mathrm{pc}$ we
can set $x_{NR}=0$, leading to the result \be  {\delta}(k,\eta) =
-9\,\phi_i(k)\Bigg\{ \left[ \frac{x \cos(x)-\sin(x)}{x^3}\right]+
\frac{\sin(x)}{x} -\frac{1}{2} - Ci(x)+\ln(x) +\gamma_E
\Bigg\}\label{Wimpdelta}\ee where $\gamma_E = 0.577216\cdots$ and
$Ci(x)$ is the cosine-integral function.  Fig. (\ref{fig:dofx})
displays $\td(x)/\td(0)$ vs. $x=k\eta/\sqrt{3}$, where
$\td(0)=-3\phi_i(k)/2$. The density perturbation receives a ``kick''
upon entering the horizon at $k\eta \sim 1$. We find numerically
that \be \frac{\td(x)}{\td(0)} \simeq 6\Bigg(\ln(x)+ \gamma_E
-\frac{1}{2} \Bigg) ~~\mathrm{for}~~x \gtrsim 10 \,.
\label{asydelNRRD}\ee

We can now estimate the crossover scale at which the Newtonian
potential is determined by radiation or CDM perturbations. For
$\tilde{a}\ll 1$ deep in the RD dominated era and for subhorizon
modes $k\eta \gg 1$ Einstein's equation (\ref{Einstein00})
determines that  \be \Bigg(\frac{\delta \rho }{\rho}\Bigg)_r \sim
6\phi_i(k)\,  \cos(x) \label{deltarad}\ee Taking the asymptotic
behavior (\ref{asydelNRRD}) for $\delta$, the Newtonian potential
determined by Einstein's equation (\ref{Einstein00}) begins to be
dominated by matter density perturbations when \be \frac{3}{2}\, \ta
\,\ln(x) > 1\,. \label{CDMdom}\ee For comoving scales smaller than a
few Mpc we find that the crossover scale from radiation to matter
\emph{perturbations} dominating the gravitational potential is \be
\ta^* \lesssim 0.1 \,.\label{crosa}\ee

During RD, $x \sim  \ta \sqrt{2}\,k/\sqrt{3}\,k_{eq}$, therefore for
all comoving scales smaller than a few $\mathrm{Mpc}$ the crossover
to the domination of the Newtonian potential by DM density
perturbations occurs within the RD dominated era.

 Passing to the variable $u$ defined by eqn. (\ref{sdef}),
  for $u > u^*$ Gilbert's eqn. (\ref{gileqn}) now becomes  \be
\td(\vk,u)  =   - \frac{9}{4}\frac{k^2_{eq}}{k^2\,\ta}\,\td(\vk,u) +
6\int_{u^*}^u du' (u-u')\,\ta(u')\,\td(\vk,u') +
{I}[k,u]\label{gilbertII}\ee where \be
{I}[k,u]=3\phi_r(k,u)-\frac{8\,k^2}{k^2_{eq}} \int_{u_{NR}}^{u^*}
du'   \ta^2(u') \phi_r(k,u')\,(u-u')-\frac{9}{2}\,\phi_i(k) \,.
\label{Iks} \ee

For $k \gg k_{eq}$ and $k \ta \gg k_{eq}$ which is valid for modes well inside
the horizon when DM density perturbations dominate, we can safely
neglect the first term   (\ref{gilbertII}) and because during radiation domination
$k\eta =\sqrt{2}k\ta/k_{eq}$ and for modes deep inside the horizon $\phi_r \sim \cos(k\eta)/k^2\eta^2$
we can also neglect the $3\phi_r$ in $I[k,u]$. We then notice that $I[k,u]$ is linear in $u$ and
 (\ref{gilbertII}) can be turned into an ordinary homogenous differential equation,
  \be \frac{d^2  }{du^2}\,\td(k,u)   - 6 \, \ta(u) \td(k,u) =0 \,,\label{2ndord}\ee with
   the initial conditions \be \td(k,u^*)=I[k,u^*] ~~;~~\frac{d\,\td(k,u)}{du}\Bigg|_{u=u^*} =
    \frac{d\,I[k,u ]}{du}\Bigg|_{u=u^*}\,.\label{inicondi}\ee

Since the
variable $u$ depends solely on the combination \be \zeta =
\sqrt{1+\ta(u)} = \frac{1}{\tanh[-u]} \label{zetavar}\ee (see eqn. (\ref{udef}))
it proves convenient to write the differential equation
(\ref{2ndord}) in terms of $\zeta$. We find \be
\frac{d}{d\zeta}\left[(1-\zeta^2)\, \frac{d\td}{d\zeta} \right]+ 6\,
\td = 0 \,. \label{legendre}\ee This is Legendre's equation of index
$\nu =2$ with solutions \bea P_2(\zeta)  & = & \frac{1}{2}\,\left(
3\,\zeta^2 -1\right) \label{p2} \\ Q_2(\zeta) & = & \frac{1}{4}
\left( 3\,\zeta^2 -1\right)
 \ln\Bigg[\frac{\zeta+1}{\zeta-1} \Bigg] - \frac{3}{2} \zeta \label{q2}\eea In terms of
 $\ta$ rather than $\zeta$ eqn. (\ref{legendre}) becomes
 \be \frac{d^2\,\td}{d\ta^2}  + \frac{(2+3\ta)}{2\ta(1+\ta)}\,\frac{d\td}{d\ta} - \frac{3}{2} \frac{\td}{\ta(1+\ta)} = 0 \label{meszaros}\ee
  this is Meszaros' equation\cite{mesaros,groth,peeblesbook}. We find remarkable   that in terms of the variable $\zeta$ Meszaros' equation is simply
 Legendre's equation of index $\nu=2$.

The general solution is \be \td(k,\ta) = \td_g(k) P_2(\zeta) + \td_d(k) Q_2(\zeta)~~;~~\zeta= \sqrt{1+\ta} \label{gsol}\ee  The coefficients $\td_{g,d}$ must be obtained from the initial conditions (\ref{inicondi}) and  the Wronskian  of the independent
solutions $P_2,Q_2$. However, we recognize that the asymptotic solution (\ref{asydelNRRD}) can be written as \be \td(k,\ta) \simeq 6\,\td_i \Bigg[\ln\Bigg(\frac{ \sqrt{2}\,k ~e^{\gamma_E-\frac{1}{2}}}{\sqrt{3}\,k_{eq} } \Bigg)+\ln\Big[\zeta^2 -1 \Big] \Bigg] \label{matchsol}\ee where we used the relation $\eta = \sqrt{2}\ta/k_{eq}$ valid during
the RD dominated era for $\eta \ll \eta_{eq}$ corresponding to $\ta \ll 1$.
 Matching (\ref{gsol}) to (\ref{matchsol}) for $\zeta \sim 1$ we find \be \td_d(k) =
 -12\td_i(k) ~~;~~ \td_g(k) = 6\td_i(k)\ln\Bigg[\frac{4\sqrt{2}\,k ~e^{\gamma_E-\frac{7}{2}}}{\sqrt{3}\,k_{eq} } \Bigg] \label{coefic}\ee
  For $\ta \gg 1$ the growing solution is given by $\td_g P_2(\zeta)$,
  namely \be \td(k,\ta) \simeq  9 \td_i(k)\ln\Bigg[\frac{4\sqrt{2}\,k ~e^{\gamma_E-\frac{7}{2}}}{\sqrt{3}\,k_{eq} } \Bigg]\,\ta \label{growingsolMD}\ee
   and the gravitational potential becomes for $\ta \gg 1$ \be \phi(k) = \frac{9}{10}\,\phi_i(k) \,T_{CDM} (k) \,,  \label{philarget}\ee
    where including the long-wavelength normalization (\ref{Tofkdef}) we find
      \be T_{CDM}(k) = \frac{45}{4} \frac{k^2_{eq}}{k^2}\,
      \ln\Bigg[\frac{4\sqrt{2}\,k ~e^{\gamma_E-\frac{7}{2}}}{\sqrt{3}\,k_{eq} } \Bigg]
       \label{Tofk} \ee is the CDM transfer function for $k\gg k_{eq}$. This result agrees with that of Weinberg\cite{weinberg} and Wu and Sugiyama\cite{HS}  and   numerically agrees to within few percent   with the numerical fit provided by Bardeen \emph{et.al.}\cite{bbks} for $k\gg k_{eq}$.

       An alternative derivation of this result which is relevant for   comparison with
       WDM below begins by defining a new variable \be \Delta(k,u) = \td(k,u)-I[k,u] \label{Deldef}\ee
        obeying  \be \frac{d^2}{du^2}\,\Delta(k,u)   - 6 \, \ta(u)
        \Delta(k,u) =   6 \, \ta(u) I[k,u]\,,\label{2ndordDel}\ee with   initial conditions
         \be \Delta(k,u^*)=0 ~~;~~\frac{d\,\Delta(k,u)}{du}\Bigg|_{u=u^*} = 0\,.\label{Delinicondi}\ee
          Therefore from the solution of (\ref{2ndordDel},\ref{Delinicondi}) we find \be \td(k,u) = I[k,u] + 6 \int^u_{u^*} \ta(u')I[k,u']
          \mathcal{G}(u,u') \,du'
           \label{solucdm}\ee where \be \mathcal{G}(u,u')  = \frac{1}{W}\Big[P(u)Q(u')-P(u')Q(u )\Big]
             \label{GF}\ee and
             $G[u,u']=\mathcal{G}(u,u')\Theta(u-u')$ is the retarded Green's
             function obeying \be \Bigg[\frac{d^2}{du^2}    - 6 \,
             \ta(u)\Bigg]G[u,u']     = \delta(u-u')\,.
             \label{greens}\ee
             The functions  $P(u) = P_2(\zeta(u));Q(u)=Q_2(\zeta(u))$ are
             the growing and decaying homogeneous solutions of \be  \Bigg[\frac{d^2}{du^2}    - 6 \, \ta(u)\Bigg]\Bigg\{\begin{array}{c}
                                  P(u) \\
                                  Q(u)
                                \end{array}
              \Bigg\}=0 \,,\label{PQdiff}\ee
              and $W=1$ their Wronskian. It is straightforward to prove that the
              solution (\ref{solucdm}) is exactly the same as (\ref{gsol}) after using the homogeneous
              differential equation (\ref{PQdiff}) for $P_2,Q_2$ and twice integrating by parts in $u'$.

              Since the source $I[k,u]$ remains bound as $u \rightarrow 0^-$
              ($\ta \rightarrow \infty$), it follows that asymptotically for
               $\ta \gg 1$ \be \td(k,u) \rightarrow \frac{6}{W}\,P(u)
               \int^0_{u^*} Q(u')\,\ta(u')\,I[k,u']du' = 9\, \ta(u) \int^0_{u^*} Q_2(u')\,\ta(u')\,I[k,u']du' \,.
               \label{cdmasygf}\ee

From (\ref{poissonmat}) and (\ref{Tofkdef}) we find \be T_{CDM}(k)=
-\frac{30}{4}\,\frac{k^2_{eq}}{k^2\,\phi_i(k)}\,\int^0_{u^*}
Q_2(u')\ta(u')I[k,u']du' \,. \label{Tcdm}\ee

                The main reason for describing this alternative in detail is because the   form (\ref{Tcdm})    generalizes to the WDM case.

\section{Warm Dark Matter:} Passing to the dimensionless variable $u$ in
(\ref{gilbertgen}), eqns. (\ref{gileqn},\ref{inh}) become  \bea \delta(\vk,u)   & = &
3\phi(k,u)- \frac{8k^2}{\alpha\,k^2_{eq} } \int_{u_{NR}}^u
  \ta^2(u')\,
\phi(k,u') \,\Pi\big[\alpha (u-u')\big] \,du' \, + \nonumber \\ &  & \frac{1}{N} \int_0^\infty
 y^3 dy
 \Big(\frac{df_0(y)}{dy}\Big)\, \Bigg\{ \frac{3}{2}\phi_i(k)\,j_0\big[y\,\alpha(u-u_{NR})+z_{NR} \big]  \nonumber \\ & + &
 2\int_0^{z_{NR}} dz' \phi(z') j_1\big[y\,\alpha(u-u_{NR})+z_{NR}-z' \big] \Bigg\}   \label{gilu}\eea
 where \be \Pi\big[\alpha(u-u')\big] = \frac{1}{N}\int_0^\infty y f_0(y) \sin\big[y\,\alpha \, (u-u') \big] \,dy
 \,= \alpha(u-u')K(k,u-u')\,,\label{PI} \ee and $N$ is defined in eqn. (\ref{Kj0}).

 When the DM perturbations dominate the gravitational potential for $u>u^*$ which is
 determined self-consistently as explained above,  $\delta$ obeys
    Gilbert's equation in the form \be
\delta(\vk,u)    -        \frac{6}{\alpha } \int_{u^*}^u
  \ta(u')\,
\td(k,u') \,\Pi\big[\alpha(u-u')\big] \,du' \, =  \, {I}[k;\alpha;u]
\label{PQeq}\ee where we neglected terms proportional to $k^2_{eq}/k^2$, and

\bea  {I}[k;\alpha;u]  & = & 3 {\phi}_r(k,u)- \frac{8 k^2}{\alpha\,k^2_{eq}} \int_{u_{NR}}^{u^*}
  \ta^2(u')\,
 {\phi}_r(k,u') \,\Pi\big[\alpha(u-u')\big] \,du' \, + \nonumber \\ &&  \frac{1}{N} \int_0^\infty
 y^3 dy
 \Big(\frac{df_0(y)}{dy}\Big)\, \Bigg\{ \frac{3}{2}\,\phi_i(k) \,j_0\big[y\,\alpha(u-u_{NR})+\frac{\kappa}{2} \big]  \nonumber \\ & + &
 2\int_0^{\frac{\kappa}{2}} dz'  {\phi}_r(z') j_1\big[y\,\alpha(u-u_{NR})+\frac{\kappa}{2}-z' \big] \Bigg\}\, ,
\label{inhofi} \eea  where we have used
$z_{NR}=k\eta_{NR}=\kappa/2$. For $k\gg k_{eq}$ the term $3\phi_r$
in the first line in (\ref{inhofi}) is subleading as compared to the
second term and will also be neglected in our analysis.

It is clear from the integral equation (\ref{PQeq}) that $\delta$
obeys the initial conditions \be \delta(k,u^*) =
I[k;\alpha;u^*]~~;~~\frac{d\,\delta(k,u)}{du}\bigg|_{u^*} =
\frac{d\,I[k;\alpha;u]}{du}\bigg|_{u^*}\,. \label{gilini}\ee

In the first line in (\ref{inhofi}) the kernel $\Pi$ determines the
free streaming of WDM perturbations during the  (RD)
stage during which the particle is \emph{non-relativistic}, whereas
the last two lines are the result of free streaming during the stage
when the particle is still \emph{relativistic}. In particular the
third term in (\ref{inhofi}) corresponds to the ISW contribution
(\ref{deltals}) (after an integration by parts) studied in section
(\ref{subsec:RDM}) which undergoes damping by free streaming during
the non-relativistic stage. As it will be seen below, this ISW
contribution yields an \emph{enhancement} of the transfer function
for $k < k_{fs}$.

Thus the inhomogeneity $I[k;\kappa;u]$ is completely determined by the past history
  during   stages I and II when perturbations in the radiation component dominate the gravitational potential.
   We have made explicit that the inhomogeneity depends both on $k$ and $\alpha$ (or $\kappa$).
   For fixed wavevector $k$ the CDM limit is obtained by letting $m (g_d)^\frac{1}{3} \rightarrow \infty$
    which lets  $\alpha \rightarrow 0$ (and $\kappa \rightarrow 0$) with fixed $k$ (see the definition (\ref{alfa}))
     and also $u_{NR} \rightarrow -\infty$ ($\eta_{NR} \rightarrow 0$).

At this stage one can proceed to a numerical integration of
(\ref{PQeq}), however in this article we will pursue an approximate
semi-analytic treatment valid for
 an \emph{arbitrary distribution function} postponing a full numerical study to a forthcoming article.

Before studying (\ref{PQeq},\ref{inhofi}), we analyze the asymptotic
long time behavior as $u \rightarrow 0$ of the WDM density
perturbation, which is obtained by neglecting the source term $I$
since it is bounded in time.

For $u \rightarrow 0$ it follows from (\ref{mdu}) that $\ta(u)
\simeq 1/u^2$. The integrand in (\ref{PQeq}) is dominated by the
region $u'\sim u \sim 0$, assuming that $\td(k,u) \rightarrow
\td(k,0) (-u)^{-\beta}  $ as $u \rightarrow 0 $ and using that for
$u'\sim u \sim 0$ it follows that $\Pi[\alpha(u-u')] \sim \alpha(u-u')$, and  we find \be
\frac{6}{\alpha } \int_{u^*}^u
  \ta(u')\,
\td(k,u') \,\Pi\big[\alpha(u-u')\big] \,du' \sim \td(k,0)\frac{6\,(-u)^\beta}{\beta(\beta+1)}\label{asy}\ee therefore there is a self-consistent solution  of eqn. (\ref{PQeq}) (for $I=0$) with $\beta = 2,-3$ corresponding to the growing and decaying solutions $\td_g(k,u) \propto \ta\,;\,\td_{d}(k,u) \propto 1/\ta^{3/2}$ respectively. This is an \emph{exact} result which shows that asymptotically for $\ta \gg 1$ $\delta \propto \ta$.

The Volterra equation of the second kind (\ref{PQeq}) has a solution in terms of the Fredholm-Neumann  series. However this iterative solution does not make explicit the growth factor $\ta$ exhibited by the exact solution. The analysis of the CDM case in the previous section suggests a re-organization of this series that manifestly exhibits the growth factor. For this purpose we cast Gilbert's equation (\ref{PQeq}) as an integro-differential equation by taking derivatives with respect to $u$.

The following integro-differential equation is obtained,
\be \frac{d^2}{du^2}\td(k,u)-6\ta(u)\td(k,u)-\frac{6}{\alpha } \int_{u^*}^u
  \ta(u')\,
\td(k,u') \,\frac{d^2}{du^2}\Pi\big[\alpha(u-u')\big] \,du' = \frac{d^2}{du^2}I[k,u]\,. \label{intdifeq} \ee
 Performing the same asymptotic analysis in the limit $u\rightarrow 0;\ta(u) \sim 1/u^2$
 leading to (\ref{asy}) we find
  in this limit\footnote{This can be found self-consistently by proposing $\td(k,u)\propto (-u)^{-\beta}$ and following the
steps leading to (\ref{asy}.)} \be -\frac{6}{\alpha } \int_{u^*}^u
  \ta(u')\,
\td(k,u') \,\frac{d^2}{du^2}\Pi\big[\alpha(u-u')\big] \,du' \sim \alpha^2\,\overline{y^2} \,\td(k,u)~~;~~\overline{y^2}= \frac{1}{N}\int_0^\infty y^4 f_0(y) dy \,. \label{asyderPi}\ee

This leading asymptotic behavior can be incorporated in
(\ref{intdifeq}) by writing \be
\frac{d^2}{du^2}\Pi\big[\alpha(u-u')\big] = - \alpha^2
\overline{y^2}\,\Pi\big[\alpha(u-u')\big] +\alpha^2
\widetilde{\Pi}\big[\alpha(u-u')\big]\label{pisub} \ee where \be
\widetilde{\Pi}\big[\alpha(u-u')\big] = \frac{1}{N}\int_0^\infty y
f_0(y)(\overline{y^2}-y^2) \sin\big[y\,\alpha \,(u-u') \big] dy
\label{tilPI}\ee

Using the original integral equation (\ref{PQeq}) we obtain
\bea  && \frac{d^2}{du^2}\td(k,u)-6\ta(u)\td(k,u)+ \kappa^2 \td(k,u) -  6\,\alpha  \int_{u^*}^u
  \ta(u')\,\widetilde{\Pi}\big[\alpha(u-u')\big]
\td(k,u')   \,du' \nonumber \\ &  = &   \frac{d^2}{du^2} {I}[k,u] + \kappa^2  {I}[k,u]   \label{intdifeq2} \eea were
  we used the
 definition (\ref{kapadef}).

The last term in the first line in (\ref{intdifeq2})  can be
interpreted as a non-local potential with a memory kernel
$\widetilde{\Pi}\big[\alpha(u-u')\big]$. It is straightforward to
show that $\widetilde{\Pi}\big[\alpha(u-u')\big] \propto (u-u')^3$
as $u'\rightarrow u$ and   from the results for the kernels (\ref{Piscdm},\ref{pisdw},\ref{pisd}) that it falls off as a high power (or exponential) of the
argument for the distribution functions considered here.

Furthermore, we have already established that asymptotically
$\td(k,u) \propto \ta \propto 1/u^2$, implementing the same analysis
leading to (\ref{asy}) and replacing this asymptotic behavior in the
memory integral in (\ref{intdifeq2}) we find that asymptotically as
$u \rightarrow 0$ it behaves  as \be \int_{u^*}^u
  \ta(u')\,\widetilde{\Pi}\big[\alpha(u-u')\big]
\td(k,u')   \,du' \propto \ln(-u) \propto \ln(\ta)\,, \ee therefore its contribution is  \emph{subleading}
in the asymptotic limit $\ta \rightarrow \infty$ as compared to all the other terms in the first line of (\ref{intdifeq2}).

 Hence, we conclude from this analysis that the memory
integral in (\ref{intdifeq2}) can be considered as a \emph{perturbation}.

Again, it is convenient to introduce the combination $\Delta(k,u)$
given by (\ref{Deldef}) that satisfies \be
\frac{d^2}{du^2}\Delta(k,u)-6\ta(u)\Delta(k,u)+ \kappa^2 \Delta(k,u)
= 6\ta(u) I[k,u] + \mathcal{J}[\td;u]\label{intdifeqDelta} \ee with
the initial conditions given by (\ref{Delinicondi}), where \be
\mathcal{J}[\td;u] =  6\,\alpha  \int_{u^*}^u
  \ta(u')\,\widetilde{\Pi}\big[\alpha(u-u')\big]
\td(k,u')   \,du' \,. \label{jota}\ee The solution of
(\ref{intdifeqDelta}) with the initial conditions
(\ref{Delinicondi}) is completely determined by the retarded Green's function obeying
\be \Bigg[\frac{d^2}{du^2} -6\ta(u) + \kappa^2   \Bigg]G[u,u'] =\delta(u-u') \,.\label{GFunc}\ee

 The formal solution of (\ref{intdifeq2}) with initial conditions (\ref{gilini}) is
\be  \td(k,u)    = I[k,u]+   \int_{u^*}^u \mathcal{G}(u,u') \Bigg[
6\ta(u') I[k,u'] + \mathcal{J}[\td;u']\Bigg] du' \label{tdforsol}\ee
where \be \mathcal{G}(u,u') =
\frac{1}{W}\Big[P(u)\,Q(u')-P(u')\,Q(u) \Big]\label{Gi}\ee where
$P,Q$ are the linearly independent growing and decaying
homogeneous solutions of the fluid-like equation \be
\Bigg[\frac{d^2}{du^2} -6\,\ta(u) + \kappa^2\Bigg] \Bigg\{ \begin{array}{c}
                                                            P(\kappa;u) \\
                                                           Q(\kappa;u)
                                                          \end{array}
\Bigg\}  =
0 \label{fluideqn}\ee and $W$ is their (constant) Wronskian. The
formal solution (\ref{tdforsol}) is again an integral equation,
however it is a re-summed form of the Fredholm-Neumann solution of
(\ref{PQeq})   that displays the asymptotic growth factor explicitly
since asymptotically the growing solution of (\ref{fluideqn})
$P(\kappa;u)$ features the growth factor $\propto \ta$ (see below).

 From
the analysis above, we note that the inhomogeneity $\mathcal{J} $ is
\emph{subleading}  compared to the first term $\ta \, I[k,u]$ for the following reasons:

\begin{itemize}
\item{At early times $u \sim u^*$, $\mathcal{J}$ vanishes as $(u-u^*)^3$
whereas $\ta \, I[k,u]$ remains finite. }

\item{Asymptotically at long
time ($u\rightarrow 0;\ta \rightarrow \infty$) $\ta \, I[k,u] \sim \ta I[k,0] \propto \ta$ whereas
$\mathcal{J} \propto \ln(\ta)$. }

\item{At long wavelengths $k \rightarrow 0$ for which $\alpha \rightarrow 0$ ($\kappa \rightarrow 0$)  it follows that $\mathcal{J} \rightarrow 0$. This is the CDM limit. }

    \item{ For short wavelengths free streaming suppresses density perturbations, this is manifest in
    the expression (\ref{inhofi}). In an iterative solution $\td$ is suppressed by free streaming and
     the term $\mathcal{J}$ involves a further suppression by
     the kernel $\widetilde{\Pi}$ with respect to $I$. }

\end{itemize}  Hence the term $\mathcal{J}$ can be
treated perturbatively as argued above, giving rise to a systematic
Fredholm-Neumann iterative solution of (\ref{tdforsol}) formally in
powers of the free streaming kernels $\Pi$ which for (WDM) are strongly
suppressed by large inverse powers of $\kappa$   at small wavelength
(see the expressions (\ref{Piscdm}-\ref{pisd}) or exponentially
suppressed as for (MB) (see (\ref{Piscdm}))
   \be \td(k,u) = \td^{(0)}(k,u)+ \td^{(1)}(k,u) +\cdots \label{itersol}\ee where
   \bea \td^{(0)}(k,u)  & = &   I[k,u]+   6 \int_{u^*}^u \mathcal{G}(u,u')   \ta(u') I[k,u'] \label{Born}
    \\ \td^{(n)}(k,u) & = &  \int_{u^*}^u \mathcal{G}(u,u') \mathcal{J}[\td^{(n-1)};u']  du' \; ;
     n \geq 1 \label{Beborn}\eea note that $\td^{(0)}(k,u)$ is first order in the free streaming
      kernels, $\td^{(1)}(k,u)$ second order, etc.

We refer to the zeroth-order solution (\ref{Born}) as the \emph{Born
approximation} because of its similarity to
 quantum scattering theory. In references
\cite{boysnudm,darkmatter} it was shown that the Born approximation
is reliable in a wide range of scales. In what follows we will study
the transfer function in the Born approximation as a prelude to a
full  numerical study of (\ref{PQeq}) and its comparison to the Born
and higher approximations to be reported elsewhere.

We note that the Born approximation is \emph{exact} for CDM since in
this case $\alpha = 0$ (consequently $\kappa=0$).

It remains to obtain the homogeneous solutions $P,Q$ of the fluid-like equation (\ref{fluideqn}), which becomes more familiar when written in terms of cosmic time $t$,
\be \Bigg[ \frac{d^2 }{dt^2} + 2H \frac{d }{dt} + \Bigg(\frac{k^2\,\langle V^2(t)\rangle}{a^2(t)}- 4\pi\rho_m(t) \Bigg)\Bigg] \Bigg\{\begin{array}{c}
                                            P \\
                                            Q
                                          \end{array} \Bigg\}
 = 0 \label{jeanseq}\ee where $\rho_m(t);\langle V^2(t)\rangle $ are the density and the velocity squared velocity
dispersion of the DM particle given by (\ref{V2},\ref{sigdisp}). This is equivalent to the Jean's fluid equation for non-relativistic matter recognizing that $k/a(t)= k_{phys}(t)$ is the physical wavevector, and replacing the (adiabatic) speed of sound by the DM particle's velocity dispersion. The term proportional to $k^2$ plays the role of a pressure term and its origin is traced back to the free-streaming kernel $\Pi$ in Gilbert's equation (\ref{PQeq}).

We emphasize that whereas the fluid equation (\ref{jeanseq})
suggests acoustic-like oscillations and is familiar, it is only
\emph{half the story}, it has \emph{no information} on the suppression of perturbations by free streaming.
The solution of Gilbert's equation (\ref{itersol},\ref{Born},\ref{Beborn}) is
completely determined by the inhomogeneity and initial conditions,
these are determined by the past history and describe  the
suppression of density perturbations by free-streaming.

\subsection{Meszaros' equation for WDM}

Just as in the CDM case (see equations
(\ref{2ndord},\ref{legendre})), it is convenient to pass to the
variable $\zeta$, in terms of which the homogeneous equation
(\ref{fluideqn}) becomes \be \Bigg[ (1-\zeta^2)\, \frac{d^2
}{d\zeta^2} -2\,\zeta\,\frac{d  }{d\zeta }+ \nu(\nu+1)\,    -
\frac{(i\kappa)^2}{1-\zeta^2}\, \Bigg]\Bigg\{\begin{array}{c}
                                               P(\kappa,\zeta) \\
                                               Q(\kappa,\zeta)
                                             \end{array}
 \Bigg\} = 0 ~~;~~\nu=2
\label{assoclegendre} \ee this is the associated Legendre equation
with indices $\nu=2\,;\,i\kappa$. We choose the growing and decaying
solutions respectively
 as \be  {P}(\kappa,\zeta) =
\mathrm{Re}\Bigg\{\Big(\frac{\zeta-1}{\zeta+1}\Big)^\frac{-i\kappa}{2}\,
F\Big[-2,3;1-i\kappa;\frac{1-\zeta}{2}\Big]
\Bigg\}\label{Psol}\ee

\be  Q(\kappa,\zeta) = \frac{\sinh(\pi\kappa)}{2\pi\kappa}
\mathrm{Re}\Bigg\{\Gamma(3-i\kappa)\Gamma(i\kappa)\,\Big(\frac{\zeta-1}{\zeta+1}\Big)^\frac{-i\kappa}{2}\,F\Big[-2,3;1-i\kappa;\frac{1-\zeta}{2}\Big]
\Bigg\}\label{Qsol}\ee where $F[a,b;c;z]$ is the hypergeometric
function. We find   \be  P(\kappa,u)   = \cos(\kappa
\,u)\,F_R(\kappa,\zeta(u)) + \kappa
\sin(\kappa\,u)\,H(\kappa,\zeta(u)) \label{Pu}\ee \be Q(\kappa,u) =
-\frac{1}{2} \Bigg\{ 3 P(\kappa,u) + (\kappa^2-2) \bigg[
\cos(\kappa\,u)\,H(\kappa,\zeta(u))-
\frac{\sin(\kappa\,u)}{\kappa}\,F_R(\kappa,\zeta(u))\bigg] \Bigg\}
\label{Qu}\ee where \be F_R(\kappa,\zeta(u)) =
1-\frac{3(1-\zeta)}{(1+\kappa^2)}+
\frac{3(2-\kappa^2)(1-\zeta)^2}{(1+\kappa^2)(4+ \kappa^2)}
\label{F}\ee \be H(\kappa,\zeta) =
-\frac{3(1-\zeta)}{(1+\kappa^2)(4+ \kappa^2)}
\bigg[1+\kappa^2+3\,\zeta \bigg] ~~;~~\zeta(u) = \frac{1}{\tanh[-u]}
\label{Hz}\ee

It is straightforward to confirm that $ P(0,\zeta) =
P_2(\zeta)~~;~~Q(0,\zeta)=Q_2(\zeta)$ are the Legendre functions
solutions of Meszaros's equation (\ref{p2},\ref{q2}) for CDM
perturbations. In fact, in terms of the variable $\ta$ equation
(\ref{fluideqn}) (or alternatively eqn. (\ref{assoclegendre}))
becomes \emph{Meszaro's equation for WDM}, \be \Bigg[\frac{d^2
}{d\ta^2} + \frac{(2+3\ta)}{2\ta(1+\ta)}\,\frac{d }{d\ta} - \frac{3
}{2\ta(1+\ta)} + \frac{\kappa^2 }{4\ta^2(1+\ta)}\Bigg]\Bigg\{\begin{array}{c}
                                                               P \\
                                                               Q
                                                             \end{array}
\Bigg\} = 0 \label{meszaroswdm}\ee whose growing and decaying
solutions are given by (\ref{Pu},\ref{Qu}) respectively.

The asymptotic behavior of the growing and decaying solutions for
$\ta \gg 1~;~ u\rightarrow 0$ are \bea P(\kappa,u) & \rightarrow &
\frac{3(2-\kappa^2)}{u^2\,(1+\kappa^2)(4+ \kappa^2)}
\label{pasy}\\Q(\kappa,u) & \rightarrow &
\frac{-u^3\,(1+\kappa^2)(4+ \kappa^2)}{30} \label{qasy} \eea from
which we extract the Wronskian \be W = \frac{2-\kappa^2}{2}\,.
\label{wronks}\ee Therefore we find \be \mathcal{G}(u,u') =
\frac{2}{2-\kappa^2} \Bigg[ P(\kappa,u)\,Q(\kappa,u')-
P(\kappa,u')\,Q(\kappa,u)\Bigg ] \,.\label{Gofuup}\ee

\begin{figure}[h]
\begin{center}
\includegraphics[height=7cm,width=7cm,keepaspectratio=true]{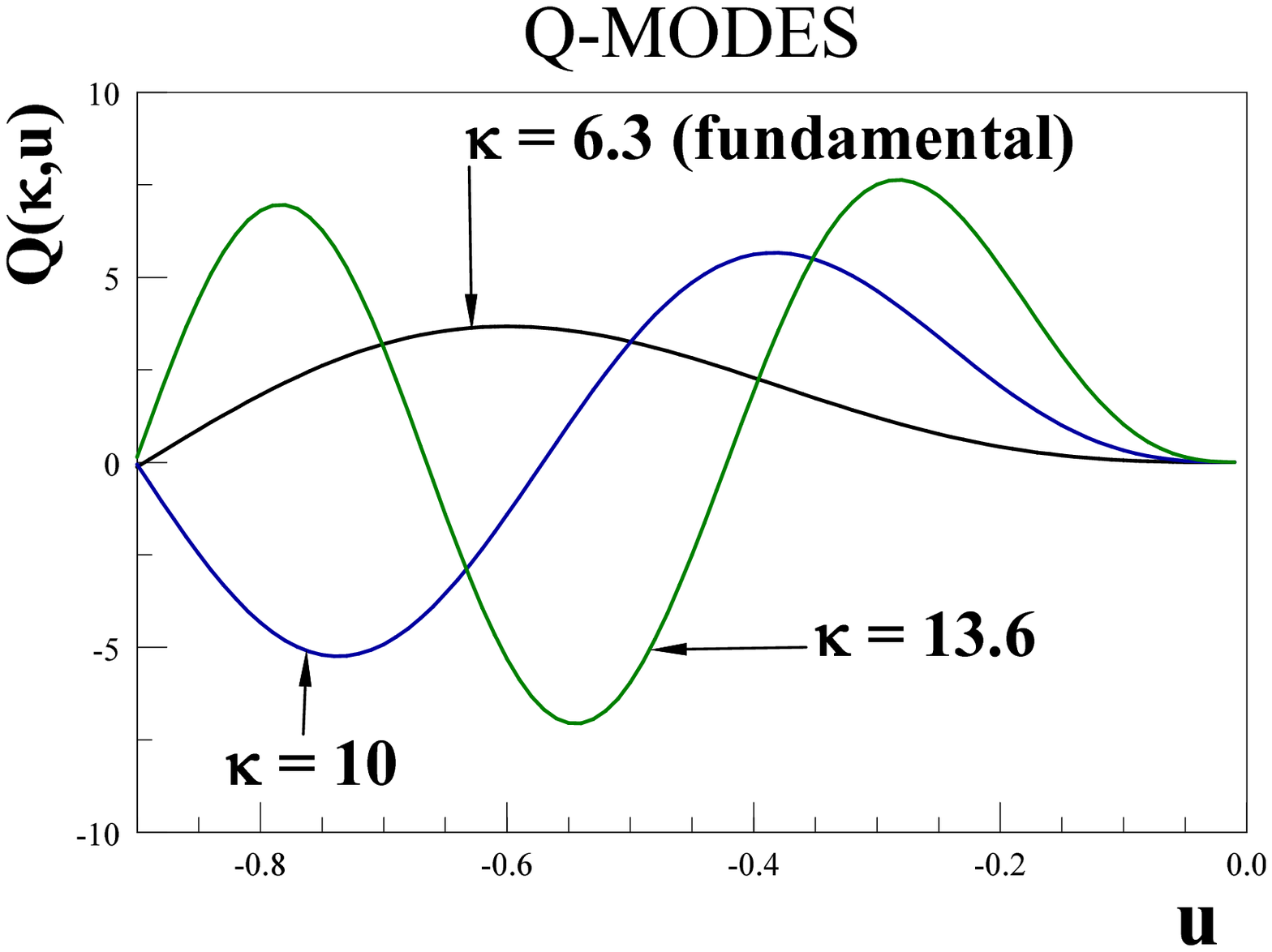}
\includegraphics[height=7cm,width=7cm,keepaspectratio=true]{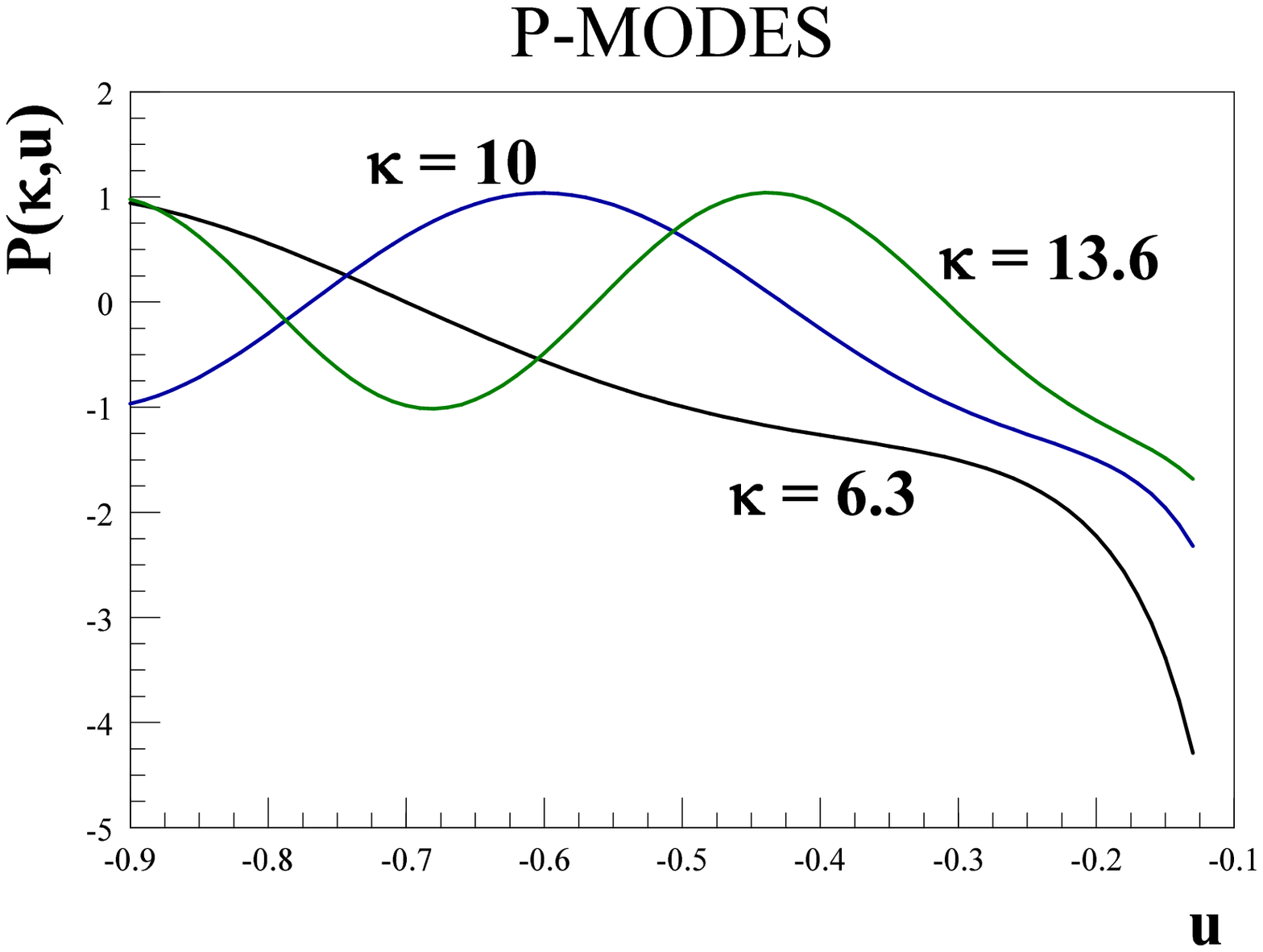}
\caption{Mode functions of fluid equation (\ref{fluideqn}).
$Q(\kappa,u)$ are the decaying and $P(\kappa,u)$ the growing
solutions. The ``fundamental'' decaying solution features a node at
matter-radiation equality. } \label{fig:modes}
\end{center}
\end{figure}

For $\ta \gg 1$ when the gravitational potential is determined by DM
perturbations, using Poisson's equation (\ref{poissonmat}), the
definition of the transfer function (\ref{Tofkdef}) and the solution
for $\td$ (\ref{tdforsol}) along with the asymptotic behavior
(\ref{pasy}) of the growing solution $P(\kappa,u)$   leads to an
\emph{exact} expression for the transfer function \be
T_{WDM}(k;\kappa)= \frac{-5\, k^2_{eq}}{k^2(1+\kappa^2)(4+
\kappa^2)\,\phi_i(k)}\, \int_{u^*}^0 Q(\kappa,u') \Bigg[ 6\ta(u')
I[k;\kappa;u'] + \mathcal{J}[\td;u']\Bigg] du'\,.\label{finTofk}\ee
The CDM transfer function $T_{CDM}(k)$  corresponds to setting
$\alpha=0;\eta_{NR} \rightarrow 0 $ which sets $
\kappa=0;u_{NR}\rightarrow -\infty$ and $\mathcal{J} =0$. In the
Born approximation we obtain  \be T_{B}(k;\kappa)= \frac{-30\,
k^2_{eq}}{k^2(1+\kappa^2)(4+ \kappa^2)\,\phi_i(k)}\, \int_{u^*}^0
Q(\kappa,u')  \,\ta(u') I[k;\kappa;u'] du'\,\label{finTofkBA}\ee and
as explained above the Born approximation is \emph{exact} for CDM
(for $k \gg k_{eq}$).

    $T_{CDM}(k)$ is given by (\ref{Tcdm}) and its leading behavior for  $k\gg k_{eq}$
    is given by  (\ref{Tofk}). It is convenient to normalize the WDM
transfer function defining \be \overline{T}(k) =
\frac{T_{WDM}(k;\kappa)}{T_{CDM}(k)}\, \label{Tbar}\ee where WDM
refers to $\kappa \neq 0$. In the Born approximation we find \be
\overline{T}_{B}(k) = \frac{4}{(1+\kappa^2)(4+\kappa^2)}\,\Bigg[
\frac{\int_{u^*}^0 Q(\kappa,u')  \,\ta(u')
I[k;\kappa;u']\,du'}{\int_{u^*}^0 Q_2(u') \,\ta(u')
I^{CDM}[k;u']\,du'} \Bigg] \label{TbarBA}\ee where $Q_2$ is the
Legendre function given by eqn. (\ref{q2}), \be \ta(u) =
\frac{1}{\sinh^2[u]} \,,\label{tauu}\ee and \be I^{CDM}[k;u] =
I[k;0;u] \,. \label{Icdmdef}\ee The matching scale $u^*$ describes
the transition from when the gravitational potential is dominated by
the radiation fluid to when the DM perturbations dominate. In the
CDM case analyzed in section(\ref{sec:cdm}) we found that this scale
is smaller than the scale of matter-radiation equality. From the
result (\ref{asydelNRRD}) and the analysis leading to
(\ref{matchsol}) we also found that the density perturbation in CDM depends
logarithmically on the change of scale and for $k \gg k_{eq}$ taking
the matching scale \be u^* \simeq u_{eq} = \frac{1}{2} \, \ln\Bigg[
\frac{\sqrt{2}-1}{\sqrt{2}+1}\Bigg]=- 0.881 \label{matchscale} \ee
yields a correction which is of order $k^2_{eq}/k^2 \ll 1$  in the
small scale regime studied here. For WDM, free streaming makes the
dependence on this scale even weaker, and it is evident from the
expression (\ref{TbarBA}) that  the contribution from $\ta \ll 1$ is
suppressed. Hence, in our analysis we   take $u^* = u_{eq} = -0.881$. A
comprehensive numerical analysis   confirms the insensitivity on the
choice of scale for $k \gg k_{eq}$.

 It is convenient to divide the inhomogeneity
(\ref{inhofi}) by $-3\phi_i(k)$ which cancels in the ratio
(\ref{finTofkBA}). Furthermore since the integrals in
(\ref{finTofkBA}) range from $   \eta_{eq}  \leqslant \eta \leqslant
\infty$ and $\phi(k,\eta) \propto 1/(k\eta)^2$ we can safely neglect
the term $3\phi_r$ in the first line in (\ref{inhofi}) as compared
to the second term for $k \gg k_{eq}$. Thus in the ratio
(\ref{finTofkBA}) $I$ simplifies to \be \widetilde{I}[k;\alpha;u] =
\frac{1}{N}\int y^2 f_0(y)
\Big[I_1[k;y;u]+I_2[k;y;u]+I_{ISW}[k;y;u]\Big]dy\; \label{Is}\ee
where \be I_1[k;y;u] = -
\frac{8\,k^2}{k^2_{eq}}\int^{u_{eq}}_{u_{NR}}
\ta^2(u')\,\varphi(k;u') \frac{\sin[y\alpha(u-u')]}{y\alpha}
du'\,,\label{I1f}\ee \be I_2[k;y;u] = \frac{d \ln f_0(y)}{d \ln y}
\Big[-\frac{1}{2}\, j_0\Big(y\alpha(u-u_{NR})+\frac{\kappa}{2} \Big)
\Big] \label{I2f}\ee \be I_{ISW}[k;y;u] = \frac{d \ln f_0(y)}{d \ln
y} \Bigg[2 \int_0^{\frac{\kappa}{2}}
\varphi(z')j_1\Big(y\alpha(u-u_{NR})+\frac{\kappa}{2}-z'
\Big)\Bigg]dz' \label{Iisw}\ee where \be \varphi(z)= \Bigg[\frac{
\big(\frac{z}{\sqrt{3}}\big)
\,\cos(\frac{z}{\sqrt{3}})-\sin(\frac{z}{\sqrt{3}})}{(\frac{z}{\sqrt{3}})^3}
\Bigg] ~~;~~
 z= {k\,\eta} \,. \label{varfi}\ee In the CDM limit ($\alpha \rightarrow 0$)\be \frac{\sin[y\alpha(u-u')]}{y\alpha} \rightarrow (u-u')~~;~~ I_2 \rightarrow -\frac{1}{2}\frac{d \ln f_0(y)}{d \ln y} ~~;~~I_{ISW} \rightarrow 0\,,\label{cdmIs}\ee leading to \be \widetilde{I}^{CDM}[k;u] =  -
\frac{8\,k^2}{k^2_{eq}}\int^{u_{eq}}_{u_{NR}}
\ta^2(u')\,\varphi(k;u')\,(u-u')\, du'\, + \frac{3}{2}\,, \label{Icdmfin}\ee

 which along with (\ref{q2}) determines the denominator in (\ref{finTofkBA}).

In the appendix we provide an explicit form for (\ref{I1f}),  we
gather all the relevant results, and provide a concise summary
  of the Born approximation for an easy numerical
implementation.

 The  contribution $I_{ISW}$ is a result of an
integration by parts in eqn.(\ref{TetaRD})   and is the only
contribution that vanishes in the CDM limit. It
originates in  stage I during (RD) when the WDM particle is still
relativistic.

\begin{figure}[ht!]
\begin{center}
\includegraphics[height=3in,width=3in,keepaspectratio=true]{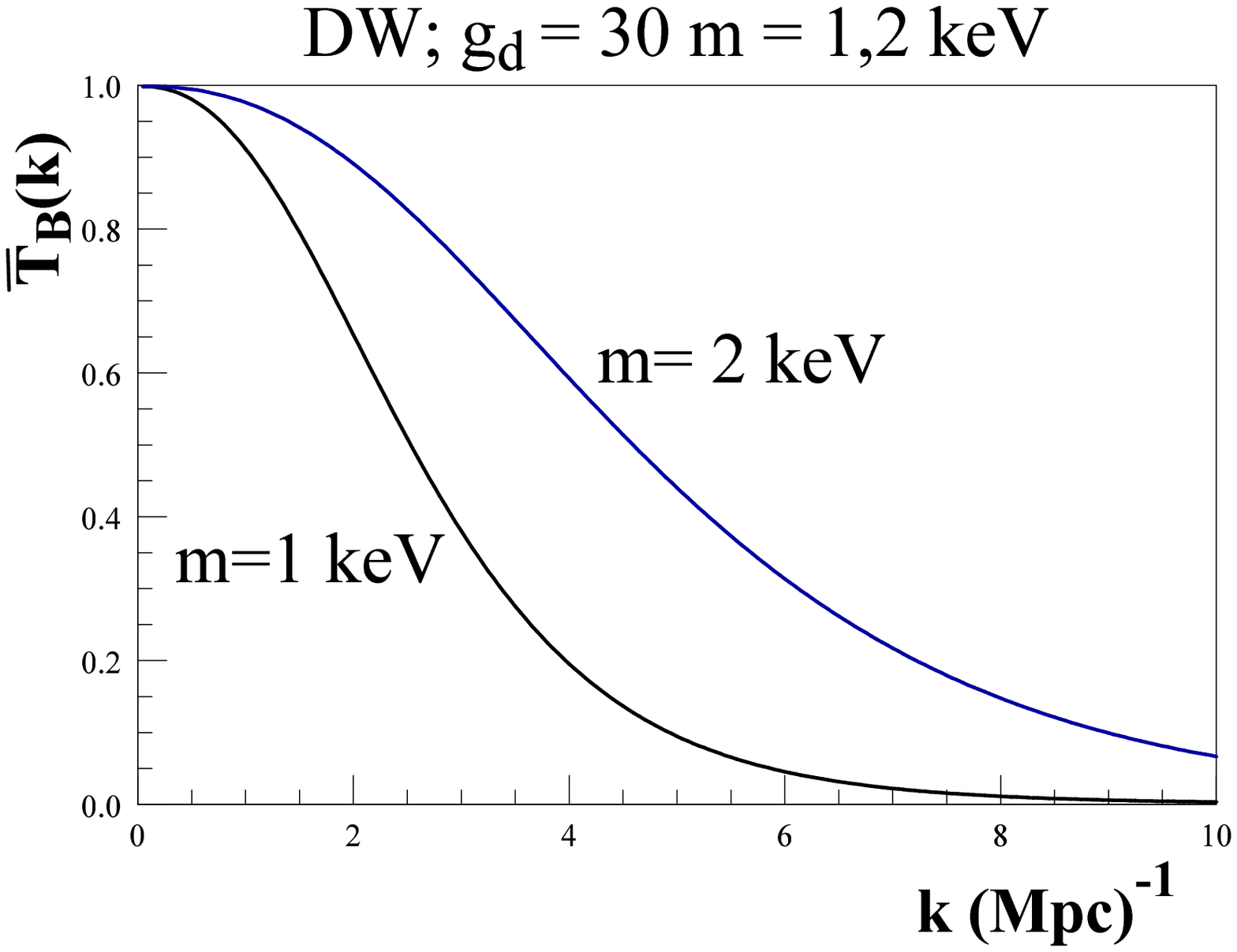}
\includegraphics[height=3in,width=3in,keepaspectratio=true]{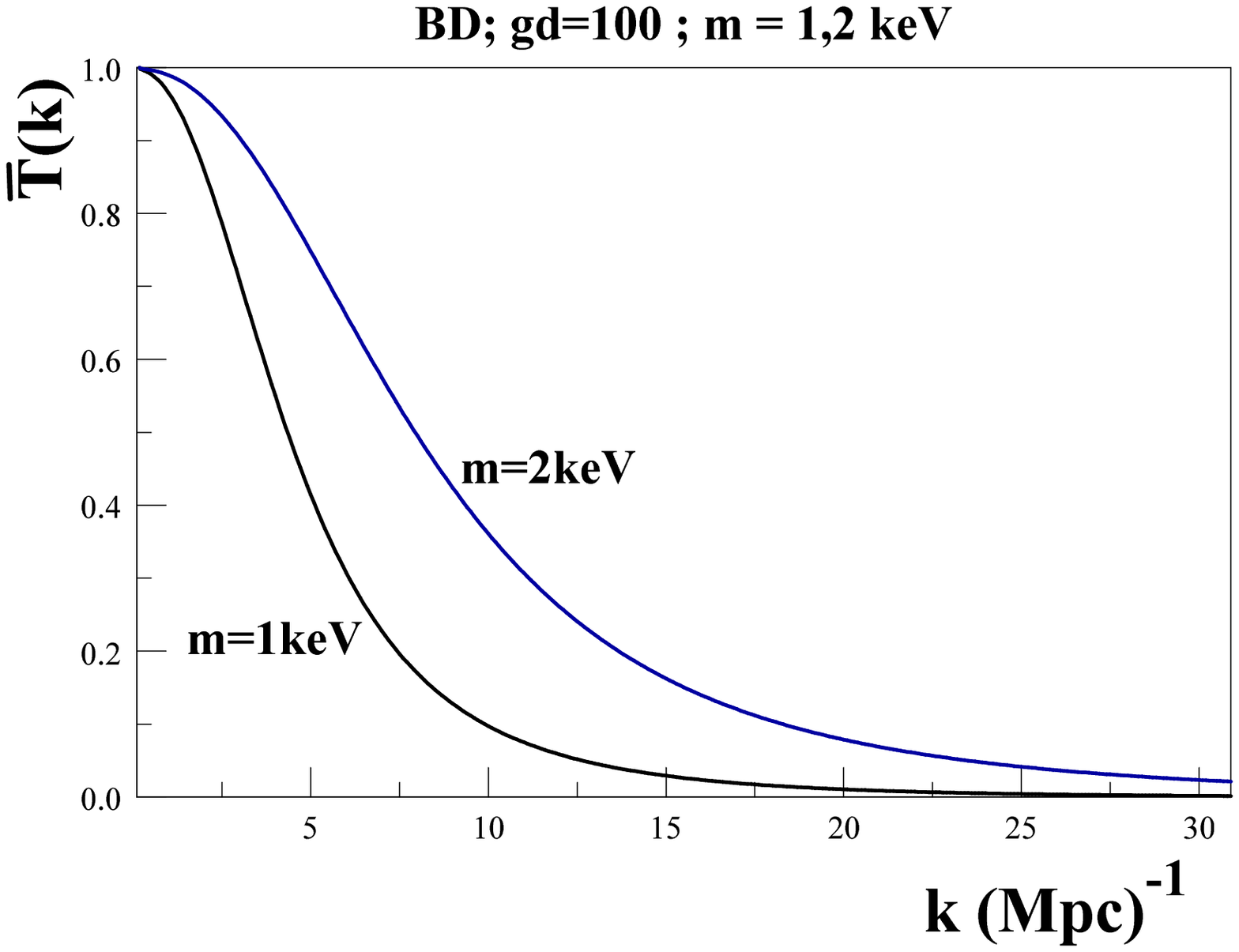}
\caption{$\overline{T}_B(k)$   for DW, and BD for  m=1,2  keV.
Sterile neutrinos produced via the BD-non-resonant mechanism are
colder for the same mass.  } \label{fig:TDW}
\end{center}
\end{figure}

\begin{figure}[ht!]
\begin{center}
\includegraphics[height=3in,width=3in,keepaspectratio=true]{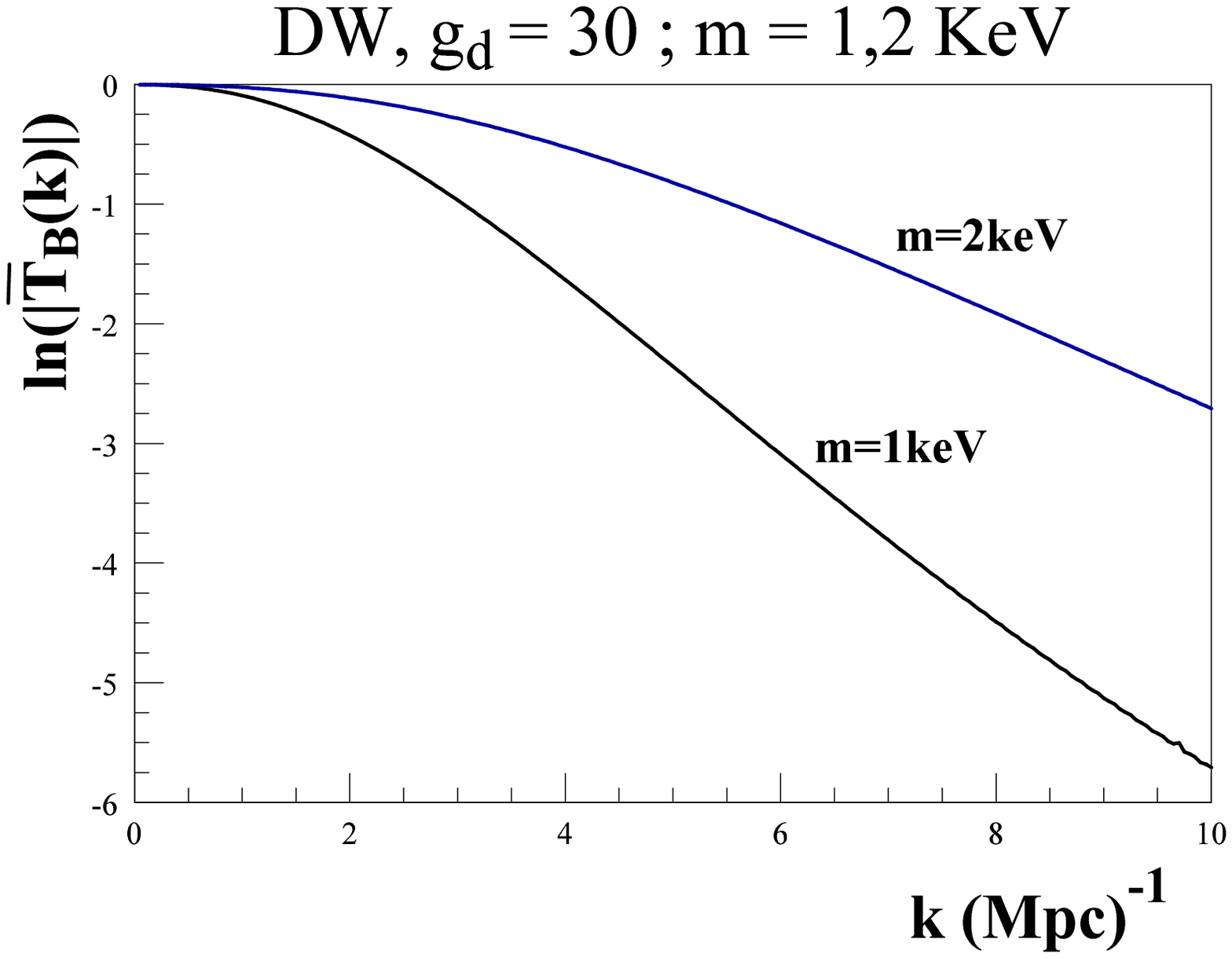}
\includegraphics[height=3in,width=3in,keepaspectratio=true]{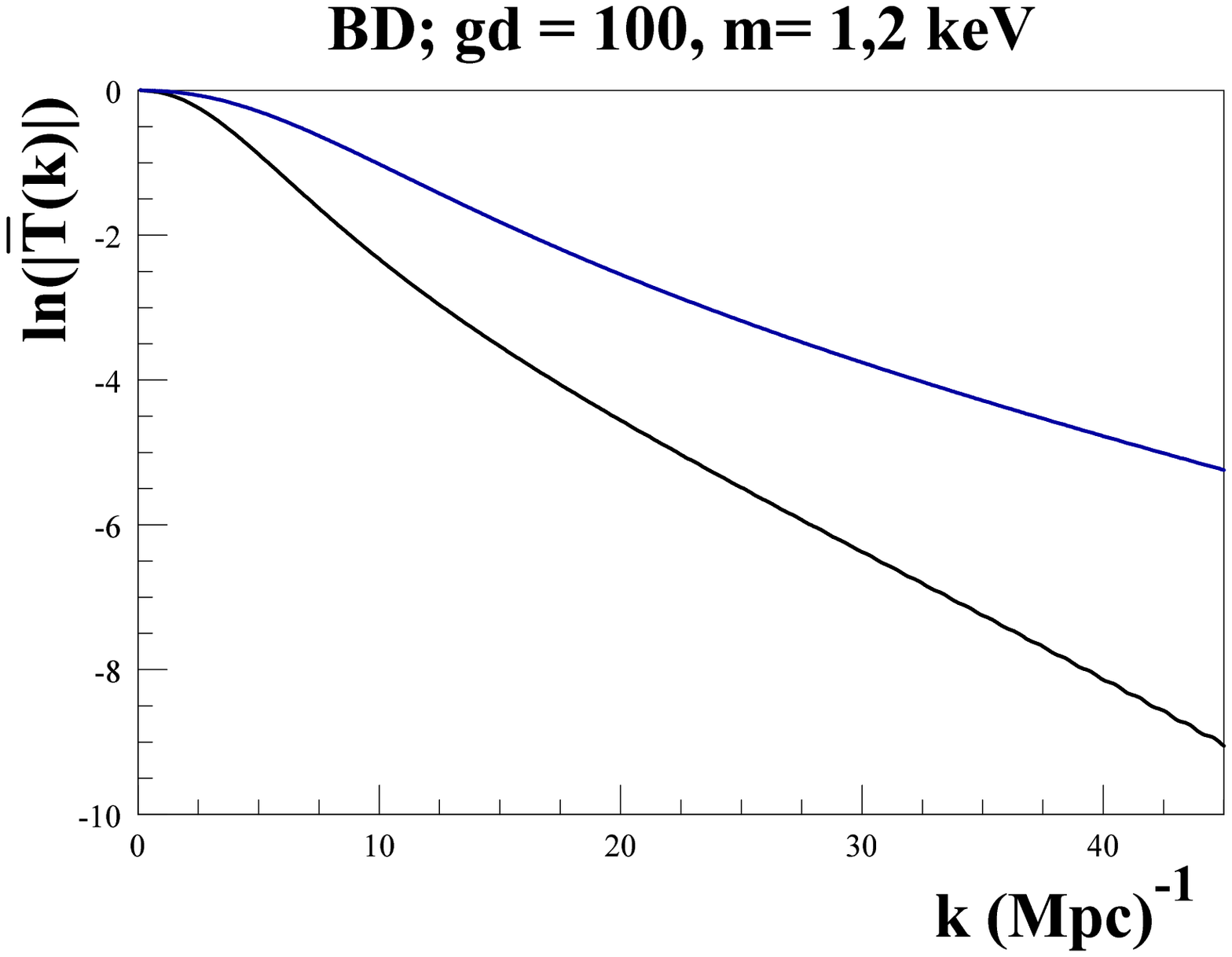}
\caption{$\ln(|\overline{T}_B(k)|)$ a  for DW   and BD, for m=1,2
keV. } \label{fig:logT}
\end{center}
\end{figure}

Figures (\ref{fig:TDW},\ref{fig:logT}) displays the ratio
$\overline{T}$ and its logarithm for both cases of non-resonant
sterile neutrino production (DW,BD). The
production via boson decay at the electroweak scale leads to a
\emph{colder} species for two reasons: i) the effective number of
degrees of freedom at decoupling $g_d$ is larger, therefore the
particle is colder today and at matter-radiation equality, and ii)
the distribution function (\ref{SDf}) favors small momenta and
yields a smaller velocity dispersion (see eqn. (\ref{ybar})). This
is manifest in the transfer functions displayed in fig.
(\ref{fig:TDW}): it is clear from this figure that the wavevector
scale of suppression for DW-produced sterile neutrinos is smaller
than for the BD-production mechanism for the same mass.

\subsection{ISW enhancement:}

As discussed above the contribution $I_{ISW}$ is a direct
consequence of the evolution of density perturbations during stage I
during the (RD) era  described by eqn. (\ref{TetaRD}), and vanishes
in the CDM limit. Therefore it is a distinct contribution to the WDM
transfer function, and only arises from the time evolution of the
Newtonian potential driven by the acoustic oscillations of the
radiation fluid, i.e. an ISW effect.

This contribution is ``out of phase'' with the first two terms
$I_{1,2}$: the Bessel functions $j_0$ of these two terms are
decreasing functions of $k$ until their arguments vanish. Instead,
the Bessel function $j_1$ grows during the initial interval when
$j_0$ decreases. As a result $I_{ISW}$ grows for small $k$. This is
precisely the behavior displayed in fig. (\ref{fig:deltao})
corresponding to the $l=0$ (monopole) component of the density
perturbation (\ref{deltals}) (integrating by parts the integral term
the $j_0$ becomes $j_1$).

\begin{figure}[ht!]
\begin{center}
\includegraphics[height=3in,width=3in,keepaspectratio=true]{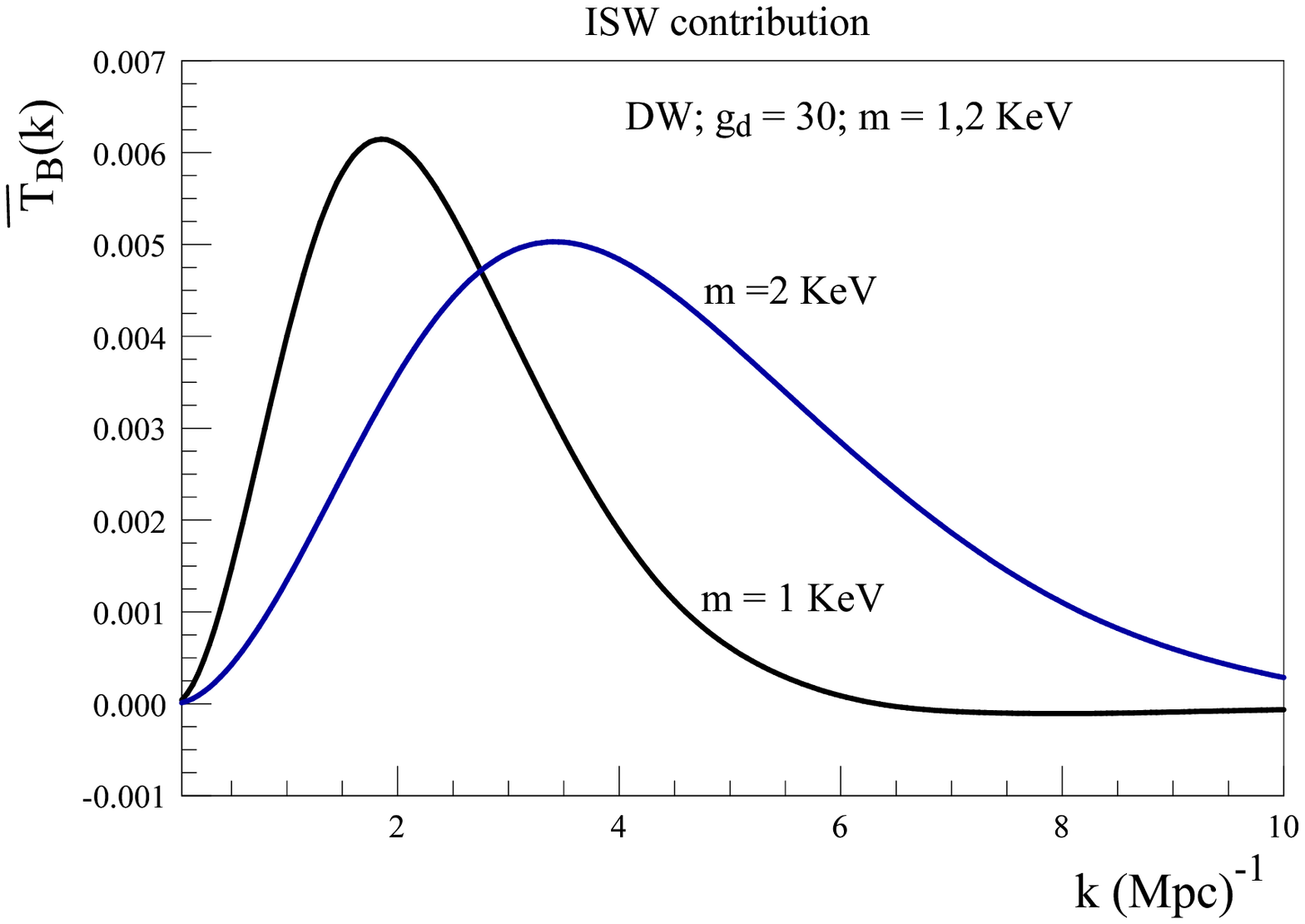}
\includegraphics[height=3in,width=3in,keepaspectratio=true]{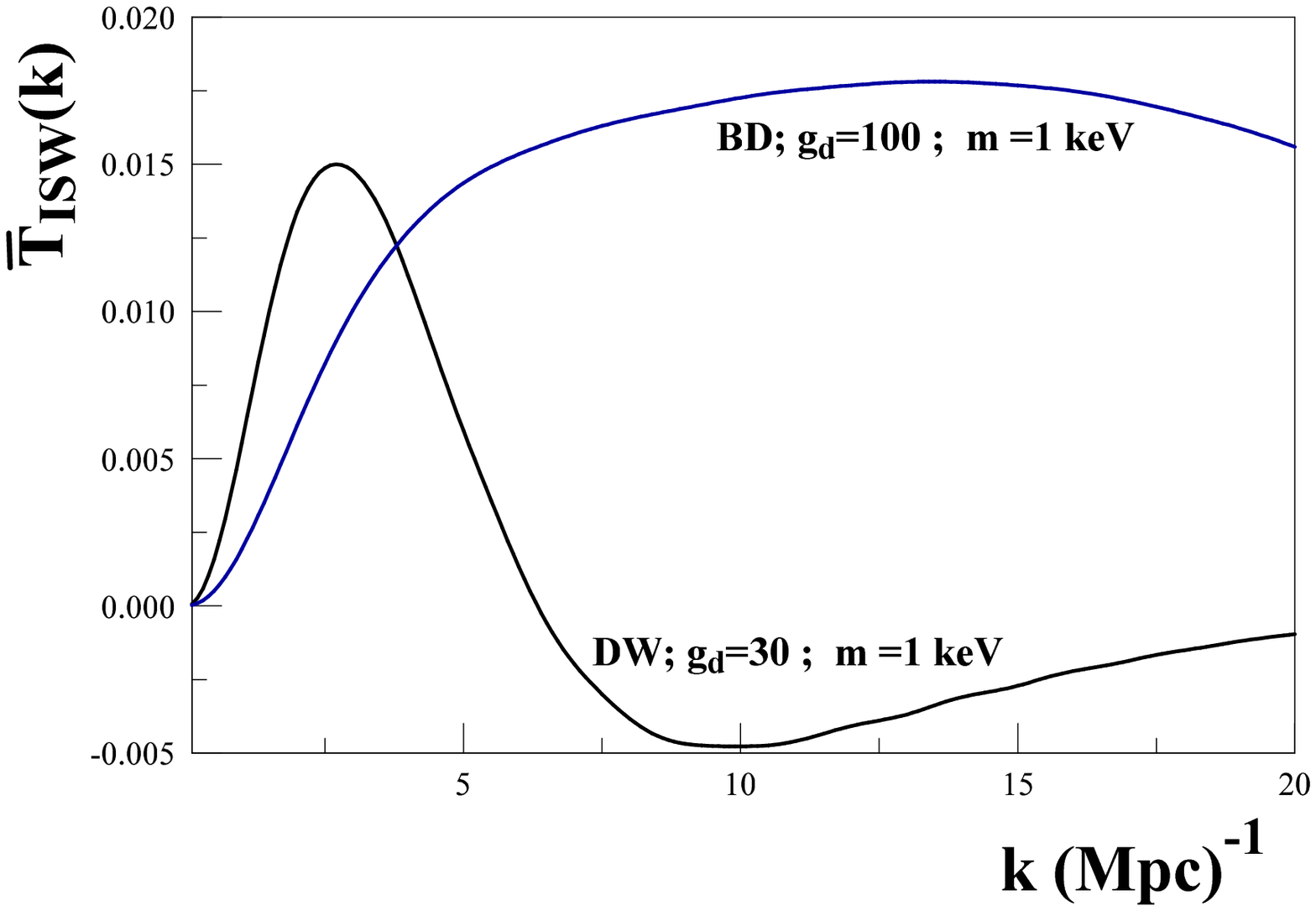}
\caption{The ISW contribution to $\overline{T}_B(k)$ for DW, m=1,2
keV and comparison with BD for m=1 keV. } \label{fig:TISW}
\end{center}
\end{figure}

Since the maximum value of $\eta$ during   stage  I is $\eta_{NR}$
and $k\eta_{NR} = \kappa/2$ the analysis following eqn.
(\ref{RDdelta}) suggests that $I_{ISW}$ features a peak when the
wavelength of the perturbation is approximately the sound horizon at
$\eta_{NR}$, namely $k\eta_{NR} \approx \sqrt{3}\pi$ or $\kappa
\approx 2\pi\sqrt{3}$. This analysis suggests that   $I_{ISW}$
features a peak at $k \lesssim k_{fs} $ because the argument of the
Bessel function is now shifted towards the positive values (since
$u-u_{NR} \geq 0$). The presence of a peak can also be gleaned from
$I_{ISW}$ directly, since for small $z'$ $\varphi(z')$ is nearly
constant but $j_1$ grows, featuring a maximum when  its argument is
$\approx 2$, which obviously suggests a peak at $k \approx k_{fs}$.
Therefore the hotter species, with smaller $k_{fs}$ must feature a
peak at a \emph{smaller} value of $k$ when compared to the colder
species which features the peak at a  larger value  $k$ because of a
larger value of $k_{fs}$. This expectation is borne out by fig.
(\ref{fig:TISW}) that displays the ISW contribution to the Born
ratio $\overline{T}_B$ (\ref{TbarBA}). The ISW enhancement extends
to larger values of $k$ for the colder species for the same mass
(BD) as a consequence of a larger value of $k_{fs}$.

For small $k$ the contributions $I_2$ and $I_{ISW}$ feature opposite
signs, therefore the ISW enhancement competes with and is partially
cancelled by $I_2$ yielding an overall suppression of the transfer
function with respect to CDM. Nevertheless, the ISW enhancement
prolongs the region in $k$ where the transfer function is closer to
that of CDM.

For $\kappa \gtrsim 30$  ($k \gg k_{fs}$)  the ISW contribution features oscillations as discussed in
section (\ref{subsec:RDM}) and shown explicitly in fig. (\ref{fig:deltao}).

\subsection{On the origin of WDM acoustic oscillations:}
The $Q$ and $P$ modes (\ref{Qu},\ref{Pu}) feature acoustic
oscillations as displayed in fig. (\ref{fig:modes}), and only the $Q$
modes enter in the evaluation of the transfer function
(\ref{finTofkBA}). This mode function always vanishes at $u=0$
(today), and there is a particular ``fundamental'' mode  that
features only one other node at matter-radiation equality, for
$\kappa \simeq 6.3$. In the integral leading to the transfer function (\ref{finTofkBA}) the mode function $Q$ multiplies the three contributions to $I$ displayed in (\ref{Is}-\ref{Iisw}). The integral over $y$ with the
distribution function leads to the dephasing of the oscillatory  functions in $I_1,I_2,I_{ISW}$ and their suppression by free streaming.  However,  we can identify some of the more obvious oscillatory contributions.
  From the study in section (\ref{subsec:RDM}) and the results displayed in fig. (\ref{fig:deltao}), the oscillations from the ISW component   begin at
$z_{NR} \gtrsim 15$ ($\kappa \gtrsim 30$) or $k \gtrsim 5\sqrt{6}\, k_{fs}$,  and are  suppressed by free streaming during stages II) and III). This suppression is encoded in
the $y$ integral with the distribution function which contributes during the stages when the particle is
non-relativistic.

 The explicit form of $I_1$ given in the appendix, (\ref{Ia2}) reveals at least two  contributions that   lead to oscillations, these are the term
$\sin[\alpha\,y\,U]/\alpha\,y$ in the first line,  and the second line in  (\ref{Ia2}). After integrating
in $y$ these contributions are proportional to the free streaming kernels (\ref{Piscdm},\ref{pisdw},\ref{pisd}), however, although these contributions   do \emph{not} feature oscillations
after the integration in $y$ by themselves, they are multiplied by the mode function $Q$. Therefore the last term in
the first line in (\ref{Ia2}) leads to oscillations for wavevectors larger than that of the ``fundamental'' Q-mode. The second line in (\ref{Ia2}) yields a contribution of the form
$$\propto \Big[1- \frac{\sin(x_{NR})}{x_{NR}}\Big] $$ times a function suppressed by free streaming. With $x_{NR}=\kappa/2$ this
contribution vanishes for $k \ll k_{fs}$, reaches the value $1$ at $\kappa = 2\pi$ and oscillates around
one for $\kappa \gg 2\pi$. Therefore this function reaches its asymptotic value $\sim 1$ for  values of $\kappa$ near  the ``fundamental'' mode. This analysis leads us to suggest that oscillations in the transfer function begin when the ``fundamental'' mode is excited, namely $\kappa \gtrsim 6.3$.

\begin{figure}[ht!]
\begin{center}
\includegraphics[height=3 in,width=3 in,keepaspectratio=true]{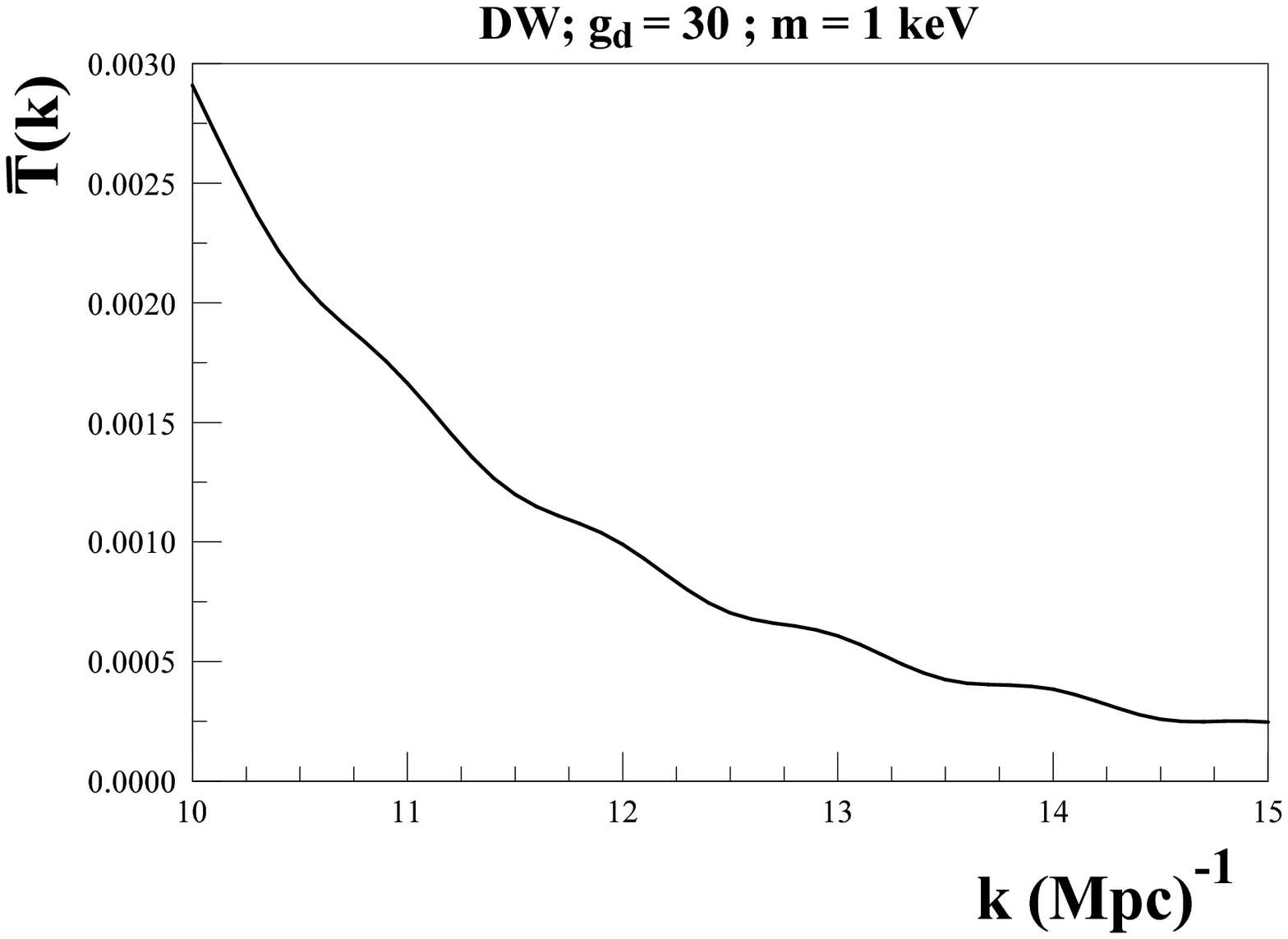}
\includegraphics[height=3 in,width=3 in,keepaspectratio=true]{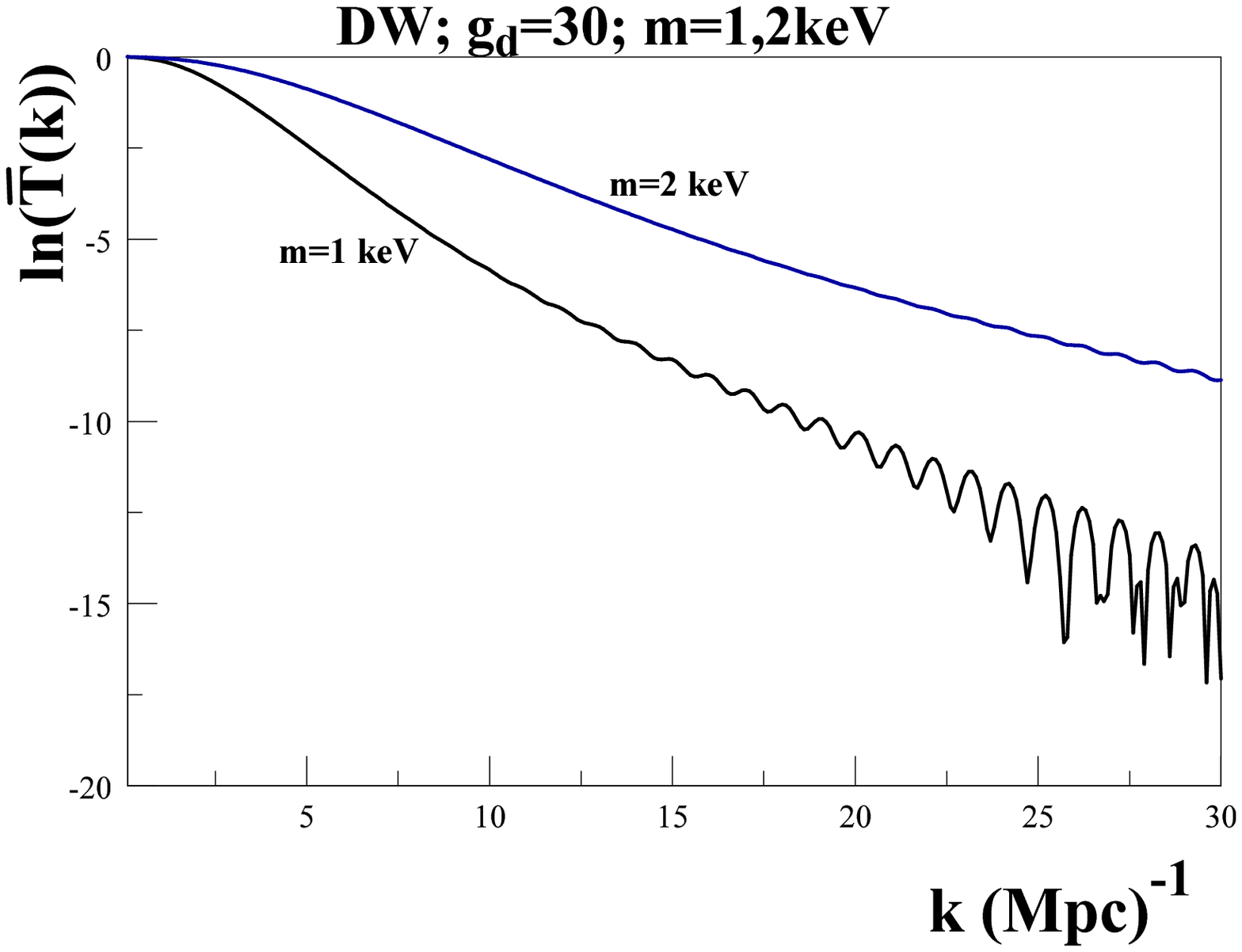}
\caption{Acoustic oscillations at small scales: (DW) species.  }
\label{fig:wdmaosdw}
\end{center}
\end{figure}

\begin{figure}[ht!]
\begin{center}
\includegraphics[height=3 in,width=3 in,keepaspectratio=true]{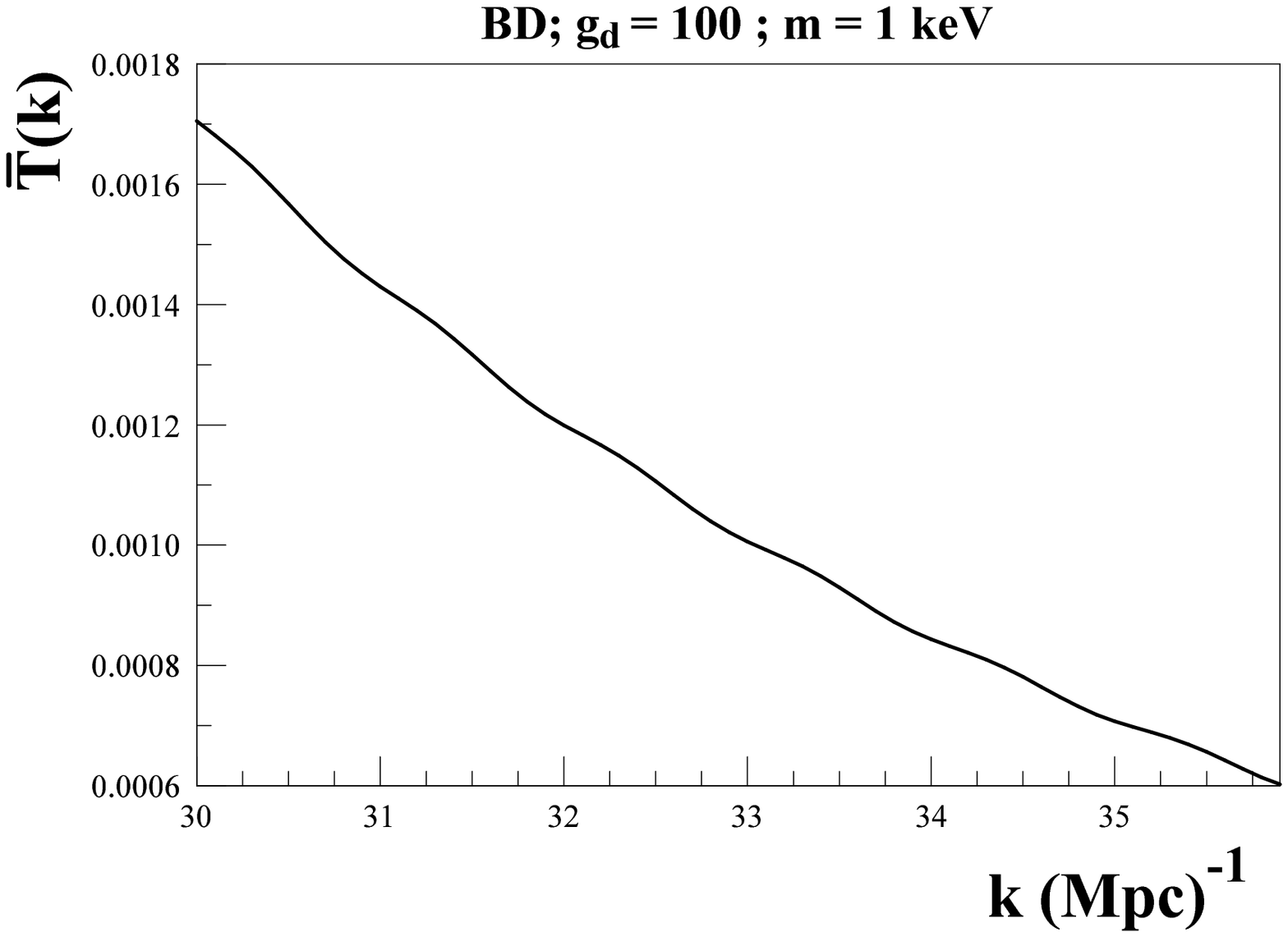}
\includegraphics[height=3 in,width=3 in,keepaspectratio=true]{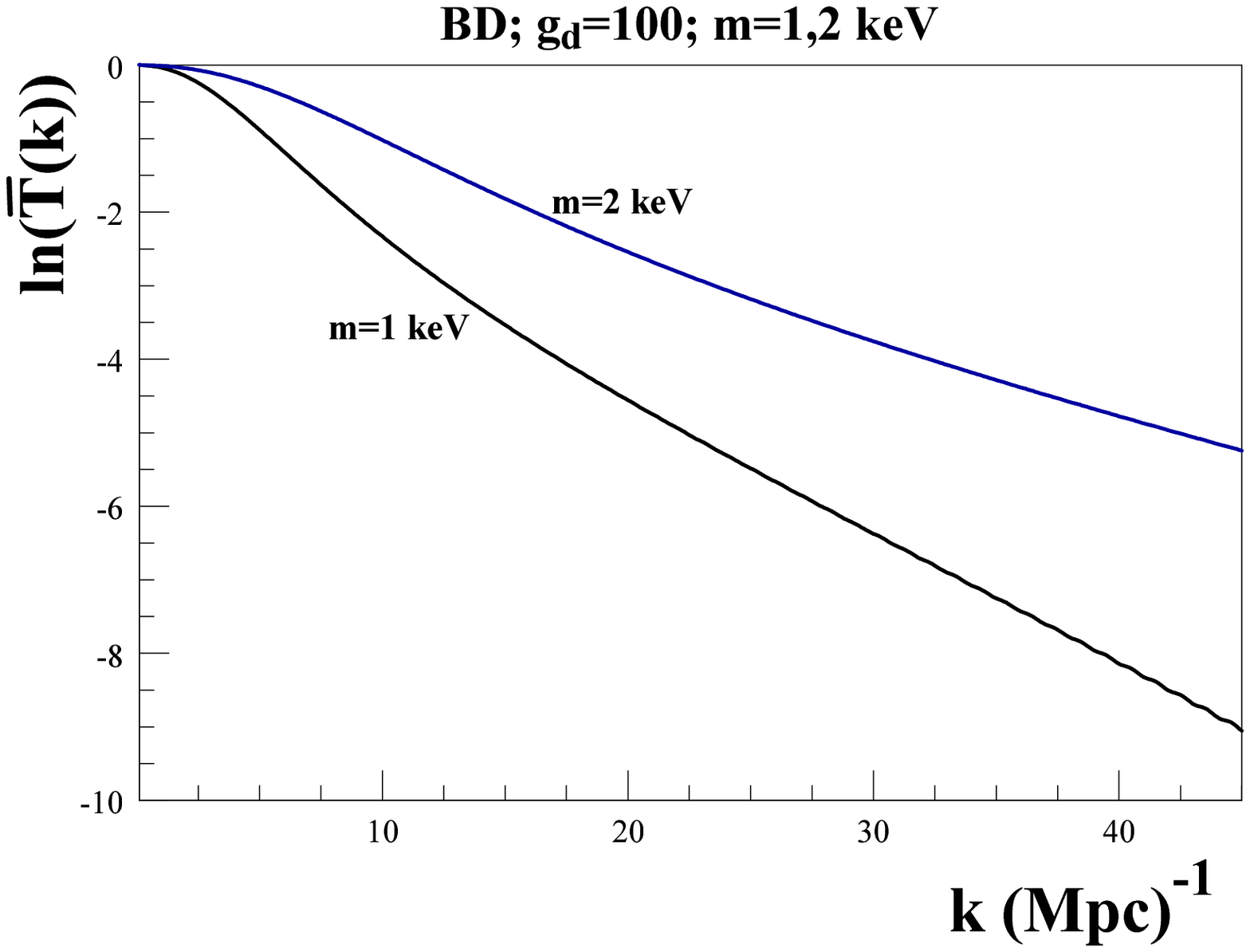}
\caption{Acoustic oscillations at small scales: (BD) species.  }
\label{fig:wdmaosbd}
\end{center}
\end{figure}

For values  of $\kappa \gtrsim  6.3$ the nodes in the mode functions
$Q$ between matter-radiation equality and today  lead to
oscillations in the transfer functions. Therefore we conclude that
oscillations are manifest for \be k \gtrsim 2 ~ k_{fs} \,.
\label{oscthres}\ee This expectation is approximately borne out, for
(DW) with $k_{fs} \sim 7.7\,(\mathrm{Mpc})^{-1} $ we see from fig.
(\ref{fig:wdmaosdw}) that oscillations begin at $k\approx 11
\,(\mathrm{Mpc})^{-1}$ and for (BD) with $k_{fs} \approx 14
\,(\mathrm{Mpc})^{-1}$,  fig. (\ref{fig:wdmaosbd}) shows
oscillations beginning at $k \approx 31.5 \,(\mathrm{Mpc})^{-1}$.
The period of the oscillations is more difficult to assess because
the various terms are out of phase leading to beating of frequencies
(a hint of this is observed in $\ln(\overline{T})$ displayed in fig.
(\ref{fig:wdmaosdw})), however, the approximate estimate $k \simeq
2\,k_{fs}$ for the emergence of oscillations is confirmed by the
numerical analysis.

It is important to recognize that both $I_1,I_{ISW}$ originate in the acoustic oscillations of the
radiation fluid, which couple to the WDM perturbations via the Newtonian potential. Therefore in this
sense, the origin of the WDM acoustic oscillations at small scales is similar to the small scale oscillations
in the CDM transfer function obtained in ref.\cite{loeb}. In that reference the oscillations originated from
the \emph{direct coupling} of the CDM particle to the radiation fluid prior to decoupling, whereas in this
work the coupling is \emph{indirect} through the gravitational potential and the \emph{past history} of the
evolution during stages I and II.

At the scale where WDM acoustic oscillations emerge the transfer function is strongly suppressed by free-streaming and as a result of this suppression in the power spectrum  the relevance of these WDM acoustic oscillations for structure formation is not clear. However, it is conceivable that the effect of the oscillations will be amplified by non-linear gravitational collapse, leading to enhanced peaks and troughs in the matter distribution at low redshift.

The (comoving) scales for these oscillations $k_{ao} \sim 11
\,(\mathrm{Mpc})^{-1}$ for (DW) and $k_{ao} \sim 31.5
\,(\mathrm{Mpc})^{-1}$ for (BD) \emph{could} lead to clumpiness in
the mass distribution with mass scales $M_{DW} \sim 3\times
10^9\,M_{\odot}$ or $M_{BD} \sim 1.8\times 10^8 \,M_{\odot}$
respectively.

The smaller amplitudes of acoustic oscillations for the (BD) species as compared to the (DW) case is consistent with the fact that (BD) sterile neutrinos are \emph{colder} and feature smaller velocity dispersions.

\subsection{Power spectra: interpolation between large and small scales.}
The power spectra normalized to CDM is given by  \be \overline{P}(k) = \Big[\overline{T}(k)\Big]^2 \,, \label{Pnormcdm}\ee and the full power spectra is therefore, \be P(k) =
P_{CDM}(k)\,\overline{P}(k)\,. \label{pofktot}\ee

   Since the transfer
function for WDM particles is indistinguishable from that of CDM for
small $k$, and as we have pointed out above the result (\ref{Tofk})
coincides within a few percent with the result by Bardeen et.
al.\cite{bbks}for $k\gg k_{eq}$,  we use   the numerical fit
provided by Bardeen \emph{et.al.}\cite{bbks} for the CDM transfer
function (without baryons) to extrapolate $P_{CDM}(k)$ to large
scales: \be P_{CDM}(k) = A\,k^{n_s}\,\Big[T_{BBKS}(k)\Big]^2
\label{Pbbks}\ee where $A$ is the overall amplitude and is
determined by the power spectrum of scalar fluctuations during
inflation\cite{dodelson}, and $n_s\simeq 0.96$ is the index of
scalar perturbations during inflation\cite{WMAP5}. Without baryons
and with three relativistic (standard model) neutrinos \cite{bbks}:
\be T_{BBKS}(k)= \frac{\ln\Big[1+2.34 \,q
\Big]}{2.34\,q}\,\Bigg[1+3.89\,q+(16.1 \, q)^2+(5.46 \, q)^3+(6.71
\, q)^4 \Bigg]^{-\frac{1}{4}}~~;~~ q = \frac{k}{\Omega_m
\,h^2}(\mathrm{Mpc})^{-1}\,. \label{tbbks}\ee

Combining eqns. (\ref{Pnormcdm},\ref{pofktot},\ref{Pbbks}) and using the Born approximation for $\overline{T}(k)$ we find the following expression for the power spectra that interpolates between large and small scales,
\be P(k) = A\,k^{n_s}\Bigg[ \frac{4\,T_{BBKS}(k)}{(1+\kappa^2)(4+\kappa^2)}\,
\frac{\int_{u_{eq}}^0 Q(\kappa,u')  \,\ta(u')
\widetilde{I}[k;\kappa;u']\,du'}{\int_{u_{eq}}^0 Q_2(u') \,\ta(u')
\widetilde{I}^{CDM}[k;u']\,du'}  \Bigg]^2 \label{Pofkinter}\ee The inhomogeneities $\widetilde{I},\widetilde{I}^{CDM}$ are given by (\ref{Is}-\ref{Icdmfin}), $u_{eq}=-0.881$ and the mode
functions $Q_2,Q$ are given by eqns. (\ref{q2},\ref{Qu}) respectively. The appendix gives a simplification of these terms along with a numerical implementation. This compact expression provides an interpolation between large and small scales that describes accurately the CDM limit for long-wavelengths  $k \ll k_{fs}$ and captures the free streaming suppression at small scales encoded in the Born approximation. Its numerical implementation is fairly straightforward for arbitrary distribution functions, mass and decoupling temperature.

This is one of our main results.

\subsection{Comparison to numerical results from Boltzmann codes:}
The (WDM) power spectrum for non-thermal   sterile neutrinos produced via the (DW) mechanism has been studied   in refs.\cite{colombi,este,abadwdm,hansendwdm,vieldwdm}. The most recent studies using the Boltzmann codes
CMBFAST\cite{cmbfast} and or CAMB\cite{camb} have been reported in refs.\cite{abadwdm,hansendwdm,vieldwdm}. The results of ref.\cite{hansendwdm} coincide with those of ref.\cite{vieldwdm} and are summarized by the fit given
by eqns. (6,7) in ref.\cite{vieldwdm}. In both refs.\cite{hansendwdm,vieldwdm} the distribution function for sterile neutrinos is that given by eqn. (\ref{DWf}) obtained in ref.\cite{dw}. However, the fitting function eqn. (6,7) given in ref.\cite{vieldwdm} (which reproduces the results of ref.\cite{hansendwdm}) fits the results of the Boltzmann code in the range $k < 5 ~ h~ \mathrm{Mpc}^{-1}$\cite{vieldwdm}.

 In ref.\cite{abadwdm} the kinetic equation for production of sterile neutrinos given in ref.\cite{dw}
  was solved numerically and the solution was input in the numerical Boltzmann codes. In this reference
  the explicit form
of the distribution function is not provided but instead a fitting formula for the transfer function
normalized to CDM is given, eqn. (11,12) in this reference. Whereas both fitting functions in
 refs.\cite{vieldwdm,abadwdm} are of the same \emph{form}, they differ in the powers of momenta:
 at large $k$ the fitting formula (11) in ref.\cite{abadwdm} falls off with a power $\simeq k^{-6.93}$
 whereas the fit given by eqn. (6) in
ref.\cite{vieldwdm} falls of with a power $\simeq k^{-10}$.
Therefore at small scales there is a large difference between these
fits, whereas at large and intermediate scales there is a
substantial agreement (see fig.4 in ref.\cite{abadwdm}). Because in
ref.\cite{abadwdm} the distribution function has been obtained
directly from the numerical integration of the kinetic equation
derived in ref.\cite{dw}, it is not clear whether the main
differences with the results of ref.\cite{vieldwdm} are a
consequence of the distribution function obtained numerically and
input in the Boltzmann code being \emph{different} from the form
(\ref{DWf}) which is the one used in
refs.\cite{hansendwdm,vieldwdm}.

Because our study relies on a pre-determined form of the distribution function and we
 neglect baryons, we can most directly compare our results with the distribution function (\ref{DWf})
  to the results in ref.\cite{vieldwdm}, which also uses the form (\ref{DWf}) and neglects baryons,
  however it includes $\Omega_\Lambda=0.7$ which our study does not.

\begin{figure}[ht!]
\begin{center}
\includegraphics[height=3 in,width=3 in,keepaspectratio=true]{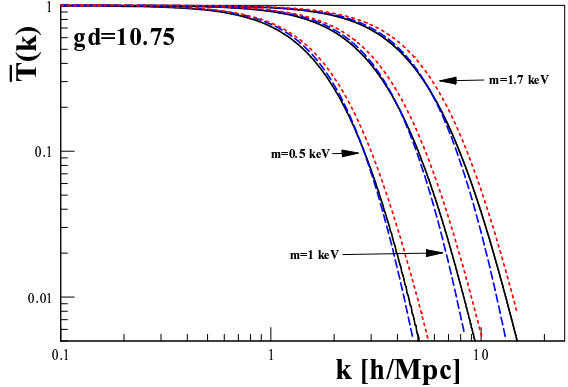}
\caption{Comparison of the transfer function for DW with $g_d=10.75$
normalized to CDM with the results from  Boltzmann codes.  The solid
line is the  semi-analytic result from eqn. (\ref{TbarBA}), the
(blue) dashed line is the result from the interpolation eqns.(6,7)
(non-thermal case) from ref.(\cite{vieldwdm}), the (red) dotted line
is the result from the interpolating fit eqn. (11,12) in ref.
(\cite{abadwdm}). For all cases $h=0.72, \Omega_{DM}h^2 = 0.133,g_d
= 10.75$.} \label{fig:compara}
\end{center}
\end{figure}

  We compare our results for
   the transfer function $\overline{T}(k)$ (normalized to CDM) given by (\ref{TbarBA})
    with those obtained from  the fit given by  eqns. (6,7) (for the non-thermal case)
    in ref.\cite{vieldwdm}, with the caveat that this fit may \emph{not} be the correct description
    of the power spectrum for $k > 5~h~\mathrm{Mpc}^{-1}$ as suggested by the discussion in
     ref.\cite{vieldwdm}. We also compare to the fit (11,12) in ref.\cite{abadwdm}, although
     this may \emph{not} be fair comparison because we \emph{assume} the distribution function
     (\ref{DWf}) whereas in ref.\cite{abadwdm} the effective distribution function may be
     different and the difference cannot be quantified in absence of a functional form.
     Furthermore, we use the ``standard'' value $g_d = 10.75$ for the comparison, whereas
     as discussed in ref.\cite{abadwdm} the actual value may differ because this species of
     sterile neutrinos is produced very near the QCD phase transition where the effective number of
      relativistic degrees of freedom vary rapidly. Recognizing all these caveats  we present the
      comparison of  the transfer functions normalized to CDM  in the range of masses and scales displayed
       in refs.\cite{abadwdm,vieldwdm} in fig. (\ref{fig:compara}), $m=0.5,1.0,1.7\,\mathrm{keV}$:
       the solid line is $\overline{T}(k)$ from the Born
       approximation (\ref{TbarBA}), the dashed line is the
       fit given by eqns(6,7) for the non-thermal case in
       ref.\cite{vieldwdm}, the dotted line is the fit (11,12) in
       ref.\cite{abadwdm}.

\begin{figure}[ht!]
\begin{center}
\includegraphics[height=3 in,width=3 in,keepaspectratio=true]{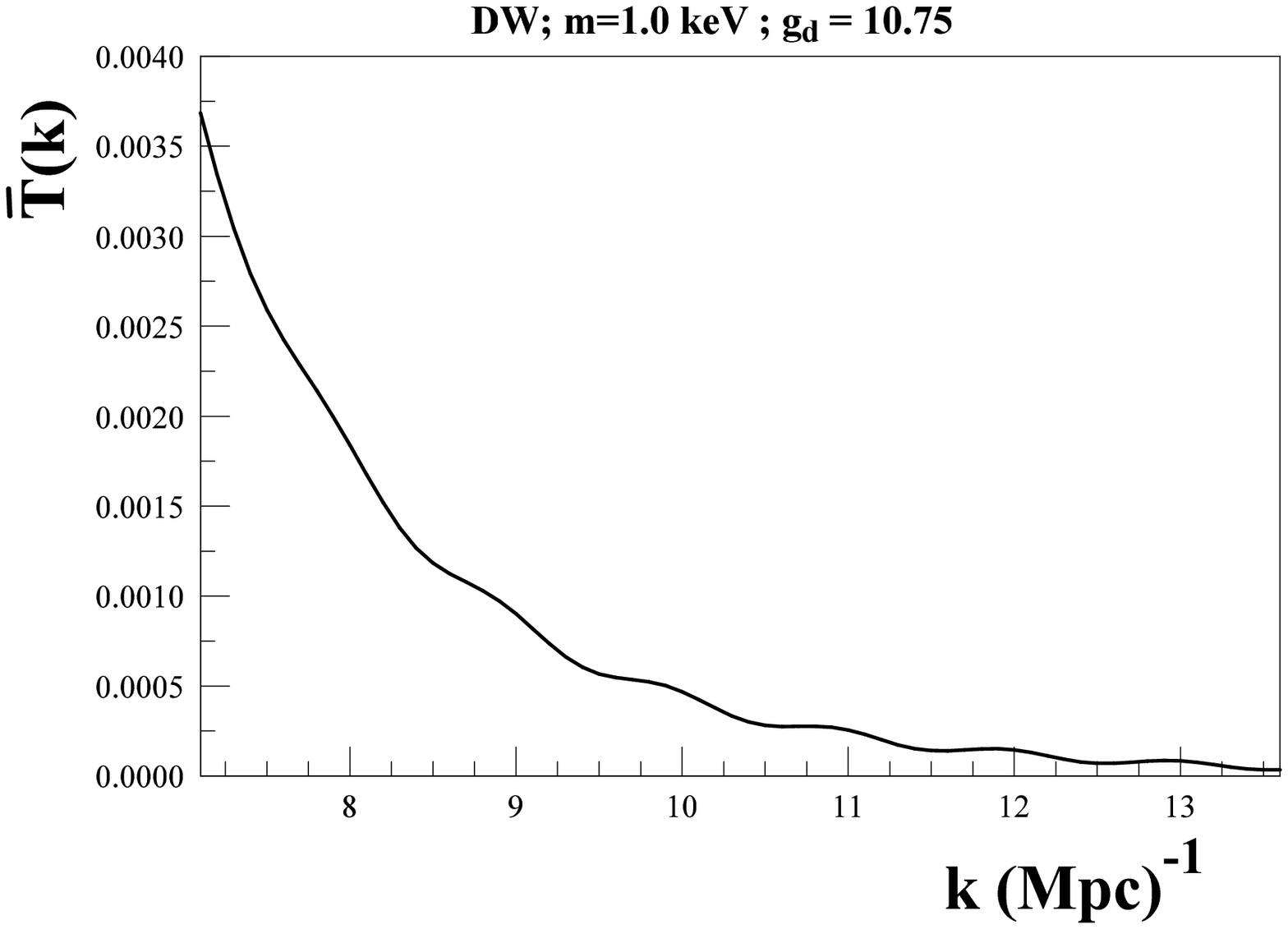}
 \includegraphics[height=3 in,width=3 in,keepaspectratio=true]{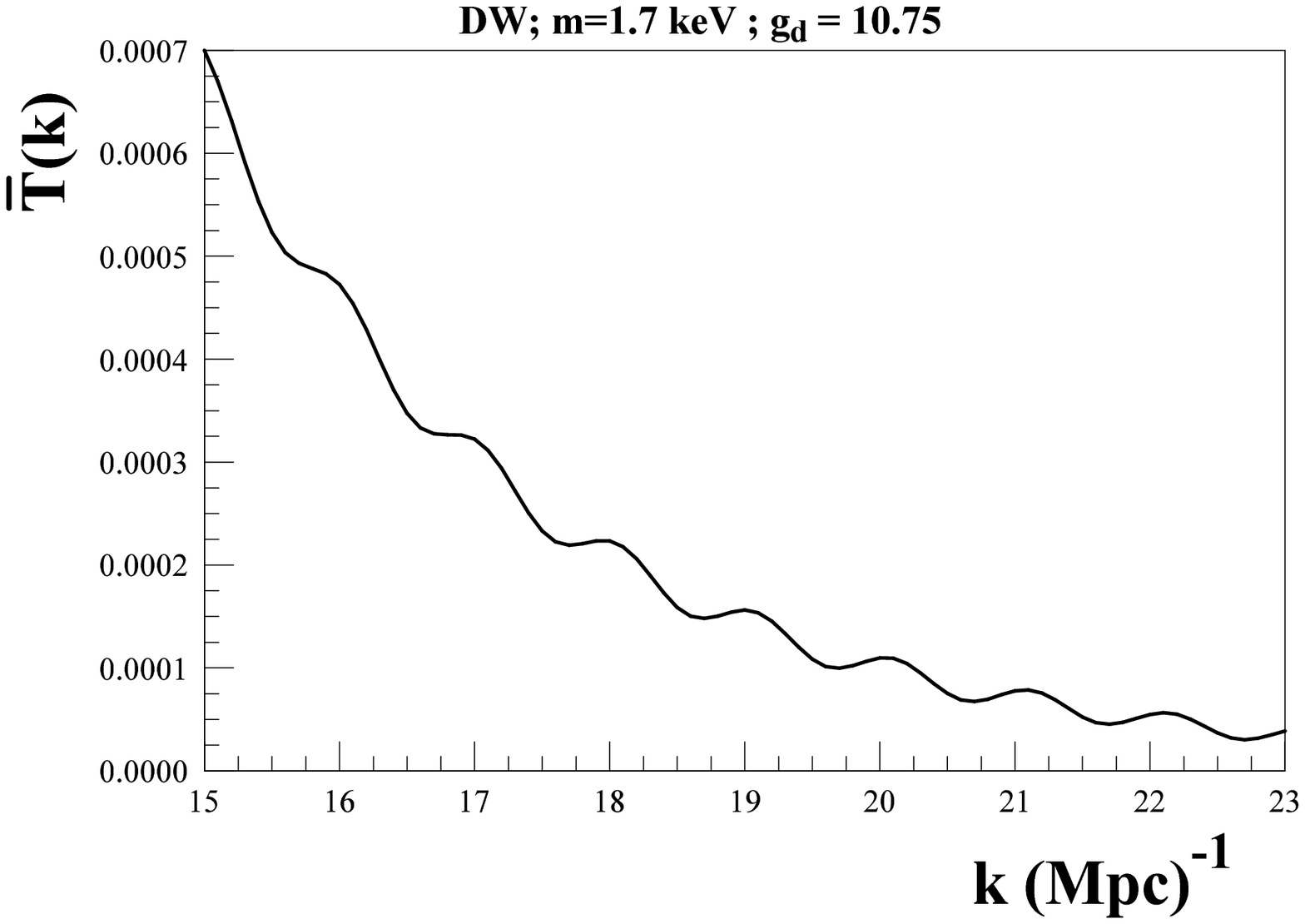}
\caption{$\overline{T}(k)$ from the semianalytic approximation (\ref{TbarBA}) displaying the acoustic oscillations at small scales $k \gtrsim 2 k_{fs} \sim 10.8, 18.5~(\mathrm{Mpc})^{-1}$ for $m =1.0,1.7 \mathrm{keV}$ respectively. Note that the horizontal scale is in $(\mathrm{Mpc})^{-1}$ and that vertical scales differ by a factor $5$ between the two figures.}
\label{fig:compaosci}
\end{center}
\end{figure}

We find a remarkable agreement, to less than $5\%$ with the fit
given by eqns. (6,7) (non-thermal case) in ref.\cite{vieldwdm} in a
wide range in which their fit is valid  (see discussion in
       ref.\cite{vieldwdm}) for $m  \gtrsim 1\,\mathrm{keV}$ the
       agreement is substantially better in a far larger range.  In fig.
(\ref{fig:compara})  the comparison is in the range displayed in
refs.\cite{abadwdm,vieldwdm} to highlight agreements and
discrepancies. In all cases reported in the literature the range
studied or displayed are for wavectors $k$ \emph{far smaller} than
the range in which the acoustic oscillations become manifest. The
\emph{approximate} estimate (\ref{oscthres}) for the threshold
suggests that
 for $m=0.5, 1.0,1.7~\mathrm{keV}$ oscillations should be manifest
 for $k  \gtrsim 5.4, 10.8, 18.5~(\mathrm{Mpc})^{-1}$
  (corresponding to $k \gtrsim 7.5, 15.0, 25.6~h~(\mathrm{Mpc})^{-1}$ respectively).
   Fig. (\ref{fig:compaosci}) displays $\overline{T}(k)$ from (\ref{TbarBA}) in a linear-linear scale
    for $k \gtrsim 2 k_{fs} $ for $m=1.0,1.7 ~\mathrm{keV}$. These figures are
     the continuation of the \emph{same} $\overline{T}(k)$  displayed as  solid lines in fig. (\ref{fig:compara})
      to the smaller scales $k  \gtrsim 2 k_{fs}$ in each case.

 This comparison, with all the caveats mentioned above, suggests that the semi-analytic formulation
 along with the Born approximation summarized by (\ref{TbarBA}) captures the
  essential physical processes and provide a reliable tool to study the transfer function and power spectra for arbitrary distribution functions.

\subsection{Impact on N-body simulations and Lyman-$\alpha$
constraints:}

State of the art N-body simulations of galaxy
formation\cite{tikho,maccio} and large high resolution data sets of
Lyman-$\alpha$  forest spectra\cite{lyman,lyman2,vieldwdm} have
been used to constrain the mass of WDM particles\cite{lyman2,vieldwdm}.

The most recent large scale N-body simulations\cite{tikho,maccio}
incorporate WDM by considering a power spectrum that is cutoff at
small scales, however, initial velocity dispersion is not yet
included in the simulations. Extracting constraints from the
Lyman-$\alpha$ forest involves also large scale numerical
simulations, and the most recent constraints\cite{lyman2} on the
mass of the WDM particle rely either on a thermal or (DW)
distribution functions. The (DW) distribution function is
proportional to a thermal distribution function and the
proportionality constant \emph{only} determines the abundance but is
irrelevant for the free streaming length or indeed the transfer
function (as can be gleaned from the previous sections).

\begin{figure}[h!]
\begin{center}
\includegraphics[height=4in,width=4in,keepaspectratio=true]{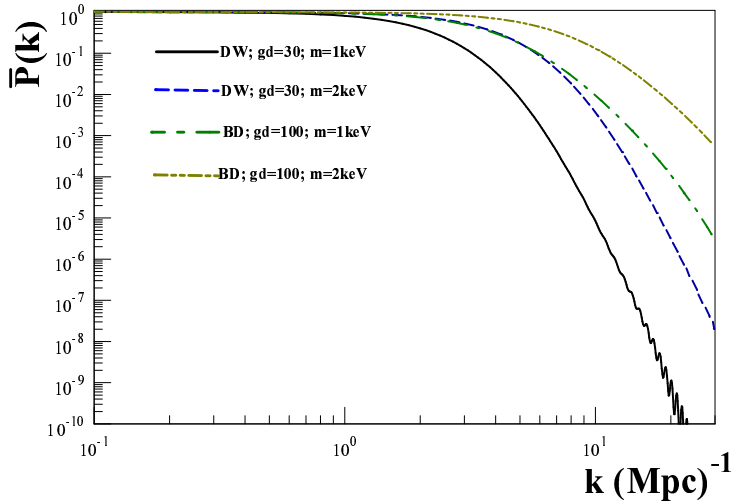}
\caption{The power spectra normalized to CDM for DW and BD with $m=1,2\,\mathrm{keV}$. Note
 that for the same mass the BD (colder species) is less suppressed than the DW (hotter species).} \label{fig:pofknor}
\end{center}
\end{figure}

Our study points out that the power spectrum features a \emph{quasi-degeneracy} in that a more massive
WDM particle with a (DW) distribution function features a similar power spectrum as a less massive one but
with a (BD) distribution function \emph{in a wide range of scales}.
To make this more explicit, fig. (\ref{fig:pofknor}) displays the power spectra normalized
 to CDM (\ref{Pnormcdm}).

From this figure it is clear that $\overline{P}(k)$ for (DW) with
$m= 2\,\mathrm{keV}$ is almost indistinguishable from
$\overline{P}(k)$ for (BD) with $m= 1\,\mathrm{keV}$ for $ k
\lesssim 6-8\,(\mathrm{Mpc})^{-1}$. This is because the (BD) sterile
neutrinos are \emph{colder} for two reasons: they decouple earlier
and their distribution function favors small momenta, therefore the
(BD) WDM particle has \emph{smaller} velocity dispersion. Therefore,
we emphasize that the mass is \emph{not} the only relevant indicator
for the power spectrum of the WDM particle, but also two important
aspects must enter in the assessment: the decoupling temperature
(the higher, the colder the particle) and the details of the
distribution function at small momenta: enhanced small momentum
behavior leads to a \emph{colder species} and a less suppressed
power spectrum, for a given mass.

Hence the \emph{quasi-degeneracy}: the current constraints on the mass of the WDM particle, either from
(quasi) WDM simulations (quasi because these simulations do not include velocity dispersion of the WDM particle, therefore miss the aspects related to the non-thermal distribution functions), or from Lyman-$\alpha$ forest analysis, which typically input thermal WDM distribution functions  or (DW) distribution function which
is indistinguishable from thermal for the purpose of the transfer function, do not
\emph{directly} apply to non-thermal WDM particles.

\begin{figure}[h!]
\begin{center}
\includegraphics[height=5in,width=5in,keepaspectratio=true]{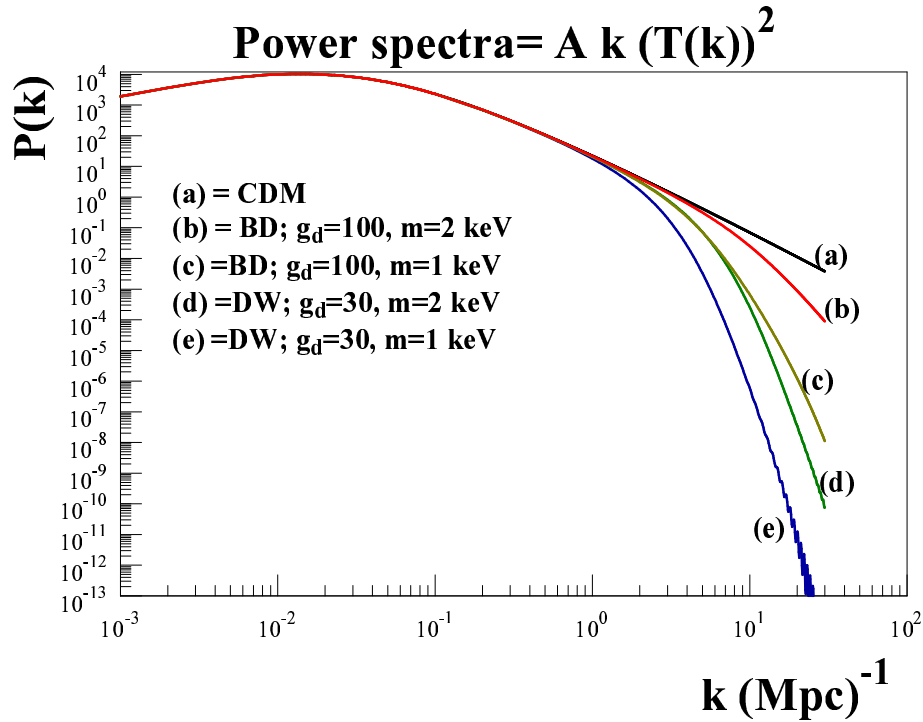}
\caption{The matter power spectra: $P(k)= A\,k \, (T(k))^2$ for
$n_s=1$ (A is the normalization amplitude) for CDM, DW and BD for
$m=1,2 \mathrm{keV}$. Note the quasi degeneracy for DW with
$m=2\mathrm{keV}$ (d) and BD with $m=1\mathrm{keV}$ (c) in a large range of
$k\lesssim 12\,(\mathrm{Mpc})^{-1}$. } \label{fig:powerspectra}
\end{center}
\end{figure}

To highlight this point, we obtain the full power spectra for the
different species considered here using the interpolating eqn.
(\ref{Pofkinter}). Fig. (\ref{fig:powerspectra}) displays $P(k)$ for
$n_s=1\,;\,\Omega_m\,h^2 = 0.134$ for the different species
considered here. Note how the two cases (c) (BD,
$m=1\,\mathrm{keV}$) and (d) (DW, $m=2\,\mathrm{keV}$) are nearly
indistinguishable for $k \lesssim 6-8 \,(\mathrm{Mpc})^{-1}$.

Therefore, we conclude that \emph{non-thermal} distribution functions may evade the constraints
 on the mass of the WDM particles both from current numerical simulations and the Lyman-$\alpha$ forest data.

\section{Conclusions and discussions}

In this article we provide a semi-analytic study of small scale aspects of  the power spectrum of WDM candidates in a radiation-matter cosmology for arbitrary mass and distribution function of the decoupled WDM particle. There are three stages in the evolution of density perturbations of WDM candidates that decouple while they are relativistic: stages I) and II) describe the evolution during the RD era when the particle
is relativistic and non-relativistic respectively but the gravitational potential is dominated by the radiation fluid, during stage III, the particle is non-relativistic and matter density perturbations dominate the
gravitational potential. We consider adiabatic initial conditions determined when all the cosmologically
relevant modes are superhorizon. The collisionless Boltzmann equation is solved in the three stages by
using the solution at the end of a stage as the initial condition for the next stage. The transfer function is characterized by two widely separated scales: $k_{eq} \simeq 0.01\,(\mathrm{Mpc})^{-1}$ corresponding to the wavevector that enters the horizon at matter-radiation equality and $$  k_{fs} = \frac{\sqrt{3}\,k_{eq}}{2\,\langle V^2_{eq}\rangle^\frac{1}{2}} $$ where $\langle V^2_{eq}\rangle^\frac{1}{2}$ is the mean square root velocity dispersion of the WDM particle at \emph{matter-radiation equality}. This latter scale also determines the size of the comoving horizon when the WDM particle becomes non-relativistic:
$$\eta_{NR} = \frac{\sqrt{3}}{\sqrt{2} \, k_{fs}}\,.$$ During stages I) and II) the acoustic oscillations in the radiation fluid dominate the gravitational potential, leading to an ISW effect that amplifies WDM density perturbations on scales larger than the sound horizon at $\eta_{NR}$. This amplification translates in a prolonged plateau in the transfer function for $k \lesssim k_{fs}$ which is more pronounced for \emph{colder} species since these feature a larger $k_{fs}$.

When the particle is non-relativistic and WDM perturbations
dominate the gravitational potential, the evolution is described by the Boltzmann-Poisson equation
which yields an integral equation for density perturbations and is equivalent to integro-differential
 equation with an inhomogeneity and initial conditions determined by the past history during stages I and II.
 This equation is amenable to a systematic Fredholm expansion valid at small scales, whose leading order is
 the Born approximation which establishes a direct relation with a fluid description of WDM perturbations.
  The resulting fluid equation is the generalization of Meszaros' equation for CDM but with an inhomogeneity
   and initial conditions that incorporate suppression by free streaming during the first two stages.
    The Born approximation lends itself to a simple numerical implementation for
     \emph{arbitrary distribution functions and mass of the decoupled WDM particle}.
     Its main ingredients are the growing and decaying solution of the generalized
     Meszaros fluid equation for WDM perturbations, and the initial conditions and
     inhomogeneity that are completely determined by the past history during the first two stages.
     The solutions of the fluid equations feature \emph{(WDM)-acoustic oscillations} which are manifest
     in the transfer function and power spectra for $k \gtrsim 2 k_{fs}$.

  An approximate form of the power spectra that interpolates
between large and small scales for arbitrary distribution functions
is given by eqn. (\ref{Pofkinter}) and a simple and concise summary of
the main elements of the Born approximation and its numerical
implementation are provided in the appendix.

We study in detail and compare the transfer functions and power spectra of sterile neutrinos with mass
 in the $\sim \mathrm{keV}$ range for two non-resonant production mechanisms: Dodelson-Widrow (DW)
 (sterile-active mixing) and Boson-decay (BD) near the electroweak scale.
 The former yields a distribution function proportional to a thermal fermion but with a
  decoupling temperature $T_d \sim 150 \,\mathrm{MeV}$, whereas the
   latter leads to a strongly non-thermal distribution with a decoupling temperature
   $T_d \sim 100 \, \mathrm{GeV}$ that  favors small momentum and yields a \emph{colder}
    species of sterile neutrinos for a given mass.
For a sterile neutrino with mass $\sim \,\mathrm{keV}$ the (DW)-species is \emph{warmer}
 with $k^{(DW)}_{fs} \simeq 7.7\,(\mathrm{Mpc})^{-1}$ and the (BD)-species is \emph{colder}
  with $k^{(BD)}_{fs} \simeq 14.12\,(\mathrm{Mpc})^{-1}$ and its transfer
  function features a longer plateau for $k \lesssim k_{fs}$ as a consequence of the ISW
  enhancement during stage I.

Although the power spectra is strongly suppressed by free streaming at the scales at
which (WDM) acoustic oscillations emerge, we \emph{conjecture} that non-linear gravitational
collapse \emph{may} amplify these oscillations into peaks and troughs in the matter distribution
 at small scales, leading to clumpiness on mass scales $\sim 10^{9} M_{\odot}$ for
 (DW) and $\sim 10^{8} M_{\odot}$ for (BD). Perhaps coincidentally this latter scale
 is of the order of the mass contained within a \emph{half-light radius} in the (DM) halos of spiral,
low surface brightness and dwarf spheroidal galaxies\cite{stacey}.

Our study also reveals a \emph{quasi-degeneracy} between the mass, properties of the
distribution function and decoupling temperature of the (WDM) candidate: particles
with the same mass but that decoupled at different temperature with very different
distribution functions may yield similar power spectra in a wide range of scales.
 As an example of this (quasi) degeneracy, the power spectra of (DW) sterile
 neutrinos with $m \sim 2 \,\mathrm{keV}$ is similar to that of a (BD) sterile
 neutrino with $m \sim 1 \,\mathrm{keV}$ for $k \lesssim 12-15\,(\mathrm{Mpc})^{-1}$.
  This result suggests caveats on the constraints on the mass of sterile
  neutrinos from current (WDM) N-body simulations and Lyman-$\alpha$ forest data that
  typically input the distribution functions of thermal or (DW) species.

  We have compared  the results for the  transfer function for sterile neutrinos produced via the (DW) mechanism from the semi-analytic formulation presented here to the results obtained in refs.\cite{hansendwdm,vieldwdm,abadwdm} from the Boltzmann codes. Although we recognized several caveats in the comparison, we find excellent agreement to $< 5\%$  between the results from the Born approximation (\ref{TbarBA}) and the    the numerical fit to the result of Boltzmann codes presented in ref.\cite{vieldwdm} in the region of scales where the fit is valid.

The next step of the program will explore a numerical solution of
the full Gilbert equation (\ref{PQeq}) along with its comparison to the
Born approximation and   will be reported elsewhere.

\acknowledgements DB and JW are supported by NSF grant award
 PHY-0852497. JW thanks support through   Daniels and Mellon Fellowships.

 \appendix
 \section{Simplification of $I_1$}

 It is convenient to introduce the variables \be x =
 \frac{k\eta}{\sqrt{3}}~~;~~ \varpi(k) =
 \frac{4\sqrt{2}}{\sqrt{3}}\,\frac{k}{k_{eq}} \gg 1
 \label{varia}\ee and change integration variable from $u'$ to $\eta$
 using eqns. (\ref{udef},\ref{taofeta})), yielding
 \bea I_1 & = &  -6 I_a \\ I_a & = & \frac{1}{\alpha y} \int^{x_{eq}}_{x_{NR}} f(x)
 \frac{d}{dx}\Big(\frac{\sin(x)}{x}\Big) \sin\Big[\alpha\,y\big(U-\frac{1}{2}\ln(x)+\frac{1}{2}\ln[f(x)]\big)\Big]
 \Big]  dx  \label{I11}\eea where \be U= u+\frac{1}{2}\ln[
 \varpi(k)]~~;~~f(x) = 1+\frac{x}{\varpi(k)} \ee \be x_{eq} =
 \frac{k\,\eta_{eq}}{\sqrt{3}} \simeq 69.3\,k\,
 (\mathrm{Mpc})~~;~~x_{NR} = \frac{\kappa}{2\sqrt{3}} \ee
 Integrating by parts and neglecting terms $\propto 1/\varpi(k) \ll
 1$ we find
 \bea I_a  & = & \frac{1}{2}\big[1+\sqrt{2} \big]
 \frac{\sin(x_{eq})}{x_{eq}\alpha \,y}\sin\Big[\alpha\,y\big(u+0.787)\big)\Big]
  \nonumber \\ & - &
  \frac{\sin(x_{NR})}{x_{NR}\alpha \,y}\sin\Big[\alpha\,y\Big(u-\frac{1}{2}\ln\Big(\frac{\langle V^2_{eq}\rangle^\frac{1}{2}}{4} \Big)\Big)\Big]
  \nonumber \\ & + & \frac{1}{2}\int^{x_{eq}}_{x_{NR}}
  \frac{\sin(x)}{x^2}\,
  \cos\Big[\alpha\,y\big(U-\frac{1}{2}\ln(x)\big)\Big]dx \,.
  \label{Ia1}\eea In the second and third line in the above
  expression we have approximated $f(x) \sim 1$ since
  $x_{NR}/\varpi(k) \sim \langle V^2_{eq}\rangle^\frac{1}{2} \ll 1$
  and the contribution from the upper limit to the integral in the
  third line (the region where $x/\varpi(k) \sim 1$ ) is suppressed
  by $\sim 1/x^2_{eq} \sim k^2_{eq}/k^2$. It is convenient to
  extract the singular term $\propto 1/x $ as $x\sim x_{NR}$ when
  $x_{NR} \ll 1$ (this is the CDM limit), integrating by parts
  again, leading to \bea I_a  & = & \frac{1}{2}\big[1+\sqrt{2}
  \big]\,
 \frac{\sin(x_{eq})}{x_{eq}\,\alpha
 \,y}\sin\Big[\alpha\,y\big(u+0.787)\big)\Big] -\frac{\sin\big[\alpha\,y\,U \big]}{\alpha\,y}
  \nonumber \\ & + &
 \Big[1- \frac{\sin(x_{NR})}{x_{NR}}\Big]\frac{1}{\alpha \,y}\,\sin\Big[\alpha\,y\Big(u-\frac{1}{2}\ln\Big(\frac{\langle V^2_{eq}\rangle^\frac{1}{2}}{4} \Big)\Big)\Big]
  \nonumber \\ & + & \frac{1}{2}\int^{x_{eq}}_{1} \frac{\sin(x)}{x^2}\,
  \cos\Big[\alpha\,y\big(U-\frac{1}{2}\ln(x)\big)\Big]dx \nonumber
  \\ & - & \frac{1}{2}\int^{x_{NR}}_{1} \frac{[\sin(x)-x]}{x^2}\,
  \cos\Big[\alpha\,y\big(U-\frac{1}{2}\ln(x)\big)\Big]dx  \label{Ia2}\eea
  The CDM limit corresponds to $\alpha \rightarrow 0,
  x_{NR}\rightarrow 0 $

  \section{Numerical implementation of the Born approximation.}

  The first step in the numerical implementation is to obtain $\sqrt{\overline{y}^2}$ for the
  given distribution function of decoupled WDM particles and to input this value into the mode
  functions $Q,P$ given by eqns. (\ref{Pu}-\ref{Hz}).

  For numerical implementation, it is convenient to take the
  wavevector $k$ in units of $(\mathrm{Mpc})^{-1}$ and to write
  \bea \alpha & = & c_1 \,k ~~;~~ c_1 = 0.22 \Bigg(
  \frac{2}{g_d}\Bigg)^\frac{1}{3}\,\Bigg(
  \frac{\mathrm{keV}}{m}\Bigg) \label{c1}\\
  \kappa & = & c_2 \,k ~~;~~ c_2 = c_1 \sqrt{\overline{y}^2}
  \label{c2}\\
  c_{22} & = & \frac{c_2}{2\sqrt{3}} = 0.289\,c_2 \label{c22}\eea
  along with eqns. (\ref{Veq},\ref{uNRvalue})), lead  to
  \bea I_1[k;y;u]   = && -6\Bigg\{\frac{1}{c_1\,k \,y} \Big[1- \frac{\sin(c_{22}k)}{c_{22}k}\Big] \,
  \sin\Big[c_1\,k\,y\Big(u-\frac{1}{2}\ln\Big(\frac{\langle V^2_{eq}\rangle^\frac{1}{2}}{4}
  \Big)\Big)\Big] \nonumber \\ &&+ \frac{0.017}{k^2 \,c_1\,y} \,\sin[69.3\,k]\,\sin\Big[c_1\,k \,y\big(u+0.787)\big)\Big]
   -\frac{\sin\big[c_1\,k\,y\,\big(u+2.904+0.5\ln(k)\big)
   \big]}{c_1\,k\,y} \nonumber \\&& + \frac{1}{2}\int^{69.3\,k}_{1} \frac{\sin(x)}{x^2}\,
  \cos\Big[c_1\,y\big(u+2.904+0.5\ln(k)-0.5\ln(x)\big)\Big]dx
  \nonumber   \\&&-\frac{1}{2}\int^{c_{22}\,k}_{1} \frac{[\sin(x)-x]}{x^2}\,
   \cos\Big[c_1\,y\big(u+2.904+0.5\ln(k)-0.5\ln(x)\big)\Big]dx
     \Bigg\}\label{I1num} \eea

     \bea I^{CDM}_1[k;u] = && -6\Bigg\{  \frac{0.017}{k } \,\sin[69.3\,k]\big(u+0.787)\big)-
      \big(u+2.904+0.5\ln(k)\big)\nonumber\\+&& 0.211 - \frac{1}{2}\int^{\infty}_{69.3\,k}
      \frac{\sin(x)}{x^2}\,dx
           \Bigg\} \label{I1cdmnum}\eea The last integral term is
           $\lesssim 10^{-3}$ for $k \geq 0.2$ and can be neglected
           for small scales.

           \vspace{2mm}

           \be I_2[k;y;u] = -\frac{1}{2}\,\Bigg(\frac{d\,\ln f_0(y)}{d\ln y}
           \Bigg)\,j_0\Bigg[c_1\,k\,y\,(u-u_{NR})+0.5 \, c_2\,k \Bigg]
           \label{I2num}\ee

           \vspace{2mm}

           \be I^{CDM}_2[k; u] = -\frac{1}{2}\,\Bigg(\frac{d\,\ln f_0(y)}{d\ln y}
           \Bigg) \label{I2cdmnum}\ee

           \vspace{2mm}

           \be  I_{ISW}[k;y;u] = \frac{12}{c_2\,k}\,\int^1_0
           \frac{dt}{t^2} \Bigg[\cos\big(c_{22}\,k\,t\big)-\frac{\sin\big(c_{22}\,k\,t\big)}{\big(c_{22}\,k\,t\big)} \Bigg]
           \,j_1\Big[c_1\,k\,y(u-u_{NR})+0.5\,c_2(1-t) \Big]
           \label{I3num}\ee

           \vspace{2mm}

           \be I^{CDM}_{ISW} =0 \ee

\end{document}